\documentclass[11pt]{article}

\usepackage{jcapmod}
\usepackage{booktabs}
\usepackage[english]{babel}
\usepackage{amsmath,amssymb,amsbsy,amstext,amsthm,simplewick}
\usepackage{multirow}
\usepackage{hyperref}
\usepackage{epsfig,multicol,bbm,simplewick}
\usepackage{graphicx}
\usepackage{amsfonts}
\usepackage{amssymb}
\usepackage{upgreek}
\usepackage{framed}
\usepackage{enumitem}
\usepackage{bm}
\usepackage{braket}
\usepackage{wrapfig}
\usepackage{graphicx}
\usepackage{amsfonts}
\usepackage{upgreek}
\usepackage{exscale,relsize}
\usepackage[makeroom]{cancel}
\usepackage{soul}
\usepackage{bbold}
\usepackage{lipsum}
\usepackage{mdframed}
\usepackage{mathtools}
\usepackage[export]{adjustbox}
\usepackage{colortbl}

\usepackage{subcaption}
\usepackage{floatrow}
\newfloatcommand{capbtabbox}{table}[][\FBwidth]

\newcommand{\circled}[2][]{%
\tikz[baseline=(char.base)]{%
\node[shape = circle, draw, inner sep = 1pt]
    (char) {\phantom{\ifblank{#1}{#2}{#1}}};%
\node at (char.center) {\makebox[0pt][c]{#2}};}}
\robustify{\circled}
\newcommand*\diff{\mathop{}\!\mathrm{d}}
\newcommand{\overbar}[1]{\mkern 1.5mu\overline{\mkern-1.5mu#1\mkern-1.5mu}\mkern 1.5mu}

\definecolor{lightgreen}{cmyk}{0.2, 0, 0.2, 0.2}
\definecolor{lightgray}{cmyk}{0.1,0.2,0,0.1}
\definecolor{lightgray2}{cmyk}{0.1,0.1,0,0.1}
\makeatletter
\newlength{\apb@width}
\newcommand{\autoparbox}[2][c]{\settowidth{\apb@width}{#2}\parbox[#1]{\apb@width}{#2}}

\newcommand{\Cen}[2]{%
  \ifmeasuring@
    #2%
  \else
    \makebox[\ifcase\expandafter #1\maxcolumn@widths\fi]{$\displaystyle#2$}%
  \fi
}
\makeatother
\newcommand{\beq}{\begin{equation}\begin{aligned}}
\newcommand{\eeq}{\end{aligned}\end{equation}}

\numberwithin{equation}{section}
\def\beq{\begin{equation}}
\def\eeq{\end{equation}}
\def\Beq{\begin{equation}\begin{aligned}}
\def\Eeq{\end{aligned}\end{equation}}
\def\bea{\begin{eqnarray}}
\def\eea{\end{eqnarray}}

\def\d{{\rm d}}

\def\beq{\begin{equation}}
\def\eeq{\end{equation}}
\def\bea{\begin{eqnarray}}
\def\eea{\end{eqnarray}}

\def\d{{\rm d}}

\def\d{{\rm d}}

\def\bp{\boldsymbol{p}}

\newcommand\kk{{\bm{k}}}
\newcommand\qq{{\bm{q}}}

\definecolor{darkviolet}{rgb}{0.58, 0.0, 0.83}

\def\eq#1{{(\ref{#1})}}

\newenvironment{bottompar}{\par\vspace*{\fill}}{\clearpage}

\newcommand*{\xdash}[1][3em]{\rule[0.5ex]{#1}{0.55pt}}

\DeclareRobustCommand{\SkipTocEntry}[4]{}
\DeclareSymbolFont{extraup}{U}{zavm}{m}{n}
\DeclareMathSymbol{\varheart}{\mathalpha}{extraup}{86}
\DeclareMathSymbol{\vardiamond}{\mathalpha}{extraup}{87}
\DeclareMathSymbol{\test}{\mathalpha}{extraup}{88}
\definecolor{darkblue}{rgb}{0.15,0.35,0.75}
\definecolor{redd}{rgb}{0.1, 0.1, 0.1}
\definecolor{vdev}{cmyk}{0.9,0,0.8,0.1}
\usepackage[linktocpage=true]{hyperref}
\hypersetup{pageanchor=false,
colorlinks=true,
citecolor=darkblue,
linkcolor=redd,
urlcolor=darkblue,
pdfauthor={},
pdftitle={},
pdfsubject={}
}

\definecolor{gbcolor}{rgb}{.1,.1,.8}
 
\definecolor{gbcolor2}{rgb}{.4,.2,.6}

\begin{document}

\begin{titlepage}

\setcounter{page}{1} \baselineskip=15.5pt \thispagestyle{empty}

\bigskip\

IFT-UAM/CSIC-22-88  \hfill DESY-22-129 

\vspace{0.3cm}
\begin{center}

{\fontsize{20.74}{24}\selectfont  
\bfseries  
Primordial black holes and gravitational waves\\ \vspace{0.15cm} from dissipation during inflation}

\end{center}

\vspace{0.2cm}

\begin{center}
{\fontsize{12}{30} \bf   Guillermo Ballesteros$^{1,2}$, Marcos A.~G.~Garc\'ia$^{3}$,\\ Alejandro P\'erez Rodr\'iguez$^{1,2}$, Mathias Pierre$^{4}$, Juli\'an Rey$^{1,2}$}
\setcounter{footnote}{0}
\end{center}

\begin{center}

\vskip 7pt 
\textsl{$^1$ Departamento de F\'{\i}sica Te\'{o}rica, Universidad Aut\'{o}noma de Madrid (UAM), \\Campus de Cantoblanco, 28049 Madrid, Spain}\\
\textsl{$^2$ Instituto de F\'{\i}sica Te\'{o}rica (IFT) UAM-CSIC,  Campus de Cantoblanco, 28049 Madrid, Spain}\\
\textsl{$^3$ Departamento de F\'isica Te\'orica, Instituto de F\'isica, Universidad Nacional Aut\'onoma de M\'exico, Ciudad de M\'exico C.P. 04510, Mexico}\\
\textsl{$^4$ Deutsches Elektronen-Synchrotron DESY, Notkestr. 85, 22607 Hamburg, Germany}
\vskip 7pt

\end{center}

\vspace{0.3cm}
\centerline{\bf Abstract}
\vspace{0.3cm}

\noindent We study the generation of a localized peak in the primordial spectrum of curvature perturbations from a transient dissipative phase during inflation, leading to a large population of primordial black holes. The enhancement of the power spectrum occurs due to stochastic thermal noise sourcing curvature fluctuations. We solve the stochastic system of Einstein equations for many realizations of the noise and obtain the distribution for the curvature power spectrum. We then propose a method to find its expectation value using a deterministic system of differential equations. In addition, we find a single stochastic equation whose analytic solution  helps to understand the main features of the spectrum. Finally, we derive a complete expression and a numerical estimate for the energy density of the stochastic background of gravitational waves induced at second order in perturbation theory. This includes the gravitational waves induced during inflation, during the subsequent radiation epoch and their mixing. Our scenario provides a novel way of generating primordial black hole dark matter with a peaked mass distribution and a detectable stochastic background of gravitational waves from inflation.

\begin{bottompar}
\noindent\xdash[15em]\\
\small{
guillermo.ballesteros@uam.es\\
marcos.garcia@fisica.unam.mx\\
alejandro.perezrodriguez@uam.es\\
mathias.pierre@desy.de\\
julian.rey@uam.es}
\end{bottompar}

\end{titlepage}

\hypersetup{pageanchor=true}

\tableofcontents

\newpage

\section{Introduction}

Primordial black holes (PBHs) with masses $10^{-16}$ to $10^{-12}$ times lighter than the Sun may constitute the totality of the dark matter of the Universe \cite{Carr_2010,Niikura:2017zjd,Katz:2018zrn,Montero-Camacho:2019jte,Green:2020jor}. Several mechanisms have been proposed to explain how such objects could have formed in the early Universe. The most popular among them is the gravitational collapse, during the radiation epoch, of Hubble-sized regions with a density contrast above an $\mathcal{O}(1)$ threshold. These regions can originate from large curvature fluctuations produced during inflation. In the simplest approximation, which assumes that these fluctuations are Gaussian, the cosmological abundance of PBHs is exponentially sensitive to the primordial power spectrum of curvature fluctuations, $\mathcal{P_R}$. Within this approximation, it is required that $\mathcal{P_R}\sim\mathcal{O}(10^{-2})$ for PBHs to be all the dark matter. Interestingly, if such a large value of $\mathcal{P_R}$ is reached at the comoving scales that correspond to the aforementioned PBH mass range, a stochastic background of gravitational waves (GWs) with frequency approximately peaking between $10^{-3}$~Hz and $10$~Hz  is also produced \cite{Saito:2009jt}. This range contains the frequencies at which LISA (to be launched in 2037) and other space-based interferometers (currently being discussed) such as DECIGO and BBO, are planned to have their best sensitivities, offering an indirect handle on the possible existence of an abundant PBH population \cite{Cai:2018dig,Bartolo:2018evs}.

In this work we discuss the generation of a large $\mathcal{P_R}$, and its associated stochastic background of GWs, from a transient dissipative phase during inflation. This mechanism of producing PBHs from large density fluctuations of inflationary origin is fundamentally different from the others that have been previously proposed. The most popular among them assumes a brief phase of ultra slow-roll, during which the acceleration of the inflaton is rather abruptly diminished, {leading to a temporarily-growing} super-horizon curvature mode, see e.g.\ \cite{Ballesteros:2017fsr,Ballesteros:2020qam}. In our case, the effect of dissipation on the homogeneous inflaton dynamics can be described under adequate conditions by introducing a dissipative coefficient $\Gamma$ in its time evolution: 
\begin{align}
\ddot\phi +(3H +\Gamma)\dot\phi +V_\phi =0\,,
\end{align}
where $\phi$ denotes the inflaton field, $H$ (which is approximately constant) is the Hubble function describing the growth of the scale factor during inflation and $V_\phi$ is the first derivative of the inflaton potential (with respect to the inflaton field). During ultra slow-roll (with $\Gamma =0$) the equation above becomes $\ddot\phi +3H \dot \phi\simeq 0$. Instead, during a strongly dissipative regime ($\Gamma\gg H$) it reads $\ddot\phi +\Gamma\dot\phi\simeq 0$. One may naively think that both regimes are analogous and the deceleration of the inflaton must have similar effects on $\mathcal{P_R}$ in the two cases, enhancing it at specific scales. This intuition is flawed. 
In the absence of other effects, the presence of $\Gamma$ alone in the equations for the fluctuations makes the primordial spectrum of curvature perturbations {\it decrease}  \cite{LopezNacir:2011kk}. However, an enhanced $\mathcal{P_R}$ can arise in a strongly dissipative regime due to a stochastic source for the fluctuations coming from a thermal bath originating from the couplings between the inflaton and other fields that are also responsible for $\Gamma$.

The framework we use to describe the dynamics of dissipation during inflation is essentially the one of warm inflation \cite{Berera:1995ie}. However, whereas in warm inflation dissipation is active throughout the complete duration of inflation (which ends when the energy density of the radiation bath overcomes that of the inflaton), we instead assume a brief duration ($\sim\mathcal{O}(1)$ $e$-folds) for the dissipative phase. PBH production in warm inflation has been considered in \cite{Arya:2019wck,Bastero-Gil:2021fac}.
 In those models the primordial power spectrum $\mathcal{P_R}(k)$ appears to grow towards the smallest comoving wavelengths ($\sim 1/k$) exiting the horizon at the end of inflation and re-entering shortly after. That leads to very light PBHs (below $\sim 10^{-27}$ Solar masses) that cannot account for the dark matter of the Universe regardless of their abundance, as they would have evaporated by now due to Hawking radiation. Given that the mass of PBHs formed during radiation domination from the collapse of large overdensities scales like $k^{-2}$, PBHs in the appropriate window for dark matter may be obtained provided that dissipation occurs only at the right time during inflation, i.e.\ around $\sim$ 30 $e$-folds after CMB scales become super-horizon, producing a primordial spectrum with a well defined peak at $k \sim 10^{14}$~Mpc$^{-1}$.

We model our scenario using a phenomenological approach, assuming a peaked $\Gamma$ as a function of the inflaton background field {(or, equivalently, any other clock)}, which is also proportional to the third power of the temperature of the radiation bath (as it is common in warm inflation). We provide an analytical understanding of the features of the primordial spectrum and solve numerically the stochastic differential equations for the perturbations of the inflaton field, the metric and the radiation density, using two different methods. One of them consists in a Langevin approach, in which we integrate the system of stochastic linearized	 equations for multiple realizations of the thermal noise, obtaining a histogram for $\mathcal{P_R}(k)$ at each $k$. The other method reduces the stochastic system of differential equations to a deterministic one by focusing on the computation of the relevant two-point functions. We refer to this second method as the matrix formalism. The results of both methods agree with high precision and are in broad agreement with the analytical understanding of the system. Both numerical methods, which we discuss in detail, can be useful in other scenarios in which dissipation (and hence stochasticity) occurs during inflation, including in particular the case of ``standard'' warm inflation. 

Having obtained the primordial power spectrum of curvature fluctuations, we then compute the stochastic background of GWs induced at second order in cosmological perturbation theory.  We take into account not only the GWs generated during the radiation phase that we assume follows after inflation, but also the GWs sourced at second order during inflation itself (which happen to be subdominant). There is also an interference term between both contributions, whose expression we derive analytically and whose value we estimate numerically. 

We start in the next section by presenting the key features of the framework and the scenario we consider, deferring to an appendix the introduction of a toy model that can lead to a peaked $\Gamma$ like the one we use in our analysis. We leave for future work a more detailed search for concrete Lagrangians.

Throughout the paper we set $c=\hbar=k_{\rm B}=1$.

\section{Dissipation during inflation}

\begin{figure}[t]
\begin{center}
$
\includegraphics[width=.32\textwidth]{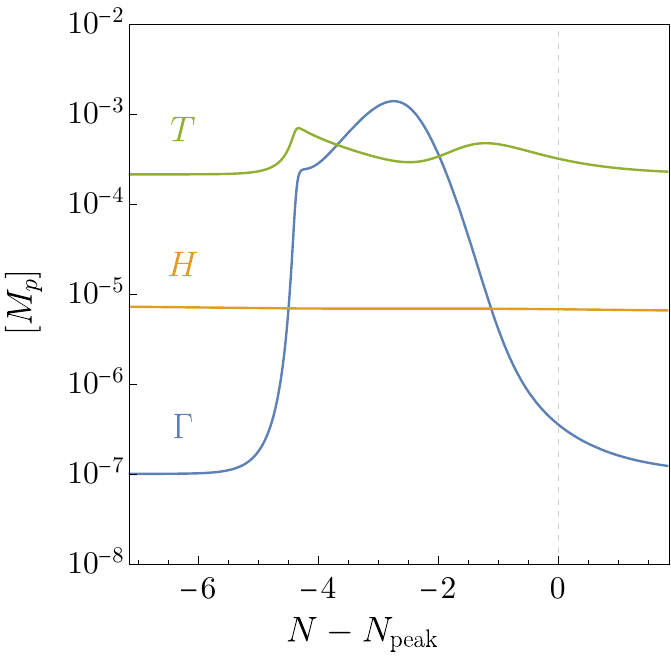}\quad\includegraphics[width=.32\textwidth]{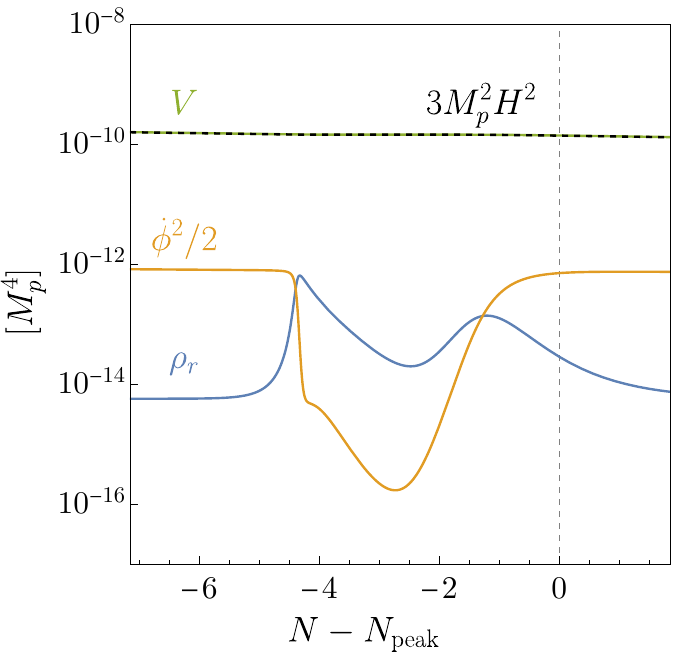}\quad
\includegraphics[width=.30\textwidth]{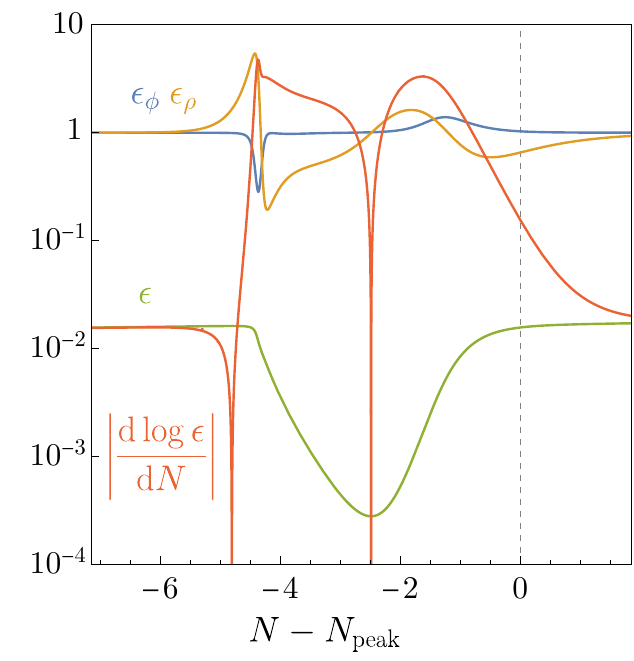}
$
\caption{\em \label{fig:bgm} {\bf Left}: $T$, $H$ and $\Gamma$ in units of $M_p$ and as functions of the number of $e$-folds $N$, near the value $N=N_{\rm peak}$ which can be approximately associated to the comoving wavenumber $k_{\rm peak}$ at which the peak in the primordial curvature spectrum (shown in Fig.~\ref{fig:PR_stochastic}) occurs. The analytical expression for $\Gamma$ as a function of the inflaton background field is given in Eq.~\eq{eq:gammapheno}. The parameters chosen for this benchmark example are given in Eq.~(\ref{eq:benchpars}). {\bf Center}: The inflaton potential $V=\lambda \phi^4/4	$ (green), its kinetic energy density, the radiation density and $3M_p^2H^2 = V+\dot\phi^2/2+\rho_r$ (black dashed). {\bf Right}: The slow-roll parameter $\epsilon$, its derivative, and the ratios $\epsilon_\phi$ and $\epsilon_\rho$ defined in Eq.~\eq{eq:attpars}.}
\end{center}
\end{figure}

We assume that for a period of time during inflation the Universe contains a non-negligible thermalized radiation component with energy density $\rho_r$ and temperature $T$ related by 
\begin{align}
\rho_r=\frac{\pi^2}{30}g_\star T^4\,, 
\label{eq:rhor}
\end{align}
where $g_\star$ quantifies the effective number of relativistic degrees of freedom. The inflaton field $\phi$ acts as the source for this thermal component. At the background level, this dissipation is assumed to be adiabatic, 	according to the following equations: 
\begin{align} \label{eq:backg1}
\ddot{\phi}+(3H+\Gamma)\dot{\phi}+V_\phi&=0,\\ \label{eq:backg2}
\dot{\rho}_r+4H\rho_r&=\Gamma\dot{\phi}^2,\\ \label{eq:backg3}
\rho_r+V+\frac{1}{2}\dot{\phi}^2&=3M_p^2H^2\,,
\end{align}
which satisfy the conservation of the total energy-momentum tensor. The dots denote derivatives with respect to cosmic time ($\cdot\equiv\d/\d t$) and $M_p \simeq 2.45 \times 10^{18}~\text{GeV}$ is the reduced Planck mass. As long as friction eventually comes to dominate the dynamics for sufficient time, the initial conditions for these equations are irrelevant due to the presence of an attractor characterized by the ratios
\begin{equation}
\label{eq:attpars}
\epsilon_\phi\equiv-\frac{V_\phi}{(3H+\Gamma)\dot{\phi}}\simeq 1, \qquad \epsilon_\rho\equiv\frac{\Gamma\dot{\phi}^2}{4H\rho_r}\simeq 1\,.
\end{equation}
We show these ratios on the right panel of Fig.~\ref{fig:bgm}, together with the slow-roll parameters
\begin{equation}
\label{eq:srpars}
\epsilon\equiv -\frac{\d\log H}{\d N}\,\quad \text{and}\quad \frac{\d\log\epsilon}{\d N}\,,
\end{equation}
where 
\begin{align}
N = \int H {\rm d}t
\end{align}
defines the number of $e$-folds of inflation. The figure corresponds to the benchmark example that we introduce in Section~\ref{sec:pheno}. The end of inflation, defined by $\epsilon =1$, would typically occur for a number of  $e$-folds after the peak scale crosses the horizon of $N-N_\text{peak} \sim 25$. For this example, the dissipation coefficient $\Gamma$ transiently grows in value for $\sim 4$ $e$-folds, becoming even $\Gamma \gg H$, as shown in the left panel of the same figure. This suppresses the kinetic energy of $\phi$ while it enhances the radiation energy density (see middle panel).

In order to describe the dynamics of cosmological fluctuations we use the Newtonian gauge,
\begin{equation}
\d s^2=-(1+2\psi)\d t^2+a^2(1-2\psi)\delta_{ij}\d x^i\d x^j,
\end{equation} 
where the two scalar perturbations of the metric are identical (and denoted as $\psi$) by virtue of one of Einstein's equations and the absence of anisotropic stress. 
We denote by $\delta\phi$ and $\delta \rho_r$ the perturbations of the inflaton field and the radiation energy density, respectively. Defining the momentum fluctuation
\begin{equation}
\delta q_r=\frac{4}{3}\rho_r\delta v_r,
\end{equation}
where $\delta v_r$ is the velocity perturbation of the radiation, the remaining Einstein's equations in Fourier space and at linear order are (see e.g.\ 	\cite{Ma:1995ey})
\begin{align} \displaybreak[0]
3H(\dot{\psi}+H\psi)+\frac{k^2}{a^2}\psi&=-\frac{1}{2M_p^2}\bigg[\delta\rho_r+\dot{\phi}(\delta\dot{\phi}-\dot{\phi}\psi)+V_\phi\delta\phi\bigg],\\ \displaybreak[0]
\label{eq:psieom}
\dot{\psi}+H\psi&=-\frac{1}{2M_p^2}\left(\delta q_r-\dot{\phi}\delta\phi\right),\\
\ddot{\psi}+4H\dot{\psi}+(2\dot{H}+3H^2)\psi&=\frac{1}{2M_p^2}\bigg[\frac{1}{3}\delta\rho_r+\dot{\phi}(\delta\dot{\phi}-\dot{\phi}\psi)-	V_\phi\delta\phi\bigg].
\end{align}

The conservation of the  energy-momentum tensor $\nabla_{\mu}T^{\mu\nu}=\nabla_{\mu}T^{\mu\nu}_{(\phi)}+\nabla_{\mu}T^{\mu\nu}_{(r)} =0$ is satisfied in agreement with the continuity equations for $\phi$ and $\rho_r$ \cite{Kodama:1984ziu}:
\begin{align}	\label{eq:conteqs}
\nabla_{\mu}T^{\mu\nu}_{(\phi)} & = Q^{\nu}\,,\\ 
\nabla_{\mu}T^{\mu\nu}_{(r)} & = -Q^{\nu}\,,
\end{align}
where $Q^{\mu}$ contains a stochastic piece induced by thermal fluctuations, whose form is determined by the so-called fluctuation-dissipation theorem (see e.g.~\cite{Gleiser:1993ea,Bastero-Gil:2014jsa} {and Appendix \ref{app:FD}}), 
\beq
\label{fdf}
Q_{\mu} = -\Gamma\, u^{\nu}\nabla_{\nu}\phi\,\nabla_{\mu}\phi + \sqrt{\frac{2\Gamma T}{a^3}}\xi_t \nabla_{\mu}\phi.
\eeq 
In this expression $u^{\nu}$ denotes the 4-velocity of the radiation component and $\xi_t\equiv \d W_t/\d t$, where $\d W_t$ is a Wiener increment\footnote{See Appendix \ref{app:Diffeqs} for the definition of a Wiener process and a quick review on stochastic differential equations.} satisfying $\langle \xi_t({\bm x})\xi_{t'}({\bm x}')\rangle_{\rm S}=\delta^{(3)}({\bm x}-{\bm x}')\delta(t-t')$. Here, $\langle\cdots\rangle_{\rm S}$ denotes a stochastic average over different realizations. The linearized equations for $\delta\phi$, $\delta\rho_r$	and $\delta v_r$ are (see e.g.\,\cite{Bastero-Gil:2014jsa})
{\fontsize{10.4}{\baselineskip}
\begin{align}
\label{eq:phieom}
\delta\ddot{\phi}+(3H+\Gamma)\delta\dot{\phi}+\left(\frac{k^2}{a^2}+V_{\phi\phi}+\dot{\phi}\Gamma_\phi\right)\delta\phi+\Gamma_T\frac{\dot{\phi}T}{4\rho_r}\delta\rho_r-4\dot{\psi}\dot{\phi}+(2V_\phi+\Gamma\dot{\phi})\psi&=\sqrt{\frac{2\Gamma T}{a^3}}\,\xi_t,\\
\label{eq:rhoeom}
\delta\dot{\rho}_r+\bigg(4H-\Gamma_T\frac{\dot{\phi}^2T}{4\rho_r}\bigg)\delta\rho_r-\frac{k^2}{a^2}\delta q_r+\Gamma\dot{\phi}^2\psi-4\rho_r\dot{\psi}-(\Gamma_\phi\delta\phi+2\Gamma\delta\dot{\phi})\dot{\phi}&=-\sqrt{\frac{2\Gamma T}{a^3}}	\dot{\phi}\,\xi_t,\\ 
\label{eq:qeom}
\delta\dot{q}_r+\frac{4}{3}\rho_r\psi+3H\delta q_r+\frac{1}{3}\delta\rho_r+\Gamma\dot{\phi}\delta\phi&=0,
\end{align}}
where $\Gamma_T \equiv \partial\Gamma/\partial T$.

Eqs.~(\ref{eq:phieom})--(\ref{eq:qeom}), together with one of Einstein's equations, for instance (\ref{eq:psieom}), form a complete set for the four variables $\delta\phi$, $\delta\rho_r$, $\delta q_r$ and $\psi$. These equations can be further simplified using the following combination of Einstein's equations,
\begin{equation}
\label{qconstraint}
\bigg(2M_p^2\frac{k^2}{a^2}-\dot{\phi}^2\bigg)\psi+\delta\rho_r+\dot{\phi}\,\delta\dot{\phi}+(V_\phi+3H\dot{\phi})\delta\phi-3H\delta q_r=0.
\end{equation}
Imposing this constraint  allows to reduce the number of equations by one, so we can eliminate, for instance, Eq.~(\ref{eq:qeom}). However, we find that not imposing this constraint can be more stable numerically, as we discuss in more detail in Section~\ref{sec:matrix} and Appendix~\ref{app:eqs}. We use the initial conditions
\begin{equation}
\label{eq:inicondsperts}
\delta q_r=0,\qquad\delta\rho_r=0,\qquad\psi=0,\qquad\delta\phi=-\frac{\dot{\phi}}{2M_p\,aH\sqrt{k\epsilon}}\exp\bigg(-ik\int\frac{\d t}{a}\bigg),
\end{equation}
where the initial condition for $\delta\phi$ comes from assuming that the field fluctuations are in the Bunch-Davies vacuum at early times. As we will show later, the choice of initial conditions is not very relevant, since the noise term leads to an attractor behaviour for the evolution of the perturbations.

\subsection{A peaked dissipative coefficient}
\label{sec:pheno}

Our main goal is the description of a peak in the curvature power spectrum arising from transient dissipation. The perturbation equations are driven by a source of noise with amplitude $\sim\sqrt{\Gamma T}$, and so they are significantly enhanced whenever $\Gamma$ is sufficiently large. If the peak of the spectrum of the curvature perturbation is localized around an adequate scale, the PBH mass function will be narrow enough so that the borders of the allowed window for PBH dark matter can be avoided. Therefore, we focus on modeling a dissipative coefficient $\Gamma$ that satisfies $\Gamma \gg H$ only for a few $e$-folds at most. Rather than building a full model of the complete inflationary history we focus on the local description of the dynamics around the relevant region. Although we remain agnostic about the details of the microphysics that gives rise to such a peaked dissipative coefficient, in Appendix~\ref{app:micro} we present a toy example of a Lagrangian with the necessary features that could potentially serve as a basis for future models (which should also fit the CMB and provide enough inflation). 

We assume the following parameterization of the dissipative coefficient
\begin{equation}
\label{eq:gammapheno}
\Gamma(\phi,T)=\frac{ T^3}{m^2+M^2\tanh^2\left[(\phi-\phi_\star)/\Lambda\right]}\,,
\end{equation}
where $m,M,\Lambda$ and $\phi_{\star}$ are free dimensionful parameters. As discussed in Appendix \ref{app:micro}, the $T^3$ dependence of $\Gamma$ arises naturally in a specific low-temperature limit, which is common in warm inflation. The temperature dependence of $\Gamma$ is not crucial for the stochastic noise to generate a peak in the primordial power spectrum. A temperature-independent $\Gamma$ that is peaked as a function of $\phi$ also produces a similar effect. However, the parameterization \eq{eq:gammapheno}  resembles more closely the actual $\Gamma$ that may be expected from a concrete Lagrangian in which $\phi$ couples to other fields.

For our benchmark example of Fig.~\ref{fig:bgm} we choose the following set of parameters:
{\fontsize{10.4}{\baselineskip}
\begin{equation}
\label{eq:benchpars} 
g_\star=8,\quad \phi_\star=22 M_p\,,\quad M=10^{-2}M_p,\quad m=1.4\times10^{-4}M_p,\quad\Lambda=0.1M_p\,,\quad\lambda=2.5\times 10^{-15}\,,
\end{equation}}
where $\lambda$ is the coupling of a quartic inflaton potential,
\begin{equation}\label{eq:quartic}
V(\phi)=\frac{\lambda}{4} \phi^4.
\end{equation}
We choose this potential for its simplicity and lack of features.~\footnote{ Notice that previous works have shown that the presence of radiation can allow inflation with a quartic potential to be compatible with Planck constraints on the tensor-to-scalar ratio and scalar index~\cite{Benetti:2016jhf}.} Since our focus is on studying the phenomenology of the dynamics generated by dissipation (at the background and perturbation levels) we could have chosen any other potential compatible with slow-roll inflation that does not have any peculiarities that would introduce spurious effects beyond those we want to analyze.  In addition, the potential $V(\phi)$ only needs to be valid a few $e$-folds before and after the region where $\Gamma\gg H$ because we are only concerned with describing the appearance of a large peak in the primordial power spectrum, which is a local feature. Nevertheless, the value of $\lambda$ is chosen in such a way that we obtain the correct amplitude for the power spectrum, $A_s\simeq 2\times 10^{-9}$ \cite{Planck:2018jri}. Our choice of parameters leads to a $\mathcal{P}_\mathcal{R}$ with a peak value of $\sim 10^{-2}$. As seen  in Fig.~\ref{fig:bgm}, the potential and the evolution of $H$ are essentially featureless, whereas the kinetic energy of the inflaton does change significantly in the relevant region. 

\begin{figure}[t]
\begin{center}
$\includegraphics[width=.48\textwidth]{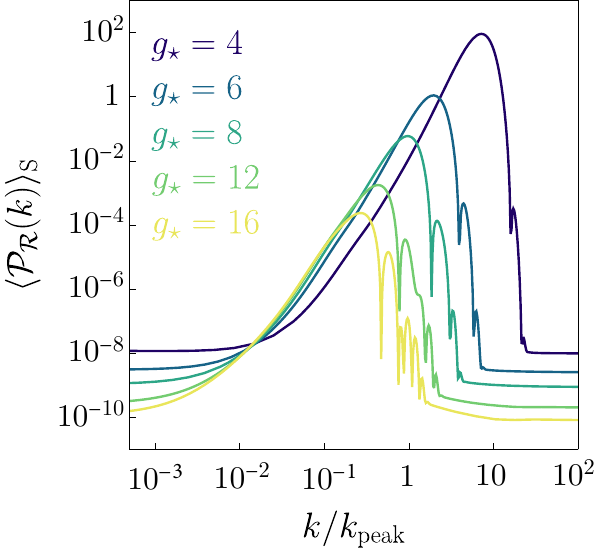}
\qquad\includegraphics[width=.48\textwidth]{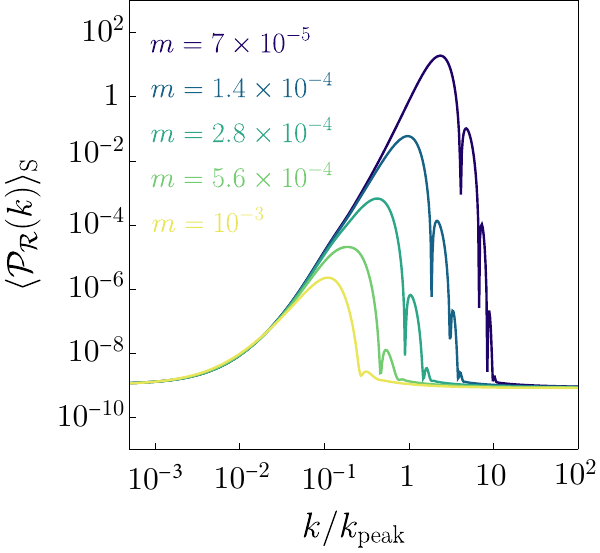}$
\caption{\em \label{fig:pars_spectrum} {Stochastic average of the power spectrum computed for the benchmark parameters of Eq.~(\ref{eq:benchpars}) by varying $g_\star$ (left panel) and $m$ in units of $M_p$ (right panel). The horizontal axis is normalized at the scale $k=k_{\rm peak}$ at which the peak in the benchmark spectrum (with $g_\star=8$ and $m=1.4\times10^{-4}M_p$) occurs.}}
\end{center}
\end{figure}

For the initial conditions of the background variables we choose 
\begin{equation}
\phi(N=0)=26 M_p,\quad\frac{\d\phi}{\d N}\bigg|_{N=0}=-\frac{2\sqrt{6}}{\phi_0}M_p^2,\quad\rho_r(N=0)=10^{-5} M_p^4,
\end{equation}
although the last two are essentially irrelevant due to the presence of the background attractor discussed in the previous section. This choice makes the background quantities converge quickly to their attractor values. The time at which the localized growth in $\Gamma$ occurs (and therefore the scale at which the peak in the power spectrum is located) can be controlled by varying $\phi_\star$. Decreasing $m$ or $M$ makes the peak of $\mathcal{P_R}$ larger. In particular, since we choose $m\ll M$, decreasing $m$ makes $\mathcal{P_R}$  increase without changing the value of $\Gamma$ far away from the wavenumbers associated to $\phi_\star$, so that $\mathcal{P_R}$ retains its normalization at small distance scales. Similarly, increasing $\Lambda$ makes $\mathcal{P_R}$ larger. Finally, decreasing $g_\star$ makes the peak of $\mathcal{P_R}$ larger. To understand why this is the case, let us determine how the coefficient in front of the thermal noise in the equation of motion for the perturbations, \eq{eq:phieom} and \eq{eq:rhoeom}, scales with $g_\star$. This can be done by isolating the temperature dependence of this quantity.
Let us define $\gamma(\phi)$ via $\Gamma(\phi,T)=\gamma(\phi)T^3$.
Then, by assuming the system is in the attractor solution (\ref{eq:attpars}) and $\Gamma\gg H$, we find
\begin{equation}
\Gamma T\propto \gamma(\phi)\bigg[\frac{1}{g_\star}\frac{V_\phi^2}{	H}\frac{1}{\gamma(\phi)}\bigg]^{4/7}.
\end{equation}
We therefore find that decreasing $g_\star$ makes the amplitude of the stochastic noise increase, thereby increasing the curvature power spectrum. The effect of varying $m$ and $g_\star$ on the spectrum is shown in Fig.~\ref{fig:pars_spectrum}.

\section{The curvature power spectrum}

In this section we present three different ways of computing the primordial power spectrum
\begin{equation}
\mathcal{P}_\mathcal{R}=\frac{k^3}{2\pi^2}|\mathcal{R}|^2\bigg|_{k\ll aH}
\label{eq:powerspectrum_curvatureperturbation}
\end{equation}
of the comoving curvature perturbation 
\begin{equation}
\label{eq:curvatureperturbation}
\mathcal{R}=\frac{H}{\rho+p}\big(\delta q_r-\dot{\phi}\delta\phi\big)-\psi,
\end{equation}
where $p$ and $\rho$ are the pressure and energy density of the total system (inflaton plus radiation).

Due to the presence of the stochastic thermal noise, our main quantity of interest is the expectation value of the power spectrum at a given comoving scale, which we denote by $\langle\mathcal{P}_\mathcal{R}(k)\rangle_{\rm S}$. The most straightforward way to compute this quantity (though not necessarily the most economical) is to determine (\ref{eq:powerspectrum_curvatureperturbation}) for a large sample of stochastic realizations, and then calculate their average. Alternatively, one can bypass this noting that $\langle\mathcal{P}_\mathcal{R}(k)\rangle_{\rm S}$ is determined by the correlation of the thermal noise, and it is therefore a deterministic quantity. The system of stochastic differential equations can then be rephrased as a deterministic system for the correlators of the scalar fluctuations, which is convenient to write in matrix form. As we show below, both approaches agree. Finally, under some approximations, the equations for the fluctuations can be solved analytically, allowing us to understand the qualitative features of the power spectrum better.

\subsection{Matrix formalism}
\label{sec:matrix}

In this section we will develop a method to find the stochastic average of the power spectrum by solving a deterministic matrix differential equation instead of the full set of stochastic differential equations. We begin by noting that the equations of motion can be written, in Fourier space, as a system of linear first-order complex stochastic differential equations. Throughout this section we will work with the number of $e$-folds as time variable and we define the `column vector'\,\footnote{The {superscript $^{\rm T}$ denotes} transpose.}
\begin{equation}
\bm \Phi	 \equiv \bigg(\psi,\delta\rho_r,\frac{\d\delta\phi}{\d N},\delta\phi\bigg)^{\rm T}.
\end{equation}
The equations of motion (\ref{eq:psieom}), (\ref{eq:phieom}), and (\ref{eq:rhoeom}) can then be conveniently written as a system of four first-order stochastic differential equations
\begin{align}
\label{sdeom}
\frac{\d \bm \Phi	}{\d N}+{\bm A}\bm \Phi	&={\bm B}\xi_N,
\end{align}
where the matrix ${\bm A}$ and the (column) vector ${\bm B}$ are real and independent of $\bm \Phi	$. Explicit expressions will be given at the end of this section. We also assume that the constraint in Eq.~(\ref{qconstraint}) has been imposed to eliminate $\delta q_r$ from the system. In this equation, $\xi_N$ denotes the Wiener increment from Eqs.~(\ref{eq:phieom}) and (\ref{eq:rhoeom}) written in terms of the number of $e$-folds\footnote{The rule for changing the time variable (in this case from cosmic time to the number of $e$-folds) in the Wiener process is derived in Appendix \ref{app:Diffeqs}.} and satisfying, in Fourier space,
\begin{equation}
\langle\xi_N({\bm k})\xi_{N'}({\bm k}')\rangle_{\rm S} = (2\pi)^3\delta(N-N')\delta^3({\bm k}-{\bm k}').
\end{equation}

We are interested in the power spectrum of the comoving curvature perturbation, which can be written as
\begin{equation}
\mathcal{R}={\bm C}^T\,\bm \Phi	,
\end{equation}
where the vector ${\bm C}$ can be read from Eq.~(\ref{eq:curvatureperturbation}) (see Eq.~(\ref{eq:BandC})). The corresponding power spectrum, averaged over stochastic realizations, can be expressed in terms of the correlation function matrix $\langle \bm \Phi	\bm \Phi^\dagger\rangle_{\rm S}$ as\footnote{The symbol $^\dagger$ denotes adjoint.}
\begin{equation}
\label{eq:pspec_projection}
\langle \mathcal{P}_\mathcal{R} \rangle_{\rm S} =\frac{k^3}{2\pi^2}{\bm C}^{\rm T}\langle \bm \Phi	\bm \Phi^\dagger\rangle_{\rm S}{\bm C}\bigg|_{k\ll aH}.
\end{equation}

It makes no difference whether we work with the real and imaginary parts of $\bm \Phi	$, or with $\bm \Phi	$ and its complex conjugate $\bm \Phi^\star$. We now choose the latter option. The probability density $P(\bm \Phi	,\bm \Phi^\star,N)$ for the system to be in state $\left\{\bm \Phi	,\bm \Phi^\star\right\}$ at time $N$ can be obtained by solving the Fokker-Planck equation\footnote{Note that the probability density $P$ is a function of two variables ($\bm \Phi	$ and $\bm \Phi^\star)$ which do not obey independent equations of motion (since the noises $\xi_N$ and $\xi_N^\star$ are correlated), so the fact that we can use the Fokker-Planck equation in its canonical form is not obvious. A derivation is performed in Appendix \ref{app:Diffeqs}.}
\begin{equation} \label{FPe}
\frac{\partial P}{\partial N}=\sum_{k\ell}\bigg[{\bm A}_{k\ell}\frac{\partial}{\partial \bm \Phi	_k}(\bm \Phi_{\ell} P)+{\bm A}_{k\ell}\frac{\partial}{\partial \bm \Phi^\star_k}(\bm \Phi^\star_{\ell} P)+({\bm B}{\bm B}^{\rm T})_{k\ell}\frac{\partial^2P}{\partial \bm \Phi_k\partial \bm \Phi	^\star_\ell}\bigg],
\end{equation}
where the two-point statistical moments are defined as
\begin{equation}
\label{moments}
{\bm Q}\equiv\langle \bm \Phi	\bm \Phi^\dagger\rangle_{\rm S}(N)\equiv \int \prod_i \diff \bm \Phi_i\, \int  \prod_j	\diff \bm \Phi^\star_j  \; P(\bm \Phi,\bm \Phi^\star,N)\;\bm \Phi	\bm \Phi^\dagger\;
\end{equation}

The equation of motion for 
${\bm Q}$
can be found differentiating the previous expression with respect to time and using the Fokker-Planck equation.\footnote{It is also necessary to integrate by parts and use the fact that the probability distribution vanishes on the integration boundaries.} The resulting deterministic differential equation for the matrix ${\bm Q}$ is
\begin{equation}
\label{matrix_eq}
\frac{\d{\bm Q}}{\d N}=-{\bm A}{\bm Q}-{\bm Q}{\bm A}^{\rm T}+{\bm B}{\bm B}^{\rm T}.
\end{equation}
By solving this deterministic differential equation we can bypass solving the full system of stochastic differential equations for the perturbations as long as we are only interested in the stochastic average of the power spectrum, which is given by Eq.~\eq{eq:pspec_projection}.

Let us give explicit expressions for each one of the matrices used in these equations with the number of $e$-folds as the time variable. The matrix ${\bm A}$ is given by
\begin{equation}
{\bm A}=
\begin{pmatrix}
f_\psi & f_\rho & f_{\d\phi} & f_\phi \\

g_\psi+4\rho_rf_\psi & g_\rho+4\rho_rf_\rho & g_{\d\phi}+4\rho_rf_{\d\phi} & g_\phi+4\rho_rf_\phi \\

h_\psi+4(\d\phi/\d N)f_\psi & h_\rho+4(\d\phi/\d N)f_\rho & h_{\d\phi}+4(\d\phi/\d N)f_{\d\phi} & h_\phi+4(\d\phi/\d N)f_\phi \\

0 & 0 & -1 & 0 

\end{pmatrix},
\end{equation}
where
\begin{align}
f_\psi&=1+\frac{k^2}{3a^2H^2}-\frac{1}{6M_p^2}\bigg(\frac{\d\phi}{\d N}\bigg)^2, & g_\psi&=\Gamma H\bigg(\frac{\d\phi}{\d N}\bigg)^2-\frac{k^2}{3a^2} \bigg[2M_p^2\frac{k^2}{a^2H^2}-\bigg(\frac{\d\phi}{\d N}\bigg)^2\bigg],\nonumber
\\
f_\rho&=\frac{1}{6M_p^2H^2}, & g_\rho&=4-\Gamma_T\frac{HT}{4\rho_r}\bigg(\frac{\d\phi}{\d N}\bigg)^2-\frac{k^2}{3a^2H^2},\nonumber
\\
f_{\d\phi}&=\frac{1}{6M_p^2}\frac{\d\phi}{\d N}, & g_{\d\phi}&=-\left(\frac{k^2}{3a^2}	+2\Gamma H\right)\frac{\d\phi}{\d N}, \nonumber
\\
f_\phi&=\frac{V_\phi}{6M_p^2H^2}, & g_\phi&=-\frac{k^2}{3a^2H^2}\left(3H^2\frac{\d\phi}{\d N}+V_\phi\right)-H\Gamma_\phi\left(\frac{\d\phi}{\d N}\right)^2,\nonumber
\\
h_\psi&=2\frac{V_\phi}{H^2}+\frac{\Gamma}{H}\frac{\d\phi}{\d N}, & h_\rho&=\frac{T\,\Gamma_T}{4H\rho_r}\frac{\d\phi}{\d N},\nonumber
\\
h_{\d\phi}&=3+\frac{\Gamma}{H}+\frac{1}{H}\frac{\d H}{\d N}, & h_\phi&=\frac{k^2}{a^2H^2}+\frac{V_{\phi\phi}}{H^2}+\frac{\Gamma_\phi}{H}\frac{\d\phi}{\d N}.
\end{align}

The vectors ${\bm B}$ and ${\bm C}$ are
\begin{align} \label{eq:BandC}
{\bm B}&=\begin{pmatrix}
0 \\
-\sqrt{2\Gamma T H/a^3}\left(\d\phi/\d N\right) \\
\sqrt{2\Gamma T/(aH)^3} \\
0 \\
\end{pmatrix},
&
{\bm C}&=\frac{1}{3H^2\left(\frac{\d\phi}{\d N}\right)^2+4\rho_r}
\begin{pmatrix}
2M_p^2k^2/a^2-4H^2\left(\frac{\d\phi}{\d N}\right)^2-4\rho_r \\
1 \\
H^2\frac{\d\phi}{\d N} \\
V_\phi\\
\end{pmatrix}.
\end{align}
Finally, the matrix of initial conditions ${\bm Q}_i \equiv {\bm Q} (N_\text{ini})$ is, in accordance with Eq.~(\ref{eq:inicondsperts}),
\begin{equation}
\label{eq:initialmatrix}
{\bm Q}_i=
\frac{1}{2ka^2(N_\text{ini})}\begin{pmatrix}
 0 & 0 & 0 & 0\\
 0 & 0 & 0 & 0\\
 0 & 0 & 1+(k/k_i)^2 & -1-i(k/k_i)\\
 0 & 0 & -1+i(k/k_i) & 1
\end{pmatrix},
\end{equation}
where $k_i$ is the scale that crosses the horizon at some initial $e$-fold value $N_{\rm ini}$. In practice, we can start integrating at some time $N_{\rm ini}$ such that $k/k_i\simeq100$, and terminate the integration a few $e$-folds after the strong dissipative phase (in which $\Gamma\gg H$) ends and the mode being computed satisfies $k\ll a H$.

\begin{figure}[t]
\begin{center}
$\includegraphics[width=.455\textwidth]{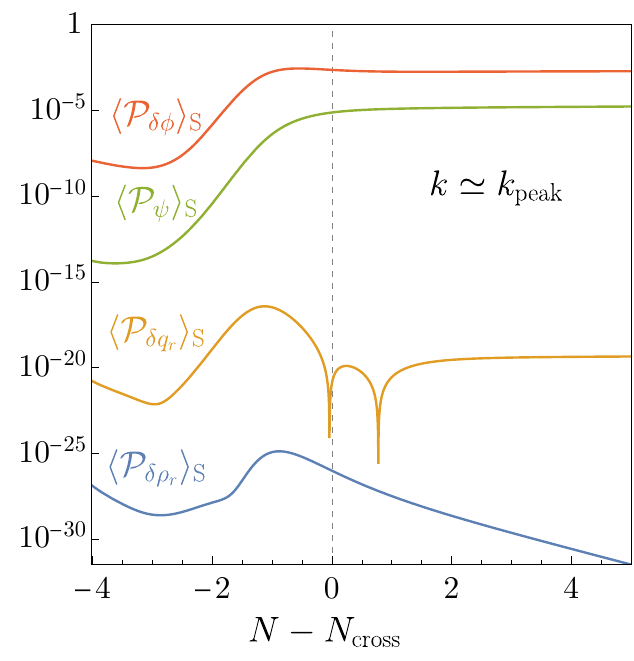}
\qquad\includegraphics[width=.48\textwidth]{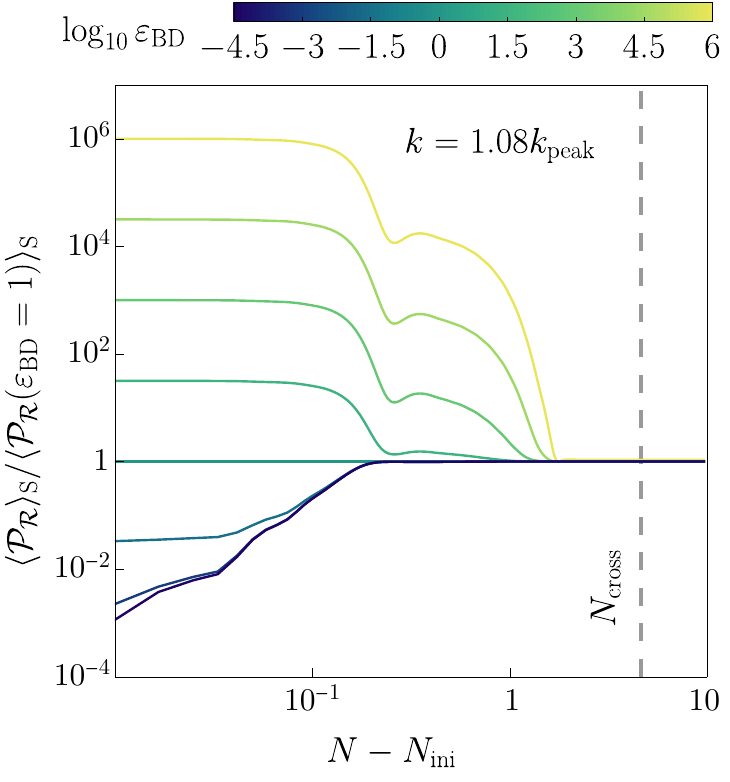}$
\caption{\em \label{fig:pertsconds} {{\bf Left:} Time evolution of the averaged power spectrum of each perturbation as a function of the number of $e$-folds. {\bf Right:} Effect of varying the initial conditions; see the discussion below Eq.~(\ref{eq:initialmatrix}). $N_{\rm cross}$ denotes the time at which the scale $k$ indicated in each panel crosses the horizon ($k = aH$). The matrix formalism of Section \ref{sec:matrix} has been used for both panels.}}
\end{center}
\end{figure}

It is worth stressing that since the stochastic source ends up dominating the evolution of the perturbations, the choice of initial conditions is actually not too relevant.
This will be made clearer in Section \ref{sec:analytical}, but for now let us illustrate it with a numerical example, redefining ${\bm Q}_i$ to model the deviation from the Bunch-Davies initial conditions by multiplying the original ${\bm Q}_i$ of \eq{eq:initialmatrix}  by some (real) number $\varepsilon_{\rm BD}$. The effect of varying $\varepsilon_{\rm BD}$ with respect to the case $\varepsilon_{\rm BD}=1$ is shown in Fig.~\ref{fig:pertsconds}. Even for very large values of this parameter, $\varepsilon_{\rm BD}\simeq 10^6$, we find that within roughly $1$ $e$-fold (and several $e$-folds before horizon crossing) the solutions converge to the same value.

As we mentioned earlier, we find that in some cases the system of differential equations is numerically more stable if we do not impose the constraint of Eq.~(\ref{qconstraint}). This gives rise to an additional equation of motion (for the variable $\delta q_r$). The matrices for this $5\times 5$ system are presented in Appendix \ref{app:eqs}. We have checked that the numerical results using either set of equations are in agreement. The power spectrum for the benchmark point of the previous section obtained by solving either system is shown as a solid line in Fig.~\ref{fig:PR_stochastic}. The evolution of the perturbations for the mode $k_{\rm peak}$ at which the power spectrum peaks is shown in Fig.~\ref{fig:pertsconds}.

\subsection{Stochastic equations}
\label{sec:stochastic}

\begin{figure}[t]
\begin{center}
$\includegraphics[width=.7\textwidth]{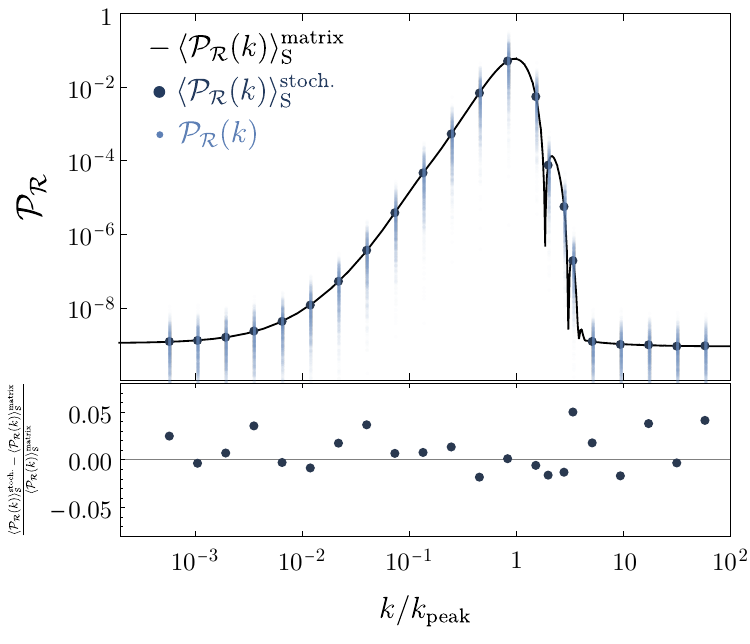}$
\caption{\em \label{fig:PR_stochastic} {\bf Top:} stochastic average of the power spectrum of the comoving curvature perturbation, $\mathcal{P_R}$, for 22 different values of the comoving wave number, $k$, (dark blue dots). The number of realizations for each value of $k$ is 2160. Each realization is represented as light blue dot. The solid black line represents the average of the power spectrum obtained via the deterministic matrix differential equation derived in Section \ref{sec:matrix}. {\bf Bottom:} Absolute value of the relative difference between the stochastic average and the matrix average of $\mathcal{P_R}$. The agreement for each $k$ is at the percent level.}
\end{center}
\end{figure}

In principle, to determine the probability distribution for the stochastic variable $\bm \Phi$, one should solve the Fokker-Planck equation \eq{FPe}. This is a rather difficult task. An alternative consists in estimating numerically the probability distribution by using a frequentist approach, i.e.\,by solving the system of Langevin equations (\ref{sdeom})
\begin{align}
\d\bm \Phi	+\boldsymbol{A}\bm \Phi	 \d N&=\frac{1}{\sqrt{2}} \boldsymbol{B}\left(\d W_N^{r}+i \d W_N^{i}\right)
\label{eq:Langevin}
\end{align}
over many different realizations, where $\d W_N^{r}\equiv\sqrt{2}{\rm Re}(\xi_N)\d N$ and $\d W_N^{i}\equiv\sqrt{2}{\rm Im}(\xi_N)\d N$ are real-valued, independent Wiener {increments.}\footnote{The factor $\sqrt{2}$ is necessary for the correlation functions of ${\rm Re}(\xi_N)$ and ${\rm Im}(\xi_N)$ to be properly normalized, as discussed in Appendix \ref{app:Diffeqs}.} This is the approach that we adopt in this section.

We impose the following initial conditions, in accordance with Eqs.~(\ref{eq:inicondsperts}) and (\ref{eq:initialmatrix})
\begin{equation}
\bm \Phi	(N_\text{ini})=\Big( 0, 0 , \dfrac{1}{a(N_\text{ini})}\dfrac{i \sqrt{k}}{\sqrt{2}k_i}+\dfrac{1}{a(N_\text{ini})}\dfrac{1}{\sqrt{2k}} ,  -\dfrac{1}{a(N_\text{ini})}\dfrac{1}{\sqrt{2k}} \Big)^{\rm T}.
\end{equation}
The limits of integration are discussed below Eq.~(\ref{eq:initialmatrix}). We solve the Langevin system with a fixed time-step Runge Kutta method.\footnote{We used Wolfram Mathematica and the It\^o Process command to simulate stochastic realizations with method ``StochasticRungeKutta".}
The convergence of the solution was checked by successively decreasing the time-step. We found that decreasing the time step below $\Delta N = 10^{-4}$ produces results for the averaged primordial spectrum that are indistinguishable at the percent level. 
The curvature perturbation and the corresponding power spectrum are determined by substituting the solution of the Langevin system into Eq.~(\ref{eq:powerspectrum_curvatureperturbation}).

Fig.~\ref{fig:PR_stochastic} shows (as light blue dots) a collection of 2160 stochastic realizations of the power spectrum for twenty different values of the wavenumber $k$. The dark blue dots represent the arithmetic average of all the realizations for each $k$. The continuous black curve, in turn, corresponds to the numerical solution of the matrix equation (\ref{matrix_eq}). The bottom panel of this same figure shows the relative difference between the frequentist approach and the matrix formalism solution. The result is a stochastic average which agrees with the matrix formalism results at the percent level.

\begin{figure}[t]
\begin{center}
$\includegraphics[width=.47\textwidth]{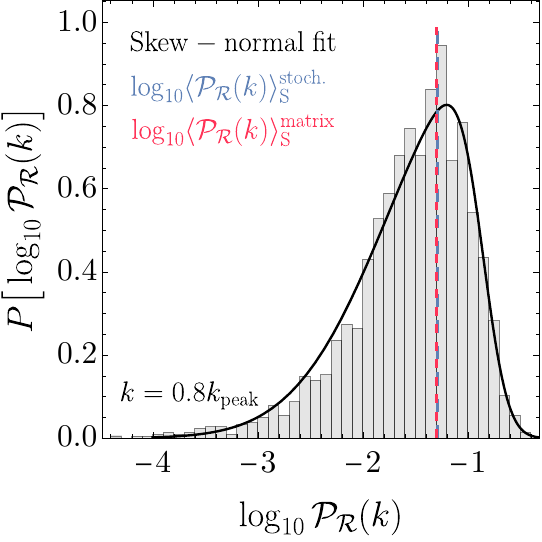}
\qquad\includegraphics[width=.47\textwidth]{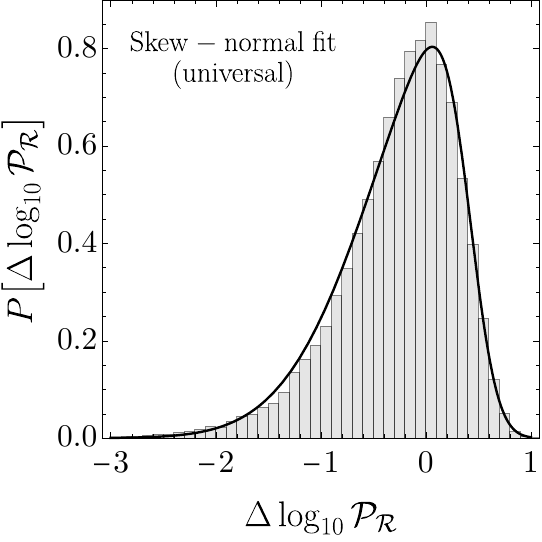}$
\caption{\em \label{fig:histograms} {{\bf Left:} Histogram of $\log_{10}\mathcal{P}_\mathcal{R}(k)$ for 2160 realizations for $k=k_{\rm peak}$, together with a skew-normal fit for the probability distribution function. {\bf Right:} Histogram of the $k$-independent variable $\Delta\log_{10}\mathcal{P}_\mathcal{R}$ defined in Eq.~(\ref{eq:kindeppower}) of $20\times 2160$ realizations, together with a universal skew-normal fit for the probability distribution function.}}
\end{center}
\end{figure}

The Langevin method provides for us not only the means to determine the first moment of the probability distribution of the power spectrum, but with enough sampling we can also determine the full distribution for $\mathcal{P}_\mathcal{R}(k)$ at each value of $k$. The left panel of  Fig.~\ref{fig:histograms} shows the normalized histogram for the 2160 realizations for $\log_{10} \mathcal{P}_\mathcal{R}(k)$ at $k=0.8\,k_\text{peak}$ for illustration. In this same panel we show as a vertical dashed blue line the corresponding expectation value over realizations, and as the vertical red dashed line the mean computed using the matrix formalism (presented in Section \ref{sec:matrix}). The continuous black curve corresponds to a skew-normal fit to the (logarithmic) data. As a reminder, a random variable $x$ is skew-normal distributed if its probability distribution function is given by
\begin{equation}
P_\text{skew-normal}(x\,|\,\mu, \sigma, \alpha)=\dfrac{1}{\sqrt{2\pi} \sigma} e^{-\frac{(x-\mu)^2}{2 \sigma^2} } \text{erfc}\left[ -\dfrac{\alpha(x-\mu)}{\sqrt{2}\sigma} \right],
\end{equation}
where $\text{erfc}(x)$ denotes the complementary error function and $\{\mu,\sigma,\alpha\}$ are free parameters. Therefore, we find that the PDF of $\mathcal{P}_\mathcal{R}$ can be modelled as a {\em skew-log-normal} distribution.\footnote{The same distribution for the power spectrum amplitude has been found for curvature fluctuations sourced by stochastically, parametrically excited spectator fields during inflation~\cite{Garcia:2019icv,Garcia:2020mwi}.} Defining for each $k$ the difference
\begin{equation}
\label{eq:kindeppower}
\Delta \log_{10} \mathcal{P}_\mathcal{R} \equiv  \log_{10} \mathcal{P}_\mathcal{R} -  \log_{10} \langle \mathcal{P}_\mathcal{R}  \rangle_{\rm S}
\end{equation}
we find that its probability distribution is
very well approximated by a $k$-{\em independent} skew-normal distribution. The right panel of Fig.~\ref{fig:histograms} shows the frequentist histogram for the full set of realizations for \eq{eq:kindeppower}. Together with it we show the corresponding universal skew-normal fit (shown in solid black), with parameters 
\begin{equation}
\label{eq:parameter_fit_universal} \{\mu,\sigma,\alpha\}=\{0.42, 0.87, -4.15 \}.
\end{equation}
A similar histogram can be created separately for each $k$, and we find that the standard deviations of the parameters $\{\mu,\sigma,\alpha\}$ for each one of these histograms with respect to the corresponding values for the universal fit shown above are of order $\{3\%,2\%,9\%\}$.

Note that the variance of the probability distribution function for the power spectrum is quite large. From Fig.~\ref{fig:histograms} it is clear that, for a specific realization in a particular Hubble patch the spectrum can reach a
value roughly one order of magnitude away from the $10^{-2}$ value required to obtain $f_{\rm PBH}\simeq 1$, leading to either overproduction or underproduction of PBHs (according to the Gaussian estimate of the abundance). This effect can always be countered by adjusting any of the parameters in $\Gamma$ that control the overall size of the average of the power spectrum, as discussed in Section \ref{sec:pheno}, as well as the threshold for the collapse (on which the abundance depends exponentially within the Gaussian estimate). 

\subsection{Analytical approximation}
\label{sec:analytical}

To get a better understanding of the evolution of the perturbations and the shape of the primordial spectrum, it is useful to simplify the equations of motion in such a way that they can be solved analytically. Let us begin by noting that at late times, the only quantity that contributes to the curvature perturbation is $\delta\phi$,
\begin{equation}
\label{eq:R_neglect}
\mathcal{R}\simeq -\frac{H\dot\phi}{\rho+p}\delta\phi\simeq-\frac{H}{\dot{\phi}}\delta\phi\bigg|_{k\ll aH}.
\end{equation}
The second observation we make is that, in the equation of motion for $\delta\phi$, \eq{eq:phieom}, we can neglect several terms and still reproduce the most important features of the spectrum,
\begin{equation}
\label{deltaphieom}
\frac{\d^2\delta\phi}{\d N^2}+\bigg(3+\frac{\Gamma}{H}\bigg)\frac{\d\delta\phi}{\d N}+\bigg(\frac{k^2}{a^2H^2}+\frac{\Gamma_\phi}{H} \frac{\d\phi}{\d N}\bigg)\delta\phi+\frac{3}{H^2}\left(\frac{\d\phi}{\d N}\right)^{-1}\delta\rho_r\simeq 0.
\end{equation}
This approximation is obtained by discarding terms involving the potential (which are slow-roll suppressed), the metric perturbation (which can be checked numerically to be a good approximation), and assuming that $\Gamma\propto T^3$ and the background remains in the attractor at all times, so that the attractor parameters defined in Eq.~(\ref{eq:attpars}) indeed satisfy $\epsilon_\phi=\epsilon_\rho=1$. This last approximation is justified by Fig.~\ref{fig:bgm}, where it can be seen that the background quantities only leave the attractor for very brief periods. In addition, we have found numerically that the stochasticity of the system can be encoded via $\delta\rho_r$ in \eq{deltaphieom}, and therefore the original noise term on the right hand side of \eq{eq:phieom} can be dropped.

Let us explain more precisely this last approximation. If we have an explicit expression for $\delta\rho_r$ as a function of time, then we can think of the $\delta\rho_r$ term in Eq.~(\ref{eq:phieom}) as a source term for $\delta\phi$, on the same footing as the $\xi_N$ term. Numerically, we find that the $\delta\rho_r$ term dominates over the $\xi_N$ term, in the sense that one can set the latter to zero and still correctly reproduce the key features of the spectrum: the location of the peak, and the size of the spectrum; the latter within an order of magnitude of the full numerical result. Note that this does not mean that the noise $\xi_N$ is irrelevant. In fact, it is precisely the noise in Eq.~(\ref{eq:rhoeom}) which determines $\delta\rho_r$, and thus in turn $\delta\phi$. In other words, the power spectrum of the comoving curvature perturbation is enhanced thanks to the source $\delta\rho_r$ (whose value is set by the thermal noise) in the equation of motion for $\delta\phi$.

\begin{figure}[t!]
        \centering
        \begin{subfigure}{.4\linewidth}
            \centering
\includegraphics[width=.87\textwidth]{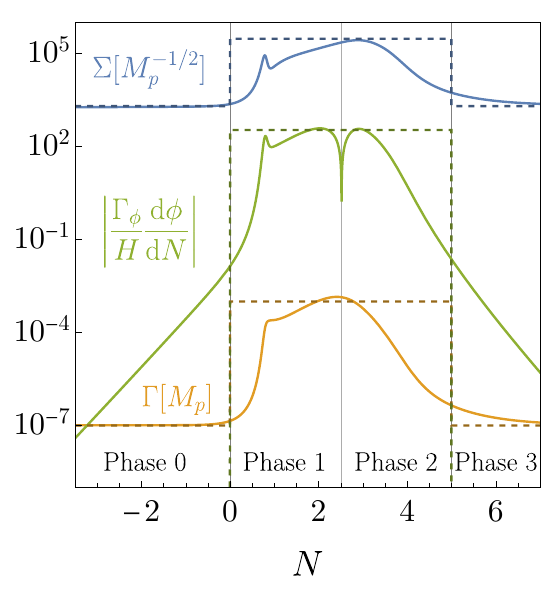}
        \end{subfigure}
        \hfill
        \begin{subfigure}{.59\linewidth}
            \centering
            \includegraphics[width=.87\textwidth]{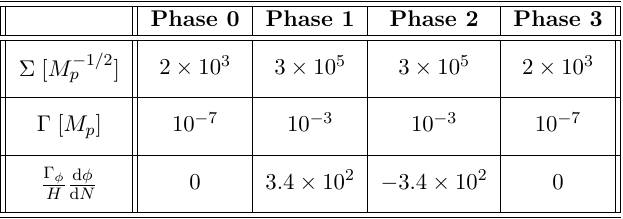}
        \begin{minipage}{.1cm}
            \vfill
            \end{minipage}
        \end{subfigure} 
        \caption{{\bf Left:  } {\it The dissipative coefficient $\Gamma$, the	quantity $\Sigma$ defined in \eq{eq:sigmadef} and the function $|\Gamma_\phi| H^{-1} |\d \phi/\d N|$ as functions of the number of $e$-folds (and in units of $M_p$)		for the model given by Eqs.~\eq{eq:gammapheno} and \eq{eq:benchpars}. The corresponding approximations as piecewise-constant functions from the table on the right are shown with dashed lines.	} {\bf Right: } {\it Benchmark parameters for the analytical calculation of the power spectrum. We take phases 1 and 2 to end at $N_1=2.5$ and $N_2=5$, respectively (we measure the number of $e$-folds from the end of phase 0, so that $N_0=0$, and we normalize $a(N_0)=1$, see the figure on the left, as well as $H=7\times 10^{-6} M_p$.}  } \label{eq:table}
\end{figure}

The strategy we will follow now is to propose a phenomenological parameterization for $\delta\rho_r$ as a function of time and use it to solve Eq.~(\ref{deltaphieom}). We will also assume that all background quantities can be approximated as piecewise-constant functions. The benchmark values we take for each quantity are shown in (the table of) Fig.~\ref{eq:table}, where we introduce the quantity
\begin{equation}
\label{eq:sigmadef}
\Sigma\equiv\sqrt{\frac{9\Gamma T}{H^3}}.
\end{equation}
These parameters have been chosen in order to obtain a primordial spectrum closely resembling in its main features the one derived for the dissipation coefficient \eq{eq:gammapheno} with the parameters \eq{eq:benchpars}. We assume the evolution proceeds in four different phases, which we label from $0$ to $3$. In phases 0 and 3 we have $\Gamma_\phi=0$ and $\Gamma\ll H$, so that we are in the weak dissipative regime. In phases 1 and 2 we have $\Gamma\gg H$. During phase 1 we have $\Gamma_\phi>0$, and during phase 2 we have $\Gamma_\phi<0$. This {parameterization} is compared to the benchmark model \eq{eq:benchpars} in Fig.~\ref{eq:table}. In addition, we parameterize the time evolution of the root mean square of $\delta\rho_r$ with the following phenomenological expression,\footnote{This expression improves over a similar one proposed in \cite{Hall:2003zp}.}
\begin{equation}
\label{eq:rho_attract}
\langle|\delta\rho_r|^2\rangle_{\rm S}\simeq\frac{2\pi^2}{15}g_\star T^5\cdot\begin{cases}
e^{-3N} &N<N_f,\\
e^{-2N}e^{\Delta N_f}(H/k) &N>N_f,
\end{cases}
\end{equation}
where the time $N_f$ at which the transition occurs is located a couple of $e$-folds before the horizon crossing time,
\begin{equation}
\label{eq:trans_time}
N_f\equiv \log(k/H)-\Delta N_f,
\end{equation}
where $\Delta N_f$ is an $\mathcal{O}(1)$, $k$-independent constant. We take $\Delta N_f = 2.1$ for definiteness. Despite its simplicity (which of course cannot capture all the details of the full numerical solution), this parameterization is enough to understand the most important features of the spectrum. Since $\delta\rho_r$ is a stochastic variable, it is not sufficient to parameterize its root mean square value, but we also need to know its correlation function. To make progress, we will assume that $\delta\rho_r$ behaves like a Wiener {increment,}
\begin{equation}
\delta\rho_r=\sqrt{\langle|\delta\rho_r|^2\rangle_{\rm S}}\,\xi^{\delta\rho}_N\,,
\end{equation}
where the correlation function for $\xi^{\delta\rho}_N$ is
\begin{equation}
\langle\xi^{\delta\rho}_N(\kk)\xi^{\delta\rho}_{\hat{N}}(\qq)\rangle_{\rm S}=(2\pi)^3\delta(N-\hat{N})\delta^3(\kk+\qq).
\end{equation}

The homogeneous solution to Eq.~(\ref{deltaphieom}) can be written as
\begin{equation}
\label{eq:phi_homo}
\delta\phi^{(h)}=e^{-\nu N}\Big[\delta\phi_+J_{\mu}\big(ke^{-N}/H\big)+\delta\phi_-J_{-\mu}\big(ke^{-N}/H\big)\Big],
\end{equation}
where $\delta\phi_{\pm}$ are constants fixed by the initial conditions, $J_\mu$ is the Bessel function of the first kind and
\begin{equation}
\nu=\frac{1}{2}\bigg(3+\frac{\Gamma}{H}\bigg),\quad\mu=\sqrt{\nu^2-\frac{\Gamma_\phi}{H}\frac{\d\phi}{\d N}}.
\end{equation}
This solution and its derivative can also be written in matrix form as 
\begin{equation}
\begin{pmatrix}
\delta\phi^{(h)} \\
\frac{\d}{\d N}\delta\phi^{(h)}
\end{pmatrix}
=
\underbrace{e^{-\nu N}\begin{pmatrix}
J_\mu(ke^{-N}/H) & J_{-\mu}(ke^{-N}/H) \\
\frac{\d}{\d N} J_\mu(ke^{-N}/H)-\nu J_\mu(ke^{-N}/H) & \, \frac{\d}{\d N} J_{-\mu}(ke^{-N}/H)-\nu J_{-\mu}(ke^{-N}/H)
\end{pmatrix}}_{M(N)}
\begin{pmatrix}
\delta\phi_{+} \\
\delta\phi_{-}
\end{pmatrix}\,.
\end{equation}
The constants $\mu$ and $\nu$ take different values in each one of the four phases. We denote their values in the $j$-th phase by $\mu_j$ and $\nu_j$. The constants $\delta\phi_{\pm}$ can be found by imposing continuity of the solution and its derivative at the end of each phase. We denote these constants by $\delta\phi_{\pm j}$ in the $j$-th phase. We use $N_j$ to refer to the time at which each phase ends. In particular, phase $0$ 
begins at $-\infty$ and ends at $N_0=0$, and phase $3$ ends at $N_3=\infty$. To be as general as possible we keep our calculations generic for $n+1$ phases, but we will eventually set $n=3$.

Following the above procedure we can find the constants in the last phase
\begin{equation}
\begin{pmatrix}
\delta\phi_{+n} \\
\delta\phi_{-n}
\end{pmatrix}
=
\Bigg[\prod_{j=1}^n M_{j}^{-1}(N_{j-1})M_{j-1}(N_{j-1})\Bigg]
\begin{pmatrix}
\delta\phi_{+0} \\
\delta\phi_{-0}
\end{pmatrix}.
\end{equation}
In this equation, terms with smaller $j$ should be placed at the end of the product.\footnote{Since these are matrices, the order of the factors is relevant.} The total solution for $\delta\phi$, including both the homogeneous and inhomogeneous solutions is
\begin{equation}
\label{phisol}
\delta\phi=\delta\phi^{(h)}+\int_{-\infty}^N\frac{\Sigma(\hat{N})}{S(\hat{N})}\frac{G(N,\hat{N})}{a(\hat{N})^{s/2}}\xi^{\delta\rho}_{\hat{N}}\d \hat{N},
\end{equation}
where $G$ is the Green's function, which we will find below, and
\begin{align}
s&=3\Theta(N_f-N)+2\Theta(N-N_f), \label{eq:smalls}\\
S&=\Theta(N_f-N)+\sqrt{\frac{k}{H}}e^{-\Delta N_f/2}\Theta(N-N_f),
\label{eq:largeS}
\end{align}
where $\Theta$ is the Heaviside step function. The constants in the homogeneous solution are obtained by imposing Bunch-Davies boundary conditions in the $0$-th region,\footnote{We can do this because in this region we are in the weak dissipative regime $\Gamma\ll H$ and thus $\mu\simeq\nu\simeq 3/2$.}
\begin{equation}
\begin{pmatrix}
\delta\phi_{+0} \\
\delta\phi_{-0}
\end{pmatrix}
=
-\frac{1}{2}\sqrt{\frac{\pi}{H}}\begin{pmatrix}
1 \\
i
\end{pmatrix}.
\end{equation}
The expectation value of the power spectrum at late times is\footnote{See Appendix \ref{app:average} for a detailed discussion on the assumptions required to arrive at this result from Eq.~(\ref{phisol}).}
\begin{equation}
\label{eq:anal_spec}
\langle\mathcal{P}_{\delta\phi}(k)\rangle_{\rm S}=\frac{k^3}{2\pi^2}\bigg[\frac{2^{\mu_n}(k/H)^{-\mu_n}}{\Gamma_{\rm E}(1-\mu_n)}\bigg]^2|\delta\phi_{-n}|^2+\frac{k^3}{2\pi^2}\underbrace{\int_{-\infty}^{\infty}\frac{\Sigma(\hat{N})^2}{S(\hat{N})^2}\frac{|G(N\rightarrow\infty,\hat{N})|^2}{a(\hat{N})^s}\d \hat{N}}_{\mathcal{I}_\mathcal{P}},
\end{equation}
where $\Gamma_{\rm E}(1-\mu_n)$ denotes Euler's Gamma function evaluated at $(1-\mu_n)$
and $\Sigma$ was defined in Eq.~(\ref{eq:sigmadef}). 

\begin{figure}[t]
\begin{center}
$\includegraphics[width=.47\textwidth]{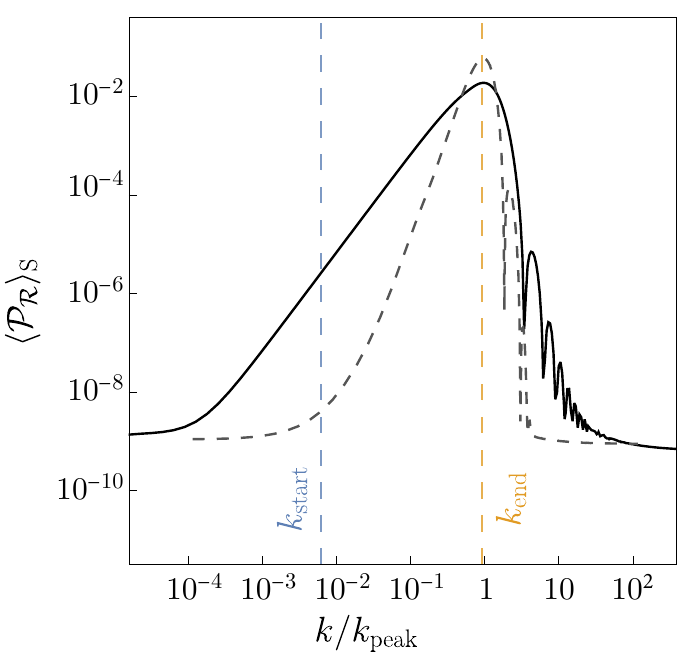}
\qquad\includegraphics[width=.47\textwidth]{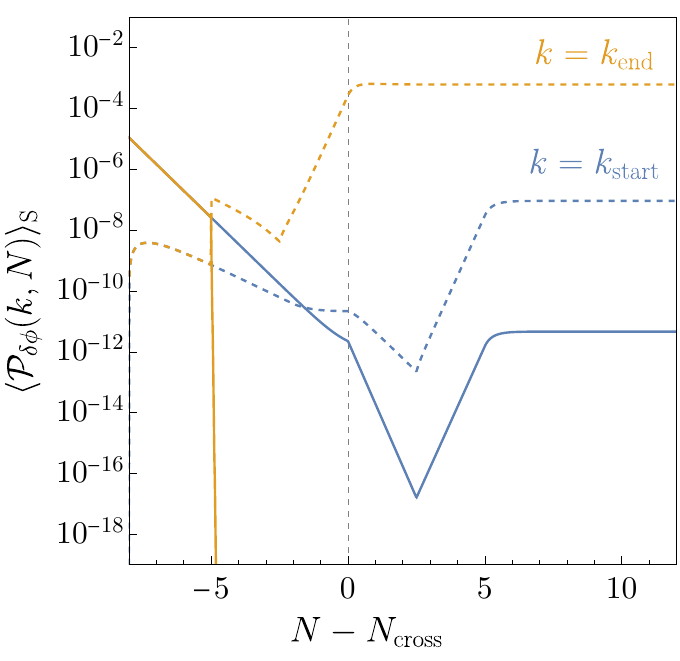}$
\caption{\em \label{fig:analplot} {{\bf Left: } the stochastic average of the power spectrum using the analytical approach is shown as a black solid line. The grey dashed line shows the result obtained with the matrix formalism of Section \ref{sec:matrix}, see also Fig.~\ref{fig:PR_stochastic}. {\bf Right:} homogeneous (solid) and inhomogeneous (dashed) solutions --the two terms in Eq.~(\ref{phisol})-- for modes leaving the horizon at the start and the end of the strongly dissipative phase in which $\Gamma\gg H$. The inhomogeneous solution, which is independent of initial conditions, always dominates at late times, indicating the presence of an attractor in the equation of motion for the perturbations. The parameters chosen for both panels are shown in (the table of) Fig.~\ref{eq:table}.}}
\end{center}
\end{figure}

The Green's function $G$ appearing in the integrand of \eq{eq:anal_spec} is
\begin{equation}
\label{eq:full_Green}
G(N,\hat{N})=\frac{\delta\phi^{(1)}(N)\delta\phi^{(2)}(\hat{N})-\delta\phi^{(1)}(\hat{N})\delta\phi^{(2)}(N)}{\frac{\d}{\d\hat{N}}\delta\phi^{(1)}(\hat{N})\delta\phi^{(2)}(\hat{N})-\frac{\d}{\d\hat{N}}\delta\phi^{(2)}(\hat{N})\delta\phi^{(1)}(\hat{N})},
\end{equation}
where $\hat{N}<N$ and $\delta\phi^{(1,2)}$ are two linearly independent solutions to the homogeneous equation. The calculation of the Green's function is simpler if instead of writing the homogeneous solutions as linear combinations of $J_\mu$ and $J_{-\mu}$, we use $J_\mu$ and $Y_\mu$ (the Bessel function of the second kind, which is itself a linear combination of $J_\mu$ and $J_{-\mu}$). We therefore write
\begin{equation}
\delta\phi^{(h)}=e^{-\nu N}\Big[\delta\hat{\phi}_+J_{\mu}\big(ke^{-N}/H\big)+\delta\hat{\phi}_-Y_\mu\big(ke^{-N}/H\big)\Big],
\end{equation}
which is completely equivalent to Eq.~(\ref{eq:phi_homo}). Since the Green's function is independent of the boundary conditions chosen for the two linearly independent solutions, we can follow a slightly different procedure from before and arbitrarily choose some linearly independent set of constants in the final region instead of the first. The constants in the previous regions can then be found by matching the solutions and their derivatives at each boundary. We choose $\big(\delta\hat{\phi}_{+n}^{(1)},\delta\hat{\phi}_{-n}^{(1)}\big)=(0,1)$ and $\big(\delta\hat{\phi}_{+n}^{(2)},\delta\hat{\phi}_{-n}^{(2)}\big)=(1,0)$ for the two solutions. 

The reason for using $Y_\mu$ instead of $J_{-\mu}$ and choosing the constants in the final region instead of the first is that we obtain the following simple limits for the two independent solutions at late times,
\begin{align}
\delta\phi^{(1)}(N\rightarrow\infty)&=-\frac{1}{\pi}\delta\hat{\phi}_{-n}^{(1)}\Gamma(\mu_n)2^{\mu_n} \left(\frac{k}	{H}\right)^{-\mu_n},\\
\delta\phi^{(2)}(N\rightarrow\infty)&=0.
\end{align}
If $\hat{N}$ is in the $i$-th region, the denominator of the Green's function becomes
\begin{equation}
{\frac{\d}{\d\hat{N}}\delta\phi^{(1)}(\hat{N})\delta\phi^{(2)}(\hat{N})-\frac{\d}{\d\hat{N}}\delta\phi^{(2)}(\hat{N})\delta\phi^{(1)}(\hat{N})}=\frac{2}{\pi}e^{-2\nu_i \hat{N}}\left(\delta\hat{\phi}^{(1)}_{+i}\delta\hat{\phi}^{(2)}_{-i}-\delta\hat{\phi}^{(2)}_{+i}\delta\hat{\phi}^{(1)}_{-i}\right)	.
\end{equation}
It is easy to show that this combination of constants is
\begin{equation}
\frac{1}{\mathcal{C}_i}=\Big(\delta\hat{\phi}^{(1)}_{+i}\delta\hat{\phi}^{(2)}_{-i}-\delta\hat{\phi}^{(2)}_{+i}\delta\hat{\phi}^{(1)}_{-i}\Big)=e^{2(\nu_{i}-\nu_{i+1}) N_{i}}\frac{1}{\mathcal{C}_{i+1}}.
\end{equation}
Since in the final region we have $\delta\hat{\phi}^{(1)}_{+n}\delta\hat{\phi}^{(2)}_{-n}-\delta\hat{\phi}^{(2)}_{+n}\delta\hat{\phi}^{(1)}_{-n}=-1$, we obtain the following expression for $\mathcal{C}_i$,
\begin{equation}
\frac{1}{\mathcal{C}_i}=-\prod_{j=i}^{n-1}e^{2(\nu_{j}-\nu_{j+1}) N_{j}}.
\end{equation}
Putting everything together, we find that the Green's function at late times is, if $\hat{N}$ is in the $i$-th region,
\begin{equation}
G(N\rightarrow\infty,\hat{N})=\underbrace{-\Gamma(\mu_n)2^{\mu_n-1} (k/H)^{-\mu_n}}_{\mathcal{B}_n}\delta\phi^{(2)}(\hat{N})e^{2\nu_i \hat{N}}\mathcal{C}_i.
\end{equation}
The integral in Eq.~(\ref{eq:anal_spec}) then reads
\begin{equation}
\label{noiseints}
\mathcal{I}_\mathcal{P}=\sum_{i=0}^{n}\Sigma_i^2\mathcal{B}_n^2\mathcal{C}_i^2\int^{e^{- N_{i-1}}}_{e^{- N_i}}\frac{v^{s-1-2\nu_i}}{S^2}\Big[\delta\hat{\phi}_{+i}^{(2)}J_{\mu_i}(kv/H)+\delta\hat{\phi}_{-i}^{(2)}Y_{\mu_i}(kv/H)\Big]^2\d v,
\end{equation}
where we have also switched variables to $v=e^{-N}$. These integrals can be found analytically in terms of hypergeometric functions.

Now that we have all the necessary ingredients we can compute the power spectrum analytically by using Eqs.~(\ref{eq:anal_spec}) and (\ref{noiseints}) and fixing the parameters as in {(the table of) Fig.\,\ref{eq:table}}. To go from $\delta\phi$ to $\mathcal{R}$ we use Eq.~(\ref{eq:R_neglect}) in the late time limit, where the ratio between the two is approximately constant, see the central panel of Fig.~\ref{fig:bgm}.
The resulting power spectrum is shown in Fig.~\ref{fig:analplot}. The overall size of the peak of the spectrum and the oscillations seen in Fig.~\ref{fig:PR_stochastic} are present in the analytical solution. The oscillatory pattern observed for scales $k>k_\text{peak}$ in the power spectrum is attributed to Bessel functions appearing in the matching conditions imposed on the homogenous solution of the $\delta \phi$ equation in the four phases. We find that, as with the numerical solution, the peak in the spectrum occurs for modes that leave the horizon around the end of the strongly dissipative phase. This is a consequence of the enhancement being an integrated effect, due to Eq.~(\ref{noiseints}), as opposed to a local one.

Having an analytical solution allows us to understand why the initial conditions for the perturbations are irrelevant. All of the information about initial conditions is contained in the homogeneous solution (\ref{eq:phi_homo}) inside the integration constants $\delta\phi_{\pm}$. However, as shown in the right panel of Fig.~\ref{fig:analplot}, this solution is completely negligible at late times. The spectrum is completely dominated by the integral in Eq.~(\ref{noiseints}), which is independent of initial conditions. This indicates the presence of an attractor in the equation of motion for the perturbations, as anticipated earlier.

The analytical approximation developed in this section is not enough to reproduce with accuracy the full averaged primordial spectrum. For instance, the actual slope of the $\log$ of the spectrum for $k< k_{\rm peak}$ is about twice the value that the analytical approximation gives. However, this approximation is going to be useful in the next section to obtain a good estimate of the peak value of the gravitational wave signal induced at second order in perturbation theory.

\section{Induced gravitational waves}
\label{sec:gw_main}

In this section we compute the gravitational wave signal induced by scalar perturbations at second order. These gravitational waves are induced both during inflation and during the subsequent radiation era \cite{Baumann:2007zm,Fumagalli:2021mpc}. The calculation is organized as follows. In Section \ref{sec:source_term} we write the equation of motion for tensor modes sourced by second order scalar perturbations (obtained by perturbing Einstein's equations) and derive explicit expressions for the source term in the two cases of interest, namely, when gravitational waves are induced in the inflationary epoch, and during radiation domination. We then solve this equation via the Green's function method. In Section \ref{sec:gw_energy} we compute the tensor power spectrum and in Section \ref{sec:gw_energy_2} relate it to the observable quantity of interest, which is the energy density of gravitational waves. In Section \ref{sec:gw_integrals} we provide a numerical estimate of the latter quantity.

\subsection{Second order scalar source}
\label{sec:source_term}

The equation of motion for tensor modes at second order in Fourier space and in terms of conformal time $\eta$ is
\cite{Baumann:2007zm}
\begin{equation}
\label{tensoreom}
h_k^{s\prime\prime}+2\mathcal{H}h_k^{s\prime}+k^2h_k^s=S_k^s,
\end{equation}
where primes denote derivatives with respect to conformal time ($\prime=\d/\d\eta$), $\mathcal{H}=a'/a$ denotes the conformal Hubble factor, and the superscript $s=(+,\times)$ stands for the polarization of the tensor mode. The source $S_k^s$ is, in the Newtonian gauge and in the absence of anisotropic stress \cite{Lu:2020diy},
\begin{equation}
\label{eq:gw_source}
S^s_k=\int\frac{\d^3p}{(2\pi)^3}{\bm{e}}^s_{ij}({\bm{k}})p_ip_j\Bigg[8\psi_{p}\psi_{k-p}+\frac{16\rho}{3(\rho+p)}\left(\psi_p+\frac{1}{\mathcal{H}}\psi'_p\right)\left(\psi_{k-p}+\frac{1}{\mathcal{H}}\psi'_{k-p}\right)\Bigg].
\end{equation}
The quantity ${\bm e}_{ij}^s$ is a symmetric transverse, traceless projector that satisfies ${\bm e}_{ij}^s({\bm k})k_i=0$ \cite{Lu:2020diy}. The second term in the equation above can be alternatively written in terms of the total momentum perturbation\footnote{The momentum perturbations $\delta q_i$ are additive, so the total momentum perturbation can be defined as the sum of the individual components, $\delta q\equiv \sum_i\delta q_i$.} $\delta q$ using Eq.~(\ref{eq:psieom}),
\begin{equation}
\psi'+\mathcal{H}\psi=-a\frac{\delta q}{2M_p^2}.
\end{equation}
In particular, during inflation, the total momentum perturbation is the sum of the radiation and inflaton components,
\begin{equation}
\label{totalmomentum}
\delta q\equiv\delta q_r+\delta q_\phi=\delta q_r-\frac{\phi'}{a}\delta\phi.
\end{equation}

The source term $S_k^s$ of Eq.~(\ref{eq:gw_source}) before ({\it pre}) and after ({\it post}) inflation ends is respectively given by:
\begin{align} \label{need1}
S^s_{k{\rm (pre)}}&=\frac{4}{3}\bigg(\frac{\rho}{\rho+p}\bigg)\bigg(\frac{\phi^{\prime 2}}{\mathcal{H}^2M_p^4}\bigg)\int\frac{\d^3p}{(2\pi)^3}{\bm e}^s({\bm k},{\bm p})\delta \phi_p\delta \phi_{k-p},\\
S^s_{k{\rm (post)}}&=\int\frac{\d^3p}{(2\pi)^3}{\bm e}^s({\bm k},{\bm p})\bigg[8\psi_p\psi_{k-p}+4\bigg(\psi_p+\frac{1}{\mathcal{H}}\psi'_p\bigg)\bigg(\psi_{k-p}+\frac{1}{\mathcal{H}}\psi'_{k-p}\bigg)\bigg],
\end{align}
where ${\bm e}^s({\bm k},{\bm p})\equiv{\bm e}_{ij}^s({\bm k})p_ip_j$. To write the first expression we have neglected the first term in the brackets of Eq.~(\ref{eq:gw_source}), as well as the $\delta q_r$ term from Eq.~(\ref{totalmomentum}). This approximation will be justified in Section \ref{sec:gw_integrals}. For the second expression we have assumed that the Universe enters a radiation-dominated era after inflation ends, so that $p=\rho/3$.

As is customary, let us fix the time at which inflation ends as $\eta=0$. The value of $\psi$ at this time, which will be the initial condition for the post-inflationary source, is, on superhorizon scales and assuming the Universe enters a radiation-dominated era after inflation ends,
\begin{equation}
\label{eq:psi_ini}
\psi_k(0)=\frac{2}{3}\mathcal{R}_k(0)=-\frac{2}{3}\frac{H\phi'}{a(\rho+p)}\bigg|_{0}\delta\phi_k(0),
\end{equation}
where we have used Eq.~(\ref{eq:R_neglect}). The post-inflationary source can then be rewritten as
\begin{equation}
S^s_{k{\rm (post)}}=\frac{4}{9}\frac{H^2\phi^{\prime 2}}{a^2(\rho+p)^2}\bigg|_{0}\int\frac{\d^3p}{(2\pi)^3}{\bm e}^s({\bm k},{\bm p})\delta\phi_p(0)\delta\phi_{k-p}(0)Q(p,|{\bm k}-{\bm p}|,\eta),
\end{equation}
where the function $Q$ is
\begin{equation}
Q=8T^\psi_pT^\psi_{k-p}+4\bigg(T^\psi_p+\frac{1}{\mathcal{H}}T^{\psi\prime}_p\bigg)\bigg(T^\psi_{k-p}+\frac{1}{\mathcal{H}}T^{\psi\prime}_{k-p}\bigg),
\end{equation}
and $T^\psi_k$ is the transfer function for $\psi_k$,
\begin{equation}
T^\psi_k=-\frac{9}{(k\eta)^3}\bigg[(k\eta)\cos\bigg(\frac{k\eta}{\sqrt{3}}\bigg)-\sqrt{3}\sin\bigg(\frac{k\eta}{\sqrt{3}}\bigg)\bigg]\,,
\end{equation}
which is defined by $\psi_k(\eta)=T^\psi_k(\eta)\psi_k(0)$ and is obtained by solving\footnote{This equation is obtained by straightforward manipulation of Einstein's equations. We have also assumed a radiation-dominated Universe with $p=\rho/3$ and $\mathcal{H}\equiv a\,H=1/\eta$.}
\begin{equation}
\psi''_k+\frac{4}{\eta}\psi'_k+\frac{k^2}{3}\psi_k=0.
\end{equation}

The Green's functions for Eq.~(\ref{tensoreom}) during inflation and during the subsequent radiation-dominated era can be found as in Eq.~(\ref{eq:full_Green}). The results are, respectively,
\begin{align}
kF_{\rm pre}(\eta,\eta')&=-\frac{1}{(k\eta')^2}\Big[k(\eta-\eta')\cos\Big(k(\eta-\eta')\Big)-(1+k^2\eta\eta')\sin\Big(k(\eta-\eta')\Big)\Big],\\
kF_{\rm post}(\eta,\eta')&=\frac{\eta'}{\eta}\sin \Big(k(\eta-\eta')\Big),
\end{align}
where we have used $\mathcal{H}=-1/\eta$ during inflation. The solution is therefore
\begin{equation}
\label{eq:g_def}
h^s_k(\eta)=T^h_k(\eta)h^s_k(0)+\underbrace{\int_0^\eta F_{\rm post}(\eta,\eta')S^s_{k{\rm (post)}}(\eta')\d\eta'}_{g^s_k(\eta)},
\end{equation}
where $T^h_k$ is the (linear) transfer function of $h^s_k$ in the radiation era,
\begin{equation}
T^h_k=\frac{\sin(k\eta)}{k\eta},
\end{equation}
and
\begin{equation}
\label{eq:hpost}
h^s_k(0)=\int_{\eta_h}^0 F_{\rm pre}(0,\eta')S^s_{k{\rm (pre)}} (\eta')\d\eta'.
\end{equation}
The lower integration limit $\eta_h$ is some early time at which we assume $h^s_k(\eta_h)=0$. In other words, we assume that no second order gravitational waves have been induced at sufficiently early times.

\subsection{The gravitational wave spectrum induced at second order}
\label{sec:gw_energy}

We now have almost all\footnote{The connection between the power spectrum of tensor modes and the energy density of gravitational waves will be completed in Section \ref{sec:gw_energy_2}.} the necessary ingredients to calculate the energy density of gravitational waves, which is related to the tensor power spectrum and is the observable quantity of interest. The expectation value of the tensor power spectrum late in the radiation era contains three terms,\footnote{In Section \ref{sec:simple_formula} we present for the first time a compact expression which includes the mixing term in standard (cold) inflation.}
\begin{equation}
\label{eq:power_h_split}
\langle\mathcal{P}_h(k,\eta)\rangle_{\rm S}=\langle\mathcal{P}_{\rm pre}(k,\eta)\rangle_{\rm S}+2\langle\mathcal{P}_{\rm mix}(k,\eta)\rangle_{\rm S}+\langle\mathcal{P}_{\rm post}(k,\eta)\rangle_{\rm S}.
\end{equation}
These three terms will in turn lead to three different contributions to the gravitational wave energy density. The first term corresponds to the gravitational waves induced during the inflationary epoch, and the third term corresponds to the gravitational waves induced during the subsequent radiation-dominated era. The middle term mixes both contributions and its value typically lies between the other two.

To perform the rest of the calculation we will use the analytical results of Section \ref{sec:analytical}. The reason for this is that to calculate the tensor power spectrum we need to take the quantum expectation value in addition to the stochastic one, in order to make the tensor power spectrum a deterministic quantity, as we did with $\mathcal{P}_\mathcal{R}$. As discussed in Appendix \ref{app:average}, when Eq.~(\ref{phisol}) is used to split $\delta\phi$ into the homogeneous and inhomogeneous solutions to its equation of motion, finding the double expectation value is straightforward, since only the homogeneous solution is quantized, and only the inhomogeneous piece is stochastic. This splitting can be done only because the equation of motion for $\delta\phi$ has been decoupled from the rest (up to the source term $\delta\rho_r$), since the full system of differential equations cannot be solved by Green's function methods. Our simplifying approach should give a reasonable estimate of the actual result.

The first term in Eq.~(\ref{eq:power_h_split}) is defined by\footnote{As discussed in Appendix \ref{app:average}, the brackets without subscripts denote a double expectation value, quantum and stochastic.}
\begin{equation}
T^{h}_k(\eta)^2\langle h_k^r(0)h_p^s(0)\rangle=(2\pi)^3\frac{2\pi^2}{k^3}\langle\mathcal{P}_{\rm pre}(k,\eta)\rangle_{\rm S}\delta^{rs}\delta^3({\bm k}+{\bm p}).
\end{equation}
The quantum expectation value is already included inside $\mathcal{P}_{\rm pre}$ on the right-hand side --see Eq.~(\ref{spectrum_stoch_app}) for the analogous scalar definition-- so we only write the stochastic average. The two-point function on the left-hand side can be computed quantizing the inflaton field perturbation. The result is, using Eqs.~\eq{need1}
and (\ref{eq:hpost}),
\begin{align}
\label{twopointpre}
\langle h_k^r(0)h_p^s(0)\rangle&=\frac{16}{M_p^4}\int_{\eta_h}^0 \frac{\phi^{\prime 2}}{a^2(\rho+p)}\bigg|_{\eta'}F_{\rm pre}(0,\eta')\d\eta'\int_{\eta_h}^0 \frac{\phi^{\prime 2}}{a^2(\rho+p)}\bigg|_{\eta''}F_{\rm pre}(0,\eta'')\d\eta''\cdot\nonumber\\
&\qquad\cdot\int\frac{\d^3q}{(2\pi)^3}{\bm e}^r({\bm k},{\bm q})\int\frac{\d^3l}{(2\pi)^3}{\bm e}^s({\bm p},{\bm l})\langle\delta\hat\phi_q(\eta')\delta\hat\phi_{k-q}(\eta')\delta\hat\phi_l(\eta'')\delta\hat\phi_{p-l}(\eta'')\rangle.
\end{align}
The four-point function of $\delta\phi$ appearing in this equation can be computed using Eq.~(\ref{phisol}), together with Wick's theorem. This calculation is done in Appendix \ref{app:average}. The resulting expression can be substituted back into the above equation and, using one of the Dirac deltas to perform the integral over ${\bm l}$, we find
\begin{align} \label{nonsymint}
\langle h_k^r(0)h_p^s(0)\rangle&=\frac{128\pi^4}{M_p^4} \frac{\delta^3({\bm k}+{\bm p})}{k^{10}}\int_{\eta_h}^0 \frac{\phi^{\prime 2}}{a^2(\rho+p)}\bigg|_{\eta'}k F_{\rm pre}(0,\eta')k\d\eta'\int_{\eta_h}^0 \frac{\phi^{\prime 2}}{a^2(\rho+p)}\bigg|_{\eta''}k F_{\rm pre}(0,\eta'')k\d\eta''\cdot\nonumber\\
&\qquad\cdot\int \d^3q\;{\bm e}^r({\bm k},{\bm q}){\bm e}^s({\bm k},{\bm q})\mathcal{Q}_\mathcal{\delta\phi}(q,\eta',\eta'')\mathcal{Q}_\mathcal{\delta\phi}(k-q,\eta',\eta'')\frac{k^3}{q^3}\frac{k^3}{|k-q|^3},
\end{align}
where we have defined\footnote{$G_k(\eta,\eta')$ denotes the Green's function for the homogeneous equation \eq{deltaphieom}, defined in Eq.~(\ref{eq:full_Green}) evaluated at $\eta=-e^{-N}/H$. We also remind the reader that we use star superscript to indicate complex conjugation.} {\fontsize{10.5}{\baselineskip}
\begin{equation}
\label{cal_q_def}
\mathcal{Q}_{\delta\phi}(q,\eta',\eta'')\equiv\frac{q^3}{2\pi^2}\bigg(\delta\phi^{(h)}_{q}(\eta')^\star\delta\phi^{(h)}_{q}(\eta'')+\int_{-\infty}^{{\rm min}(\eta',\eta'')}G_{q}(\eta',\hat{\eta})G_{q}(\eta'',\hat{\eta})\frac{9a^2}{H^2\phi^{\prime 2}}|\delta\rho_{r,q}(\hat{\eta})|^2\d\hat{\eta}\bigg).
\end{equation}}
This quantity has the following properties
\begin{equation}
\label{eq:power_id}
\mathcal{Q}_{\delta\phi}(q,\eta,\eta)=\langle\mathcal{P}_{\delta\phi}(q,\eta)\rangle_{\rm S},\qquad\mathcal{Q}_{\delta\phi}(q,\eta',\eta'')=\mathcal{Q}_{\delta\phi}^\star(q,\eta'',\eta').
\end{equation}

It is useful to make the integrand in \eq{nonsymint} manifestly real and symmetric under $\eta'\leftrightarrow\eta''$. To do so, we take Eq.~(\ref{twopointpre}), rename the dummy variables $\eta'\leftrightarrow\eta''$ and sum the result with Eq.~(\ref{twopointpre}) itself. After using the second identity in Eq.~(\ref{eq:power_id}), we find
\begin{align}
\langle h_k^r(0)h_p^s(0)\rangle&=\frac{128\pi^4}{M_p^4}\frac{\delta^3({\bm k}+{\bm p})}{k^{10}}\int_{\eta_h}^0 \frac{\phi^{\prime 2}}{a^2(\rho+p)}\bigg|_{\eta'}k F_{\rm pre}(0,\eta')k\d\eta'\int_{\eta_h}^0 \frac{\phi^{\prime 2}}{a^2(\rho+p)}\bigg|_{\eta''}k F_{\rm pre}(0,\eta'')k\d\eta''\cdot\nonumber\\
&\qquad\cdot\int \d^3q\;{\bm e}^r({\bm k},{\bm q}){\bm e}^s({\bm k},{\bm q}){\rm Re}\bigg[\mathcal{Q}_{\delta\phi}(q,\eta',\eta'')\mathcal{Q}_{\delta\phi}(k-q,\eta',\eta'')\bigg]\frac{k^3}{q^3}\frac{k^3}{|k-q|^3}.
\end{align}
The final step consists in switching to spherical coordinates in the ${\bm q}$ integral, perform one of the angular integrals, and then make the change of variables
\begin{equation}
y=q/k,\qquad z=|{\bm k}-{\bm q}|/k.
\end{equation}
This procedure amounts to the replacement \cite{Kohri_2018,Espinosa_2018}
\begin{equation}
\int \d^3q\;{\bm e}^r({\bm k},{\bm q}){\bm e}^s({\bm k},{\bm q})\longrightarrow\frac{\pi}{2}\delta^{rs}k^7\int_0^\infty \d y\int_{|1-y|}^{1+y} \d z\,\left[\frac{4y^2-(1+y^2-z^2)^2}{4yz}\right]^2y^3z^3.
\end{equation}
Thus, we finally obtain for the first term in Eq.~(\ref{eq:power_h_split}),
\begin{align}
\langle\mathcal{P}_{\rm pre}(k,\eta)\rangle_{\rm S}&=T_k^h(\eta)^2\frac{4}{M_p^4}\int_{\eta_h}^0 \frac{\phi^{\prime 2}}{a^2(\rho+p)}\bigg|_{\eta'}k F_{\rm pre}(0,\eta')k\d\eta'\int_{\eta_h}^0 \frac{\phi^{\prime 2}}{a^2(\rho+p)}\bigg|_{\eta''}k F_{\rm pre}(0,\eta'')k\d\eta''\cdot\nonumber\\
&\cdot\int_0^\infty \d y\int_{|1-y|}^{1+y} \d z\,\left[\frac{4y^2-(1+y^2-z^2)^2}{4yz}\right]^2 {\rm Re}\bigg[\mathcal{Q}_{\delta\phi}(ky,\eta',\eta'')\mathcal{Q}_{\delta\phi}(kz,\eta',\eta'')\bigg].
\end{align}

To compute the term $2\langle\mathcal{P}_{\rm mix}(k,\eta)\rangle_{\rm S}$, we use
\begin{equation}
T_k^h(\eta)\langle h_k^r(0)g_p^s(\eta)\rangle=(2\pi)^3\frac{2\pi^2}{k^3}\langle\mathcal{P}_{\rm mix}(k,\eta)\rangle_{\rm S}\delta^{rs}\delta^3({\bm k}+{\bm p}),
\end{equation}
where $g_k(\eta)$ is defined in Eq.~(\ref{eq:g_def}), and similarly for the $\mathcal{P}_{\rm post}$ term. Following a completely analogous procedure to the one we just applied to compute $\langle\mathcal{P}_{\rm pre}(k,\eta)\rangle_{\rm S}$, we obtain
\begingroup\makeatletter\def\f@size{10}\check@mathfonts
\def\maketag@@@#1{\hbox{\m@th\large\normalfont#1}}
\begin{align}
\langle\mathcal{P}_{\rm mix}(k,\eta)&\rangle_{\rm S}= T_k^h(\eta)\frac{4}{9M_p^2}\frac{H^2\phi^{\prime 2}}{a^2(\rho+p)^2}\bigg|_{0}\int_{\eta_h}^0 \frac{\phi^{\prime 2}}{a^2(\rho+p)}\bigg|_{\eta'}kF_{\rm pre}(0,\eta')k\d\eta'\int_0^\eta
kF_{\rm post}(\eta,\eta'')k\d\eta''\cdot\nonumber\\
&\cdot\int_0^\infty \d y\int_{|1-y|}^{1+y} \d z\,\left[\frac{4y^2-(1+y^2-z^2)^2}{4yz}\right]^2{\rm Re}\bigg[\mathcal{Q}_{\delta\phi}(ky,\eta',0)\mathcal{Q}_{\delta\phi}(kz,\eta',0) \bigg]Q(ky,kz,\eta'')\,,
\end{align}\endgroup
and similarly,
\begingroup\makeatletter\def\f@size{10}\check@mathfonts
\def\maketag@@@#1{\hbox{\m@th\large\normalfont#1}}
\begin{align}
\langle\mathcal{P}_{\rm post}(k,&\eta)\rangle_{\rm S}=\frac{4}{81}\int_{0}^{\eta}
kF_{\rm post}(\eta,\eta')k\d\eta'\int_0^{\eta}
kF_{\rm post}(\eta,\eta'')k\d\eta''\cdot\nonumber\\
&\cdot\int_0^\infty \d y\int_{|1-y|}^{1+y} \d z\,\left[\frac{4y^2-(1+y^2-z^2)^2}{4yz}\right]^2\langle\mathcal{P}_{\mathcal{R}}(ky)\rangle_{\rm S}\langle\mathcal{P}_{\mathcal{R}}(kz)\rangle_{\rm S}Q(ky,kz,\eta') Q(ky,kz,\eta''),
\end{align}\endgroup
where we have used Eqs.~(\ref{eq:psi_ini}) and (\ref{eq:power_id}) to relate $\mathcal{Q}_{\delta\phi}$ to $\mathcal{P}_\mathcal{R}$. This completes the calculation of the tensor power spectrum late into the radiation era.

\subsection{The energy density of induced gravitational waves}
\label{sec:gw_energy_2}
The stochastic average of the energy density of gravitational waves is \cite{Misner:1974qy,Maggiore:1900zz}
\begin{equation}
\langle\Omega_{\rm GW}(\eta,k)\rangle_{\rm S}=\frac{1}{24}\left(\frac{k}{\mathcal{H}}\right)^2\overbar{\langle\mathcal{P}_{h}(\eta,k)\rangle}_{\rm S},
\end{equation}
where the bar denotes a time average over many wavelengths. The value of this quantity at some late time in the radiation era (when the temperature is $T$) can be related to its value today (at temperature $T_0$) assuming entropy conservation \cite{Espinosa_2018},
\begin{equation}
\label{omega_def}
\langle\Omega_{\mathrm{GW}}(T_0,k)\rangle_{\rm S}=\frac{\Omega_\gamma(T_0)}{24}\frac{g_\star(T)}{g_\star(T_0)}\left(\frac{g_{\star s}(T_0)}{g_{\star s}(T)}\right)^{4/3}\left(\frac{k}{\mathcal{H}}\right)^2\overbar{\langle\mathcal{P}_{h}(\eta,k)\rangle}_{\rm S}\,,
\end{equation}
where $g_{\star s}$ is the effective number of entropic degrees of freedom.
The time average can be obtained using
\begin{align}
\overbar{T_k^h(\eta)^2}&=\frac{1}{2(k\eta)^2},\label{average1}\\
k\overbar{T_k^h(\eta)F_{\rm post}(\eta,\eta')}&=\frac{1}{2k\eta}\bigg(\frac{\eta'}{\eta}\bigg)\cos(k\eta'),\label{average2}\\
k^2\overbar{ F_{\rm post}(\eta,\eta')F_{\rm post}(\eta,\eta'')}&=\frac{\eta'\eta''}{2\eta^2}\Big[\cos(k\eta')\cos(k\eta'')+\sin(k\eta')\sin(k\eta'')\Big].\label{average3}
\end{align}
The $\mathcal{H}^2$ factor in the denominator of $\langle\Omega_{\rm GW}\rangle_{\rm S}$ will cancel out the $1/\eta^2$ in these averages, yielding a finite result in the limit $\eta\rightarrow\infty$.\footnote{The upper integration limit should be set to today, but since the scalar source decays quickly during the radiation era, the difference in the result using either limit is negligible \cite{Espinosa_2018}. This choice simplifies the integrals in Eqs.~(\ref{eq:ic},\,\ref{eq:is}). We have also neglected the evolution of the tensor modes during the late matter-dominated era, since the corresponding corrections are highly suppressed for the frequencies we are interested in, see e.g.\,\cite{Fumagalli:2021mpc}.} The energy density of gravitational waves today is therefore
{\fontsize{10.4}{\baselineskip}
\begin{align}
\langle\Omega_{\mathrm{GW}}(T_0,k)\rangle_{\rm S}=\frac{\Omega_\gamma(T_0)}{24}\frac{g_\star(T)}{g_\star(T_0)}\left(\frac{g_{\star s}(T_0)}{g_{\star s}(T)}\right)^{4/3}\int_0^\infty \d y\int_{|1-y|}^{1+y} \d z\,\left[\frac{4y^2-(1+y^2-z^2)^2}{4yz}\right]^2K(ky,kz),
\end{align}}
where the dimensionless integration kernel $K(ky,kz)$ is
\begin{equation}
\label{eq:kernel}
K(ky,kz)=K_{\rm pre}(ky,kz)+K_{\rm mix}(ky,kz)+K_{\rm post}(ky,kz),
\end{equation}
with
\begin{align}
K_{\rm pre}&=\frac{2}{M_p^4}\int_{\eta_h}^0 \frac{\phi^{\prime 2}}{a^2(\rho+p)}\bigg|_{\eta'}k F_{\rm pre}(0,\eta')k\d\eta'\int_{\eta_h}^0 \frac{\phi^{\prime 2}}{a^2(\rho+p)}\bigg|_{\eta''}k F_{\rm pre}(0,\eta'')k\d\eta''\cdot\nonumber\\
&\qquad\cdot{\rm Re}\bigg[\mathcal{Q}_{\delta\phi}(ky,\eta',\eta'')\mathcal{Q}_{\delta\phi}(kz,\eta',\eta'')\bigg],\\
K_{\rm mix}&=\mathcal{I}_s(y,z)\frac{4}{9M_p^2}\frac{H^2\phi^{\prime 2}}{a^2(\rho+p)^2}\bigg|_{0}\int_{\eta_h}^0 \frac{\phi^{\prime 2}}{a^2(\rho+p)}\bigg|_{\eta'}kF_{\rm pre}(0,\eta')k\d\eta'\cdot\nonumber\\
&\qquad\cdot{\rm Re}\bigg[\mathcal{Q}_{\delta\phi}(ky,\eta',0)\mathcal{Q}_{\delta\phi}(kz,\eta',0)\bigg],\\
K_{\rm post}&=\frac{2}{81}\langle\mathcal{P}_\mathcal{R}(ky)\rangle_{\rm S}\langle\mathcal{P}_\mathcal{R}(kz)\rangle_{\rm S}\Big[\mathcal{I}_c^2(y,z)+\mathcal{I}_s^2(y,z)\Big].
\end{align}
and  \cite{Espinosa_2018}
{\fontsize{10.4}{\baselineskip}
\begin{align}
\label{eq:ic}
\mathcal{I}_c(y,z)&=\int_0^\infty k\d\eta'(k\eta')\sin(k\eta')Q(ky,kz,\eta')=36\pi\frac{(s^2+d^2-2)^2}{(s^2-d^2)^3}\Theta(s-1)\,,\\
\mathcal{I}_s(y,z)&=\int_0^\infty k\d\eta'(k\eta')\cos(k\eta')Q(ky,kz,\eta')=-36\frac{(s^2+d^2-2)}{(s^2-d^2)^2}\bigg[\frac{(s^2+d^2-2)}{s^2-d^2}\log\frac{1-d^2}{|s^2-1|}+2\bigg]\,,
\label{eq:is}
\end{align}}
where
\begin{equation}
s=\frac{y+z}{\sqrt{3}},\qquad d=\frac{|y-z|}{\sqrt{3}}.
\end{equation}

\subsection{Induced gravitational waves in the noiseless limit} \label{sec:simple_formula}
For completeness, we present here the expression for the tensor power spectrum valid in the standard cold inflation case; that is, in the absence of the second term in the parentheses of Eq.~(\ref{cal_q_def}). In this case the spectrum can be written as the square of a sum,
{\fontsize{10}{\baselineskip}
\begin{align}
\mathcal{P}_h(k,\eta)=&
\int_0^\infty \d y\int_{|1-y|}^{1+y} \d z\;\left[\frac{4y^2-(1+y^2-z^2)^2}{4yz}\right]^2\mathcal{P}_\mathcal{R}(ky)\mathcal{P}_\mathcal{R}(kz)\nonumber\\
&\bigg|6T_k^h(\eta)\int_{\eta_h}^0 \frac{\rho+p}{\rho}\bigg|_{\eta'}k F_{\rm pre}(0,\eta')S(ky,kz,\eta')k\d\eta'
+\frac{2}{9}\int_{0}^\eta
kF_{\rm post}(\eta,\eta')Q(ky,kz,\eta')k\d\eta'\bigg|^2,
\end{align}}
where {
\begin{equation}
S(ky,kz,\eta)\equiv\frac{\mathcal{R}(ky,\eta)}{\mathcal{R}(ky,0)}\frac{\mathcal{R}(kz,\eta)}{\mathcal{R}(kz,0)}.
\end{equation}
The first term inside the square gives the gravitational waves induced during inflation \cite{Fumagalli:2021mpc} and the second one those produced during radiation domination. The mixing term obtained by developing the square has not been presented before in the literature.} 
The gravitational wave energy density can then be found by using Eq.~(\ref{omega_def}) and computing the time average with Eqs.~(\ref{average1}), (\ref{average2}, (\ref{average3}). {This result is thus useful for computing the full induced (inflation plus radiation) second order gravitational waves in standard (non-dissipative) inflation.}

\subsection{Approximating the time integrals}
\label{sec:gw_integrals}

In this section we estimate the inflationary and post-inflationary contributions to the energy density of the induced gravitational waves in our scenario with a transient dissipative phase. 

Let us focus first on the $K_{\rm pre}$ term of Eq.~(\ref{eq:kernel}). 
This contribution depends on the lower integration limit $\eta_h$ (and as noted in \cite{Fumagalli:2021mpc} is formally divergent in the limit $\eta_h=-\infty$). We deal with this problem by integrating from a finite value of $\eta_h$ that we identify with the time at which the strongly dissipative phase begins. The assumption here is that the contribution from the source prior to {this time is negligible}. This assumption is reasonable since up to that time inflation proceeds as in the standard slow roll scenario (up to the presence of a weak dissipative term that does not alter the dynamics significantly), and we do not expect the corresponding gravitational wave signal to be peaked at any particular scale or exhibit any special features, in contrast to the piece arising due to the strongly dissipative phase.

In addition, we notice that the inflationary contribution to the energy density of gravitational waves diverges in the $y=\infty$ (infinite comoving wave number) limit. In principle, this divergence should be renormalized away by properly computing the induced gravitational wave signal using the in-in formalism. However, we just impose a cut-off which renders the result finite. 

We have verified that our results do not depend on the cutoffs in conformal time and wave number unless unreasonably large values are chosen. We remark that only the $K_{\rm pre}$ and $K_{\rm mix}$ kernels can be affected by these ambiguities, unlike the post-inflationary contribution (which is finite). We also stress that the results of this section should be regarded as an accurate order of magnitude estimate of the overall size of the signal.

We can choose the time cutoff around the time at which the dissipative coefficient $\Gamma$ begins to increase, which corresponds to the departure from standard cold inflation. In terms of the analytical calculation of Section \ref{sec:analytical}, this corresponds to the beginning of phase 1. For the momentum cutoff we can choose $k_{\rm cutoff}\sim\mathcal{O}(10-100)\times k_{\rm peak}$. The four-dimensional integral in $\langle\Omega_{\rm pre}\rangle_{\rm S}$ is quite difficult to perform. 
Fortunately, 
$K_{\rm pre}$ is heavily peaked around a specific time, so the strategy we adopt is to approximate the time integrals by evaluating the integrand at this time and multiplying it by an appropriately chosen integration area.

\begin{figure}[t]
\begin{center}
$\includegraphics[width=.5\textwidth]{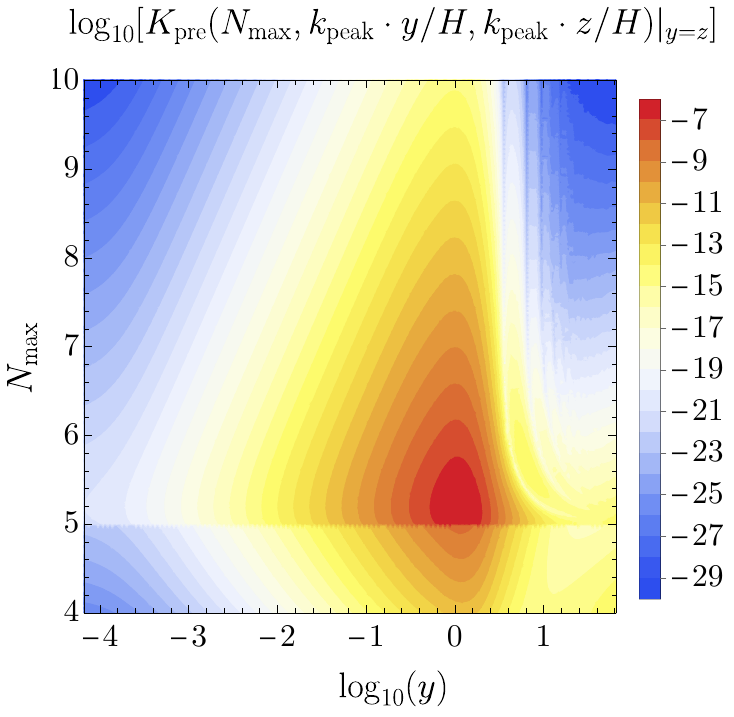}
\qquad\includegraphics[width=.455\textwidth]{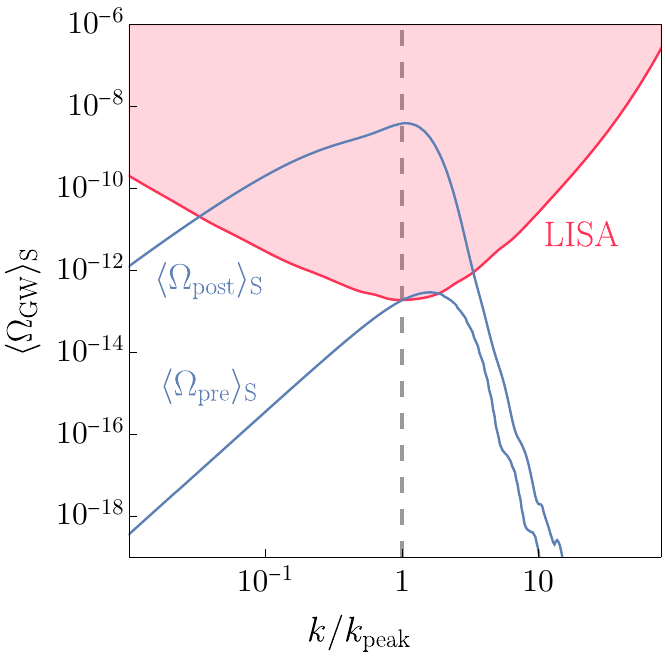}$
\caption{\em \label{fig:gwplot} {{\bf Left}: Maximum value of the integration kernel in Eq.~(\ref{eq:kpre_approx}) along the surface $y=z$ for the parameters in {(the table of) Fig.\,\ref{eq:table}}. {\bf Right}: Gravitational wave signals induced during and after inflation compared with the LISA sensitivity curve \cite{LISA:2017pwj}.}}
\end{center}
\end{figure}

To determine the point in parameter space at which the integrand is peaked, we use the fact that the integrand is symmetric under $\eta'\leftrightarrow \eta''$ and $y\leftrightarrow z$, so the set of maxima of the function must be symmetric under this transformation. If the function has a unique global maximum in some region (we do not prove that this is the case, but we have checked it numerically), then it follows that this maximum must be located along the surface with $\eta'=\eta''$ and $y=z$. On 
this surface
\begin{equation}
K_{\rm pre}\simeq\frac{1}{2M_p^2}\frac{\phi^{\prime 2}}{a^2(\rho+p)}\bigg|_{\eta_{\rm max}}(k\Delta \eta_{\rm max})^2\Big[k F_{\rm pre}(0,\eta_{\rm max})\Big]^2\mathcal{P}_\mathcal{\delta\phi}(ky,\eta_{\rm max})\mathcal{P}_\mathcal{\delta\phi}(kz,\eta_{\rm max}),
\end{equation}
where $\eta_{\rm max}$ is the value of $\eta$ at which the local maximum occurs, and $\Delta \eta_{\rm max}$ is the integration area, which must be appropriately chosen as a small square around $\eta_{\rm max}$ requiring, for instance, that the integrand does not decrease by more than an order of magnitude or so, in such a way that the approximation holds. The integration area is, in terms of the number of $e$-folds,
\begin{equation}
k\Delta\eta_{\rm max}=\frac{k}{H}(e^{-N_a}-e^{-N_b}),
\end{equation}
where $\Delta N=N_b-N_a$ is the range over which the integrand is large, which spans
a couple of $e-$folds at most. Let us write $N_a=N_{\rm max}-\Delta N_{\rm max}$ and $N_b=N_{\rm max}+\Delta N_{\rm max}$, with $\Delta N_{\rm max}\sim\mathcal{O}(1)$; and where $N_{\rm max}$ is the time in $e$-folds corresponding to $\eta_{\rm max}$. Then
\begin{equation}
k\Delta\eta_{\rm max}=2\frac{k}{H}e^{-N_{\rm max}}\sinh(\Delta N_{\rm max}).
\end{equation}
The function we need to maximize is therefore\footnote{Note that the value $N_{\rm max}$ at which $K_{\rm pre}$ peaks is really a function of $y$, as can be seen in the left panel of Fig.~\ref{fig:gwplot}. However, since the largest contribution to the integral comes from the region around $ky=k_{\rm peak}$, to make the calculation numerically less demanding we can simply take $N_{\rm max}$ as the value at which the integrand, evaluated at $ky=k_{\rm peak}$, is peaked, and use the same value $N_{\rm max}$ for all $y$. We have explicitly checked that the peak of the signal remains unchanged if the $y$-dependence of $N_{\rm max}$ is taken into account.}
\begin{equation}
\label{eq:kpre_approx}
K_{\rm pre}=\frac{2k^2}{M_p^2H^2}\frac{\phi^{\prime 2}}{a^2(\rho+p)}\bigg|_{\eta_{\rm max}}\frac{\sinh^2(\Delta N_{\rm max})}{e^{2N_{\rm max}}}\Big(k F_{\rm pre}(0,\eta_{\rm max})\Big)^2\mathcal{P}_{\delta\phi}(ky,\eta_{\rm max})\mathcal{P}_{\delta\phi}(kz,\eta_{\rm max}).
\end{equation}
The quantity in Eq.~(\ref{eq:kpre_approx}) is shown in the left panel of Fig.~\ref{fig:gwplot} for the parameters in {(the table of) Fig.\,\ref{eq:table}}. The discontinuity around $N_{\rm max}=5$ $e$-folds is due to the fact that, as mentioned in Section \ref{sec:analytical}, we take the background parameters as piecewise-constant functions for this calculation. Specifically, we take $\phi^{\prime 2}/a^2(\rho+p)=1$ in phases $0$ and $3$, and $\phi^{\prime 2}/a^2(\rho+p)=0.02$ in phases $1$ and $2$. In this figure we also take $\Delta N_{\rm max}=2$, which is clearly enough to account for the region in which the integrand is large. Changing this number by a factor of $\mathcal{O}(1)$ does not change our results.

To obtain the induced gravitational wave signal, we find the time $\eta$ at which $K_{\rm pre}$ is peaked for each
$k$ 
and perform the momentum integrals numerically. The time-dependent power spectrum $\mathcal{P}_{\delta\phi}(k,\eta)$ is calculated using the analytical formalism of Section \ref{sec:analytical}, as we anticipated earlier. Specifically, it can be found by keeping the full expression for the Green's function in Eq.~(\ref{eq:full_Green}) instead of taking the $N\rightarrow\infty$ limit. The resulting signal is shown in the right panel of Fig.~\ref{fig:gwplot}. We find that the energy density of gravitational waves induced during inflation is much smaller than that of the gravitational waves induced during the radiation era. We do not show in this figure the mixed term from Eq.~(\ref{eq:kernel}), but we find that it is well approximated by $\langle\Omega_{\rm mix}\rangle_{\rm S}\sim\sqrt{\langle\Omega_{\rm post}\rangle_{\rm S}\langle\Omega_{\rm pre}\rangle_{\rm S}}$, and is therefore also suppressed with respect to the post-inflationary contribution.

\begin{figure}[t]
\begin{center}
$\includegraphics[width=.47\textwidth]{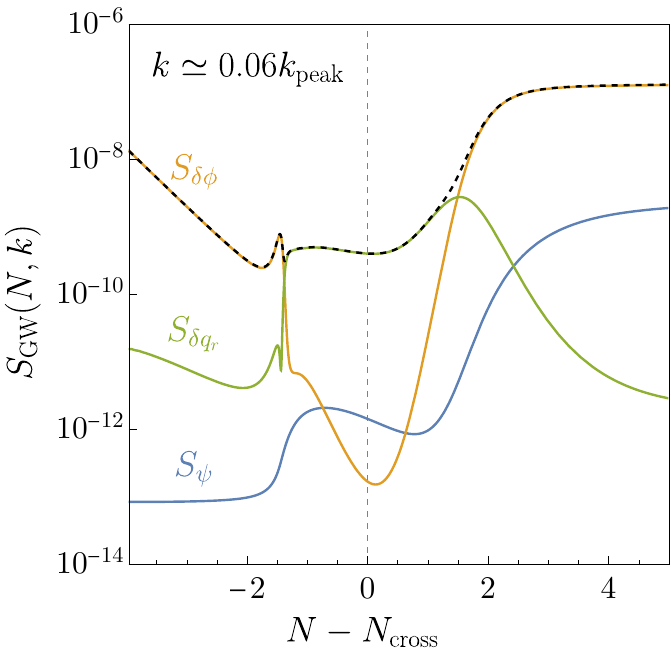}
\qquad\includegraphics[width=.47\textwidth]{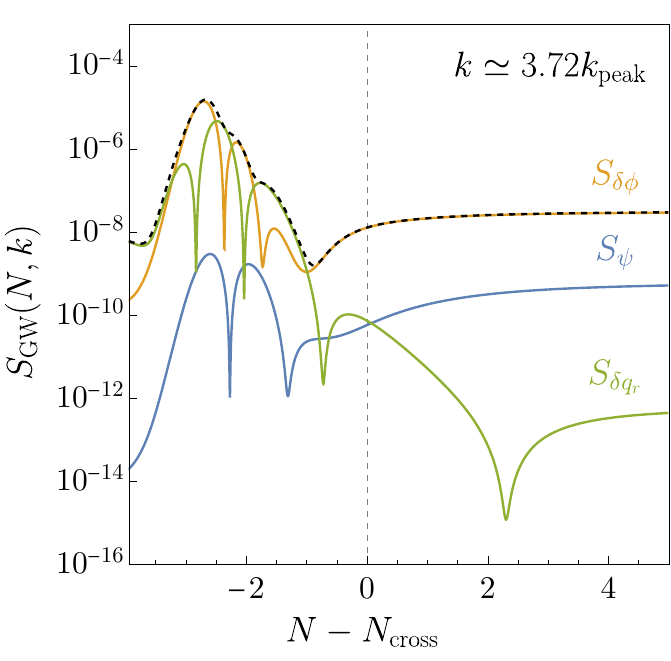}$
\caption{\em \label{fig:gwsource} {The different components entering into the quantity $S_{\rm GW}$ defined in Eq.~(\ref{eq:source_approx}) are shown as a function of time for two different values of $k$. We use the numerical results of Section~\ref{sec:matrix}.}}
\end{center}
\end{figure}

We stress that  approximating the integrand by its peak value is what allows us to neglect the subdominant terms involving $\psi$ and $\delta q_r$ in Eq.~(\ref{eq:gw_source}). Let us show that this is a good approximation by estimating the relative contribution of each term in this equation during the inflationary epoch. Since the integrand in Eq.~(\ref{eq:gw_source}) is symmetric under ${\bm p}\leftrightarrow {\bm k}-{\bm p}$, for the purpose of finding out which terms contribute the most at the point at which this integrand reaches its largest value, we can simply evaluate it at $|{\bm p}|=|{\bm k}-{\bm p}|$ as we did for $K_{\rm pre}$. We can then define
\begin{equation}
\label{eq:source_approx}
S_{\rm GW}(N,k)\equiv \underbrace{8\langle\mathcal{P}_\psi(N,k)\rangle}_{S_\psi}+\underbrace{\frac{4}{(\rho+p)M_p^2}\langle\mathcal{P}_{\delta q_r}(N,k)\rangle}_{S_{\delta q_r}}+\underbrace{\frac{4\phi^{\prime 2}}{a^2(\rho+p)M_p^2}\langle\mathcal{P}_{\delta\phi}(N,k)\rangle}_{S_{\delta\phi}},
\end{equation}
where we have ignored the mixed terms in Eq.~(\ref{eq:gw_source}), since they are always subdominant. This quantity is shown in Fig.~\ref{fig:gwsource} for two different Fourier modes, both of which become super-Hubble near the end of the strongly dissipative phase. The figure illustrates that the time integral of the scalar source of noise is dominated by the $\delta\phi$ term. Since the other contributions are at most of the same order and will therefore not change the size of the peak in the gravitational wave signal, we can neglect them.

Let us briefly summarize the results of this section. We have made three different approximations in this calculation. The first is that we approximated the time integrals as their peak value times an appropriately chosen area. The second is that we have neglected the contribution of the radiation and metric terms to the gravitational wave source during the inflationary epoch, and we have checked explicitly that the contribution of these terms is indeed subdominant. These two approximations are very good and should not change the order of magnitude of the result.  If the time integrals are performed numerically and the metric and radiation perturbations are included in the source term, we expect that the size of the signal will change at most by a factor of $\mathcal{O}(1)$. The third and final approximation is related to the divergences in the far past and for large momenta due to the behaviour of scalar modes in the Bunch-Davies vacuum. We have assumed that this effect can be taken into account correctly by imposing reasonable cutoffs. 
We have checked that our results are independent of the cutoffs unless an unreasonably large choice is made. The uncertainty due to this approximation is difficult to quantify, but
our results should be correct as an order of magnitude estimate.

\section{Conclusions}

We have shown that a period of dissipation lasting a few $e$-folds during inflation can lead to a large and peaked primordial spectrum of curvature perturbations $\mathcal{P_R}$ at specific scales. This may provide the seeds of an abundant PBH population capable of accounting for all the dark matter of the Universe. The large power spectrum that is required in the most naive estimates ($\mathcal{P_R}\sim 10^{-2}$) leads to a stochastic background of gravitational waves that we have computed using second order cosmological perturbation theory. As it is well known, given the current astrophysical bounds on the PBH abundance, this background of gravitational waves falls within the sensitivity reach of the future LISA interferometer.

The growth of $\mathcal{P_R}$ is due to the stochastic nature of the dissipation. Thermal fluctuations of the radiation background that originates from the inflaton transferring its kinetic energy to lighter degrees of freedom act as a stochastic source for the curvature perturbation. This makes the mechanism very different from those operating in other inflationary setups that can produce abundant PBH dark matter (in particular, ultra slow-roll from an inflection point in single-field inflation).

We have described a method for computing the curvature power spectrum that allows to overcome the numerical difficulties associated to the stochastic character of the system of equations governing the time evolution of the perturbations. The method consists in formulating the problem as a system of deterministic differential equations for the two-point correlation functions in Fourier space. We have verified the validity and accuracy of the method by solving the stochastic equations for many realizations and computing statistical averages. The method we have presented can be useful in more general contexts in cosmological perturbation theory where stochastic sources are present, beyond the specific scenario we have discussed in this work. We have also shown that a single, simplified, stochastic differential equation with an analytical solution can be used to understand qualitatively the enhancement of the primordial spectrum.

Our analysis of the stochastic background of gravitational waves includes two aspects that make it richer than that of previous works. First, there is the unavoidable stochastic origin of the primordial curvature spectrum, which propagates into the calculation of tensor modes at second order and must be appropriately taken into account. In addition, we have computed not only the gravitational waves from modes entering the horizon during radiation domination, but also those generated at second order during inflation, and we have written an explicit expression for the term that describes the mixing between the two contributions. {As a particular case, we also write for the first time  an expression including the mixing term in standard (non-dissipative) inflation.} These formulas could be useful in contexts beyond the scenario presented here. {Applying} the analytical approximation that we derive for the primordial curvature spectrum, we estimate the gravitational wave signal and find that it is 
dominated by the gravitational waves induced during the radiation-dominated era that follows inflation.

We have used a phenomenological parameterization to model the transfer of energy from the inflaton to the radiation background. Although there have been serious efforts on inflationary model-building aimed at obtaining significant dissipation throughout inflation (specifically, within the framework of warm inflation), we are not aware of any previous works proposing a localized dissipative period akin to the one we have explored. We have discussed, in Appendix \ref{app:micro}, a potential route to describe the microphysical origin of a peaked dissipative coefficient such as the one we propose by starting from a particular Lagrangian, but further work is needed to find concrete, well-motivated models.

The stochastic nature of the peak in the primordial spectrum is a general property of our scenario. We find that the spectrum presents a series of oscillations after the peak, a feature that is not commonly found in inflection-point models of PBH production. Similarly, the dip that is commonly seen before the peak of the spectrum in this class of models is absent in our scenario. Other details of the spectral shape and, consequently, the gravitational wave signal, are model dependent. It would be interesting to investigate whether there may be any features that could help distinguish our mechanism from other models should PBHs and a stochastic background of gravitational waves be detected in the adequate masses and frequencies of interest for dark matter.

\begin{acknowledgments}
The authors thank Mar Bastero-Gil for discussions. GB thanks J.\ Gamb\'in Egea for discussions about second order induced gravitational waves. The work of GB has been funded by a Contrato de Atracci\'on de Talento (Modalidad 1) de la Comunidad de Madrid (Spain), 2017-T1/TIC-5520 and 2021-5A/TIC-20957. The work of GB, JR and APR has been funded by
the IFT Centro de Excelencia Severo Ochoa Grants SEV-2016-0597 and CEX2020-001007-S and by MCIU (Spain) through contract PGC2018-096646-A-I00. JR and APR have been both supported by Universidad Aut\'onoma de Madrid via PhD contracts {\it contratos predoctorales para formaci\'on de personal investigador (FPI)}, calls of 2020 and 2021, respectively. MP acknowledges support by the Deutsche Forschungsgemeinschaft (DFG, German Research Foundation) under Germany's Excellence Strategy – EXC 2121 “Quantum Universe” – 390833306. MG acknowledges the support from the Instituto de F\'isica, UNAM in procuring computational resources. This work was made possible with the support of the Institut Pascal at Universit\'e Paris-Saclay during the Paris-Saclay Astroparticle Symposium 2021, with the support of the P2IO Laboratory of Excellence (program “Investissements d’avenir” ANR-11-IDEX-0003-01 Paris-Saclay and ANR-10-LABX-0038), the P2I axis of the Graduate School Physics of Universit\'e Paris-Saclay, as well as IJCLab, CEA, IPhT, APPEC, the IN2P3 master projet UCMN and EuCAPT ANR-11-IDEX-0003-01 Paris-Saclay and ANR-10-LABX-0038.
\end{acknowledgments}

\appendix

\section{Microphysics of the dissipative coefficient}
\label{app:micro}

In this appendix we {discuss} a particular microphysical realization of a localized dissipative coefficient $\Gamma$ during inflation. 
The purpose of the present discussion, however, is not to propose a definitive model but simply to show that obtaining a peaked dissipative coefficient that satisfies all the necessary constraints is, in principle, possible. Specifically, we introduce a Lagrangian which reproduces the form of the coefficient (\ref{eq:gammapheno}), assuming that the fields that couple to the inflaton are part of a thermalized bath. In order to keep the field content to a minimum we will consider a scenario where, besides the inflaton $\phi$, only two additional degrees of freedom participate in the dynamics. The first one, denoted by $\sigma$ and assumed to be a scalar, corresponds to 
{a} light radiation field in equilibrium. The second one, also a scalar and denoted by $\chi$, corresponds to a heavy catalyst field. The large effective mass of $\chi$ arises via its coupling to the slowly rolling inflaton field with a non-zero vev.  Through the coupling of the (unstable) $\chi$ with $\sigma$, the inflaton energy density can be efficiently dissipated into radiation. 
Indirect decay scenarios like this one are among the preferred mechanisms for realistic warm inflation models~\cite{Berera:2001gs,Berera:2002sp,Bastero-Gil:2009sdq}. One of the advantages of introducing {a heavy catalyst field} is to prevent the inflaton potential from receiving strong temperature corrections~\cite{Hall:2004zr,Bastero-Gil:2009sdq}.\footnote{In addition, in a simpler construction in which the inflaton directly couples to the radiation field $\sigma$, the dissipation rate is determined by the strength of the inflaton self-coupling. This effectively suppresses the value of $\Gamma$, due to the requirement of the normalization of such self-coupling by the amplitude of the primordial curvature power spectrum~\cite{Berera:2001gs}.}

Consider the Lagrangian 
\beq\label{eq:fundlag}
\mathcal{L} \;=\; -\frac{1}{2 \mathcal{K}(\varphi)}\partial^{\mu}\varphi \partial_{\mu}\varphi 
 - \frac{1}{2}\partial^{\mu}\chi \partial_{\mu}\chi 
- \frac{1}{2} \partial^{\mu}\sigma \partial_{\mu}\sigma 
  - \frac{1}{2}g^2 \varphi^2\chi^2 - \frac{1}{2}\tilde{g}^2 \varphi\chi\sigma^2 - \tilde{V}(\varphi) + \cdots\,.
\eeq
Here $\varphi$ denotes a non-canonically normalized inflaton, due to the presence of the function $\mathcal{K}(\varphi)$ in its kinetic term. In this basis, the inflaton and the mediator $\chi$ interact via a four-legged contact term with coupling $g$. On the other hand, the term with coupling $\tilde{g}$ connects the three fields. For a non-vanishing inflaton vev, this term induces the decay of $\chi$ into $\sigma$. The ellipsis corresponds to the interactions which are necessarily present in order to thermalize the light sector $\sigma$. If this sector is indeed in equilibrium, the presence of the $\phi\rightarrow \chi\rightarrow \sigma$ channel modifies the equation of motion of the inflaton. In terms of the canonically normalized inflaton $\phi$, related to $\varphi$ via
\beq\label{eq:fieldredef}
\frac{\diff \phi}{\diff \varphi} \;=\; \frac{1}{\sqrt{\mathcal{K}(\varphi)}}\,,
\eeq
the effective equation of motion for $\phi$ can be written as
\beq
\ddot{\phi} + (3H+\Gamma)\dot\phi + V_{\phi}(\phi,T) \;=\; 0\,,
\eeq
where the potential $V(\phi,T)$ includes the thermal correction due to the presence of a bath of $\sigma$ particles (fluctuation), while $\Gamma$ encodes the production of $\sigma$ quanta (dissipation). The appearance of a local dissipative term relies on the assumption that the microphysical processes which determine $\Gamma$ operate at time-scales much smaller than those characteristic of the macroscopic slow roll of the inflaton and the expansion of the Universe 
(the adiabatic-Markovian approximation~\cite{Berera:1998gx,Berera:2001gs}). Additionally, we assume that the typical interaction time-scale between the constituents of the thermal bath is much shorter than the time-scales associated to the variation of the background quantities. In this approximation, the dissipative coefficient can be evaluated as~\cite{Hosoya:1983ke,Berera:2008ar,Bastero-Gil:2010dgy}
\beq\label{eq:gammagen}
\Gamma(\phi,T) \;=\; \frac{g^4(\partial_\phi \varphi^2)^2}{2T}\int \frac{\diff^4p}{(2\pi)^4}\, n(\omega)\big[n(\omega)+1\big]\rho_\chi^2(\omega)\,,
\eeq
where $\omega$ denotes the 0-th component of the 4-momentum $p$, $n_{\chi}(\omega) = (e^{\omega/T}-1)^{-1}$ is the Bose-Einstein distribution, and 
\begin{equation}
\rho_\chi \;=\; \frac{4\omega_p\Gamma_\chi}{(\omega^2-\omega_p^2)^2+4\omega_p^2\Gamma_\chi^2}\,,
\end{equation}
is the spectral density of the catalyst field $\chi$. In this expression $\omega_p^2=|{\bm p}|^2+m^2_\chi$ is the on-shell frequency and $\Gamma_\chi$ is the decay width of $\chi$. Note that our expression for $\Gamma$ originates from the $\phi-\chi$ interaction term in (\ref{eq:fundlag}). In principle, the term proportional to $\tilde g^2$ would also contribute to the dissipation rate, although this contribution is loop-suppressed \cite{Bastero-Gil:2010dgy}. 

In order to evaluate the dissipation coefficient, we need the decay width $\Gamma_{\chi}$. {In general, this width} must be computed in non-zero temperature QFT, and a general expression can be found in e.g.~\cite{Bastero-Gil:2010dgy}. For simplicity, we will restrict our discussion to the so-called low-temperature limit in what follows, which corresponds to assuming that $T\ll m_{\chi}$. We remark however that in principle none of our assumptions forbid a peaked dissipative coefficient at higher temperatures. In the low-temperature limit the decay rate for the process $\chi \rightarrow \sigma \sigma $ may be written as
\beq
\Gamma_{\chi}(p_0,\bp) \;\simeq\; \frac{\tilde{g}^4\varphi^2}{8\pi\omega_{p}(\bp)}\,,
\eeq
in a frame boosted with respect to the rest frame of $\chi$. In terms of this rate, the strong dissipative condition, necessary to maintain thermal equilibrium between the light $\sigma$ and the heavy $\chi$, and the adiabaticity requirement correspond to
\begin{equation}
\label{eq:conditions}
\Gamma_\chi \;\gg\; H \quad{(\rm thermalization)},\qquad \Gamma_\chi \;\gg\; \frac{\dot{\phi}}{\phi}\,, \  \frac{\dot{T}}{T}  \quad{(\rm adiabaticity)}\,.
\end{equation}
Moreover, in this low-temperature regime, the thermal correction to the inflaton potential can be neglected, $V(\phi,T)\simeq \tilde{V}(\varphi(\phi))$~\cite{Hall:2004zr,Bastero-Gil:2009sdq}. Introducing the dimensionless quantities    
\begin{equation}
u=\frac{\omega}{T},\qquad \mathfrak{p}=\frac{|\bp|}{m_\chi},\qquad\tau=\frac{T}{m_\chi},\qquad\gamma_\chi=\frac{\Gamma_{\chi,0}}{m_\chi}\,,
\end{equation}
where $\Gamma_{\chi,0}=\Gamma_\chi(m_{\chi},\boldsymbol{0})$, and $m_{\chi}^2=g^2\varphi^2$, c.f.~(\ref{eq:fundlag}), the dissipation rate (\ref{eq:gammagen}) can be written as
\begin{equation}
\Gamma\;=\;\frac{16g^4(\partial_\phi \varphi^2)^2}{(2\pi)^3m_\chi}\int_0^\infty \d \mathfrak{p}\int_{\mathfrak{p}/\tau}^\infty \d u\bigg[\frac{\mathfrak{p}\gamma_{\chi}}{(u^2\tau^2-1-\mathfrak{p}^2)^2+4\gamma_{\chi}^2}\bigg]^2\bigg(\frac{1}{e^u-1}\bigg)\bigg(\frac{1}{e^u-1}+1\bigg)\,.
\end{equation}
At low temperatures $\tau\ll 1$, the first term of the integrand can be simplified disregarding the $u$ and $\gamma_{\chi}$-dependent terms in the denominator (equivalent to approximating the spectral function as $\rho_\chi \simeq 4 \Gamma_\chi / m_\chi^3$~\cite{Bastero-Gil:2010dgy}). The dissipation rate then simplifies to
\begin{equation}
\Gamma \;\simeq\; \frac{16g^4(\partial_\phi \varphi^2)^2\gamma_{\chi}^2}{(2\pi)^3m_\chi}\int_0^\infty \d \mathfrak{p}\bigg(\frac{\mathfrak{p}^2}{1+\mathfrak{p}^2}\bigg)\bigg(\frac{1}{e^{\mathfrak{p}/\tau}-1}\bigg) \;\simeq\; \frac{40g^4(\partial_\phi \varphi^2)^2 \Gamma_{\chi,0}^2T^3}{(2\pi)^3m_\chi^6}\,,
\label{eq:approximateGammamodel}
\end{equation}
where we have used the fact that the $\mathfrak{p}$-integral can be well-approximated by $5\tau^3/2$ for $\tau \ll 1$.  

{In order to recover now} the phenomenological peaked dissipation rate (\ref{eq:gammapheno}), consider the following form of the non-canonical kinetic term function for $\varphi$, 
\begin{equation}
\mathcal{K}(\varphi)=\frac{4\varphi^2 }{\Lambda^2}\bigg[1+\frac{4M^2\varphi}{m(\varphi+m)^2}\bigg] \,,
\end{equation}
where $m,M$ and $\Lambda$ are arbitrary dimensionful parameters. The corresponding solution to (\ref{eq:fieldredef}) can then be written as
\begin{equation}
\label{eq:field_redef}
\varphi(\phi)=m\bigg(\frac{1 + f(\phi)}{1 - f(\phi)}\bigg),\qquad f(\phi) \equiv \left( \frac{M^2+m^2}{M^2+m^2\coth^2\big[(\phi-\phi_\star)/\Lambda \big]} \right)^{1/2}\,,
\end{equation}
where $\phi_{\star}$ is a free parameter, which we assume is necessary to fix the position and value of the scalar potential $V(\phi)$ at its minimum. By injecting this transformation in Eq.~(\ref{eq:approximateGammamodel}), in the limit where a hierarchy is present among parameters $m \ll M$, the dissipation rate reads
\begin{equation}
\Gamma \;\simeq\; \frac{5\tilde{g}^8 M^2}{4\pi^5g^4\Lambda^2}\bigg\{\frac{T^3}{m^2+M^2\tanh^2\big[(\phi-\phi_\star)/\Lambda\big]}\bigg\},
\end{equation}
which possesses the same analytical scaling as the dissipation rate of Eq.~(\ref{eq:gammapheno}) upon parameter redefinition.

\begin{figure}[t]
\begin{center}
$\includegraphics[width=.485\textwidth]{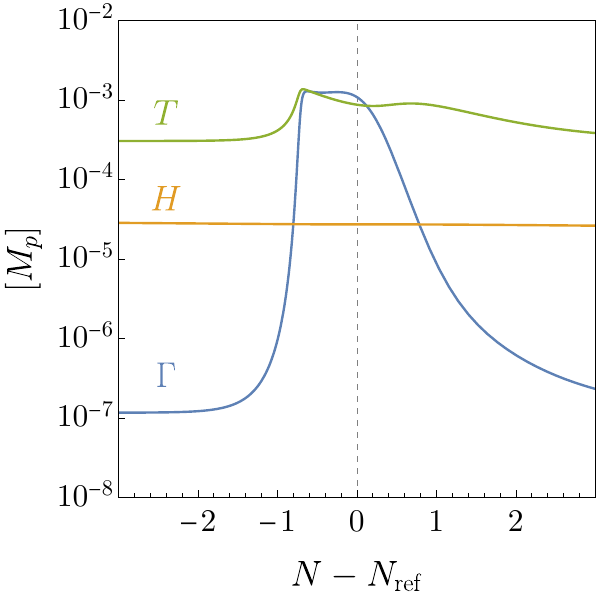}
\qquad\includegraphics[width=.47\textwidth]{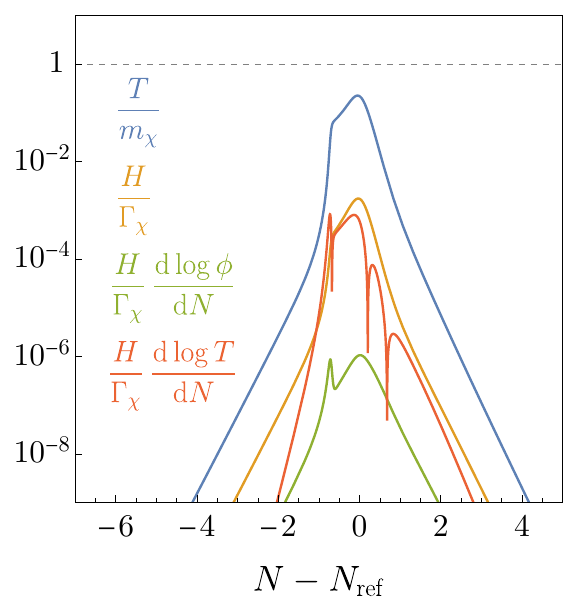}$
\caption{\em \label{fig:model} {{\bf Left:} Time evolution of the background quantities $T$, $H$ and $\Gamma$ for the dissipative coefficient using parameters of Eq.~(\ref{eq:model_parameters}). {\bf Right:}  Constraints from Eqs.~(\ref{eq:conditions}) as well as the low-temperature limit $\tau\ll1$, for the same parameters as the left panel. The reference time $N_{\rm ref}$ is chosen as the time at which $T/m_\chi$ reaches its maximum value.}}
\end{center}
\end{figure}
Fig.~\ref{fig:model} depicts the numerical solution of the background equations of motion (\ref{eq:backg1})-(\ref{eq:backg3}) for the  following choice of parameters which determine the dissipation rate,
\begin{equation}
\label{eq:model_parameters}
\Lambda\;=\;0.1M_p,\quad M\;=\;M_p,\quad m\;=\;0.05M_p,\quad \phi_\star\;=\;21 M_p,\quad g\;=\;0.08,\quad \tilde{g}\;=\;0.28\,,
\end{equation}
In addition, $\lambda=4.6 \times 10^{-14}$, $g_\star=10$ in (\ref{eq:rhor}), (\ref{eq:quartic}). The left panel shows the time evolution of the rate $\Gamma$, together with the expansion rate $H$ and the temperature $T$ in terms of the number of $e$-folds $N$. The right panel shows our consistency checks for the conditions in Eq.~(\ref{eq:conditions}), as well as the small temperature condition $T/m_\chi \ll 1$. The self-consistency of our set-up is therefore guaranteed.

As a final remark, we note that we assume that the potential $V(\phi)=\tilde{V}[\varphi(\phi)]$ supports {slow-roll inflation, and in our particular example it is a quartic polynomial} (\ref{eq:quartic}). Owing to the relation between the canonical and non-canonical forms of the inflaton field, given by (\ref{eq:field_redef}), $\tilde{V}(\varphi)$ would then be a relatively complicated function of $\varphi$. In any case, as we discuss in Section~\ref{sec:pheno}, the precise shape of the inflaton potential does not affect our conclusions. This is true as long as {the potential} does not possess any features that interfere {with} 
the effect of the dissipation rate $\Gamma$, which is the main quantity that determines the growth of fluctuations. We expect that the discussion presented in this appendix encourages further efforts to search for inflation scenarios which lead to peaked dissipative coefficients.

\section{Temperature-independent dissipative coefficient}

\begin{figure}[t]
\begin{center}
$\includegraphics[width=.7\textwidth]{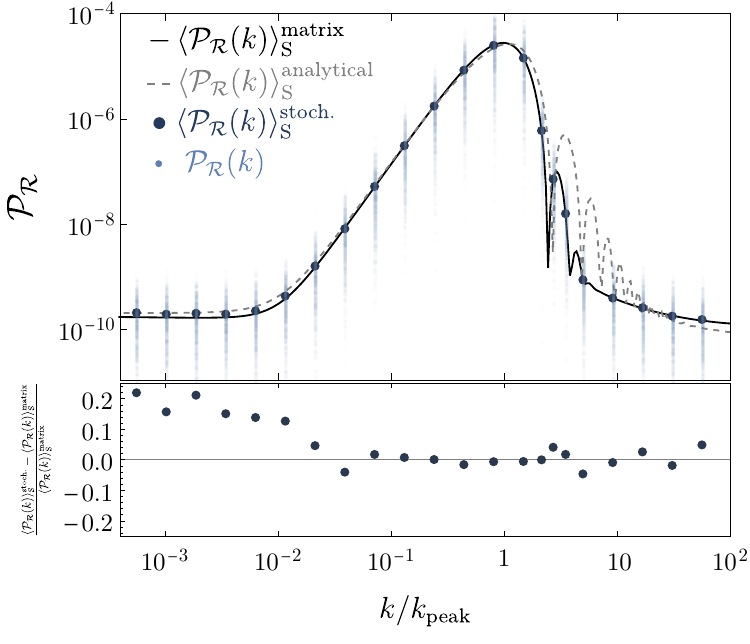}$
\caption{\em \label{fig:PR_stochastic2} {{\bf Top:} stochastic average of the power spectrum for 22 different values of $k$ (dark blue dots). The number of realizations for each value of $k$ is 900 (represented as the small, light blue dots). The solid black line represents the average of the power spectrum obtained via the deterministic matrix differential equation and the grey dashed line the results from the analytical approach. {\bf Bottom:} Absolute value of the relative difference between the stochastic average and the matrix average of the power spectrum. The agreement for each $k$ is at the percent level around the peak and of order $10-20 \%$ away from the peak.}}
\end{center}
\end{figure}

In this appendix, we investigate the possibility {that the dissipation coefficient is temperature-independent} and 
compute the power spectrum using the three methods presented previously. In order to assess the effects of the temperature dependence of the dissipative coefficient on the power spectrum, we consider a coefficient with the same dependence on the inflaton field
as 
in Eq.~(\ref{eq:gammapheno}): 
\begin{equation}\label{eq:gammapheno2}
\Gamma(\phi)= \frac{\Gamma_0}{\alpha^2+\tanh^2\left[(\phi-\phi_\star)/\Lambda\right]}\,.
\end{equation}
As a benchmark example, we choose the following set of parameters: 
\begin{equation}\label{eq:parametersgammapheno2}
g_\star =8,\quad \phi_\star = 22M_p, \quad \Gamma_0 = 2.7 \times 10^{-7} M_p, \quad \alpha = 1.4 \times 10^{-2}, \quad  \Lambda = 0.15 M_p, \quad  \lambda = 2.5 \times 10^{-15} \,.
\end{equation}
Given the fact that the dissipative coefficient is {temperature-independent}, in order to compute the curvature power spectrum one can directly apply the matrix formalism approach of Sec.~\ref{sec:matrix} or solve the system of Langevin equations as in Sec.~\ref{sec:stochastic} by setting terms proportional to $\Gamma_T$ to zero. The power spectrum obtained using these two approaches in represented in Fig.~\ref{fig:PR_stochastic2} showing a good agreement between averages over stochastic realizations and the matrix formalism approach, typically at the percent level around the peak and of the order of $10-20\%$ away from the peak.

In the analytical approach of Sec.~\ref{sec:analytical}, we argued that the fluctuations $\delta \phi$ were driven by the stochastic noise via the $\delta \rho_r$ term. However, this term is proportional to $\Gamma_T$ and vanishes for the temperature-independent dissipative coefficient. Therefore, the driving stochastic term in the equation for the fluctuation $\delta \phi$ is the explicit stochastic noise term, appearing on the right-hand-side of Eq.~(\ref{eq:phieom}). By setting $\Gamma_T=0$, the approximate equation for $\delta\phi$ can be written
as
\begin{equation}\label{eq:deltaphieom2}
\frac{\diff^2 \delta \phi}{\diff N^2}+\left(3+\frac{\Gamma}{H}\right)\frac{\diff \delta\phi}{\diff N} + \left(\frac{k^2}{a^2H^2}+\frac{\Gamma_\phi}{H}\frac{\diff\phi}{\diff N}\right)\delta\phi=\sqrt{\frac{2\Gamma T}{a^3H^3}}\xi_N \,,
\end{equation}
after neglecting metric perturbations and slow-roll-suppressed terms. The correlation function for $\xi_N$ is
\begin{equation}
\langle\xi_N(\bm{k})\xi_{\hat{N}}(\bm{q})\rangle_S=(2\pi)^3\delta(N-\hat{N})\delta^3(\bm{k}-\bm{q}),
\end{equation}
which makes Eq.~(\ref{eq:deltaphieom2}) formally identical to Eq.~(\ref{deltaphieom}) upon substituting $\xi_N \rightarrow \xi_N^{\delta \rho}$ and rescaling the prefactor of the right-hand-side of Eq.~(\ref{eq:deltaphieom2}) by the appropriate background quantities. Following the approach of Sec.~\ref{sec:analytical}, we consider
four time
intervals. In the case of Eq.~(\ref{eq:deltaphieom2}), the parameters $s$ and $S$ introduced in Eq.~(\ref{eq:smalls}) and Eq.~(\ref{eq:largeS}) become simply
$s=3$ and $S=1$ throughout the four phases. The dissipative coefficient and relevant background quantities used for the analytical computation are listed in Table~\ref{tab:table} for the four different phases. The resulting power spectrum is depicted in dashed-grey in Fig.~\ref{fig:PR_stochastic2}.  The analytical spectrum reproduces well the results from the matrix formalism approach at small $k/k_{\rm peak} \ll 1$ and around the peak. The main differences can be seen in the oscillation pattern at larger $k/k_{\rm peak} \gtrsim 1$. 
The analytical approximation matches the numerical result significantly better than in the model with temperature-dependent $\Gamma$. The reason is that the calculation in Sec.~\ref{sec:analytical} involves more approximations, mainly neglecting $\xi_N$ with respect to $\delta\rho_r$ and {parameterizing} $\delta\rho_r$ by Eq.~(\ref{eq:rho_attract}).  

We conclude that the analytical approach developed in Sec.~\ref{sec:analytical} is consistent with both the matrix formalism and the stochastic approach presented in this paper. This approach allows to describe and characterize the primordial power spectrum also
for a temperature-independent dissipation rate. We have shown that a temperature-dependent dissipative rate is actually not necessary in order to achieve an enhanced power spectrum.

\begin{center}
\begin{table}
\begin{tabular}{||c||c|c|c|c||}
\hline\hline &
{\bf Phase 0} &
{\bf Phase 1} &
{\bf Phase 2} &
{\bf Phase 3} \\
\hline\hline
\multirow{2}{*}{$\Sigma\;[M_p^{-1/2}]$} &
\multirow{2}{*}{$5.5\times 10^{2}$} &
    \multirow{2}{*}{$5.7\times 10^{4}$} &
\multirow{2}{*}{$5.7\times 10^{4}$} &
\multirow{2}{*}{$5.5\times 10^{2}$} \\
& & & & \\ \hline
\multirow{2}{*}{$\Gamma\;[M_p]$} &
\multirow{2}{*}{$3 \times 10^{-7}$} & 
\multirow{2}{*}{$1.4 \times 10^{-3}$} &
\multirow{2}{*}{$1.4 \times 10^{-3}$} &
\multirow{2}{*}{$3 \times 10^{-7}$} \\
& & & & \\ \hline
\multirow{2}{*}{$\frac{\Gamma_\phi}{H}\frac{\diff \phi}{\diff  N}$}  &
\multirow{2}{*}{0} & 
\multirow{2}{*}{$4\times 10^2$} &
\multirow{2}{*}{$-4\times 10^2$} &
\multirow{2}{*}{0} \\ 
& & & & \\ \hline\hline
\end{tabular}
\caption{{\it Benchmark parameters for the analytical calculation of the power spectrum. We take phases 1 and 2 to end at $N_1=2$ and $N_2=4$, respectively (we measure the number of $e$-folds from the end of phase 0, so that $N_0=0$, and we normalize $a(N_0)=1$ as well as $H=7\times 10^{-6} M_p$.}} 
\label{tab:table}

\end{table}
\end{center}

\section{Stochastic differential equations}
\label{app:Diffeqs}

In this appendix we review some basic elements of stochastic differential equations, including the definition of a Wiener process and the derivation of the Fokker-Planck equation. Part of the discussion follows the presentation in \cite{jacobs_2010}.

\subsection{Stochastic calculus}

Let us consider a discrete-time equation
\begin{equation}\label{ec:inc1}
\Delta y = f(y,t)\Delta t + g(y,t)\Delta W_t,
\end{equation}
where $\Delta t$ are the time increments between consecutive time nodes. In other words,
\begin{equation}
\Delta t = \frac{T}{N},
\end{equation}
where $T$ is the time range over which we consider the equation, and $N$ is the number of time nodes in which said time range is divided. The increment $\Delta W_t$ (known as \emph{Wiener increment}) is randomly drawn from a Gaussian distribution at each time step,
\begin{equation}
\label{wiener_dist}
P(\Delta W_t)=\frac{1}{\sqrt{2\pi\sigma^2}}e^{-\Delta W_t^2/2\sigma^2}.
\end{equation}
The $\Delta W_t$ increments at every time step are independent from each other. The different solutions to Eq.~\eqref{ec:inc1} obtained via some finite sequence of random increments $\Delta W_t$ are referred to as \emph{realizations}. Formally taking the \emph{continuum limit} $N\to\infty$ turns \eqref{ec:inc1} into a \emph{stochastic differential equation}
\begin{equation}\label{ec:inc0}
dy = f(y,t)\d t + g(y,t)\d W_t.
\end{equation}
We will discuss this limit later on. For the moment, let us continue with non-infinitesimal time increments.

An important fact is that the only way for the distribution in Eq.~(\ref{wiener_dist}) to be well-defined in the context of a problem like \eqref{ec:inc1} is if $\sigma^2=\Delta t$. Let us illustrate this with a heuristic argument {in a simplified case.} If we consider the equation $\Delta y=\Delta W_t$, a particular realization can be obtained by summing each random increment, $y=\sum_i^N \Delta W_{t(i)}$. The variance of $y$, denoted by $V(y)$, is simply the sum of the variance of each increment $\Delta W_{t(i)}$, 
since they are assumed to be independent. Let us impose $\sigma^2$ is equal to some power of $\Delta t$,
\begin{equation}
\sigma^2=\Delta t^n.
\end{equation}
Then, the variance of $y$ is
\begin{equation}
V(y)=\sum_i^NV(\Delta W_{t(i)})=N\sigma^2=N\Delta t^n=N\left(\frac{T}{N}\right)^n=N^{1-n}T^n,
\end{equation}
This implies a consistency condition in the continuum limit. Indeed, taking $N\rightarrow\infty$, $V(y)$ vanishes for $n>1$, and it diverges for $n<1$. Thus, the only acceptable value is $n=1$. 
Let us return to Eq.~\eqref{ec:inc1} and compute the expectation value of $\Delta W_t^2$,
\begin{equation}
\langle \Delta W_t^2\rangle=\langle \Delta W_t^2\rangle-\langle \Delta W_t\rangle^2\equiv V(\Delta W_t)=\Delta t,
\end{equation}
where the expectation values $\langle\cdots\rangle$ are taken with respect to the distribution in Eq.~(\ref{wiener_dist}), and we have used the fact that $\langle \Delta W_t\rangle=0$. By using Eq.~(\ref{wiener_dist}), we can also show that the variance of the sum of the {$\Delta W_{t(i)}^2$} is
\begin{equation}
V\bigg(\sum_i^N \Delta W_{t(i)}^2\bigg)=\sum_i^N2\Delta t^2=2\frac{T^2}{N}.
\end{equation}
Since this quantity vanishes as $N\rightarrow\infty$, we see that the sum of all the $\Delta W_{t(i)}^2$ \emph{in the continuum limit} is not actually random, but deterministic, and must therefore be equal to its mean. Let us write this explicitly. In the continuum limit, sums of increments are promoted to integrals of differentials. We thus have
\begin{equation}
\int_0^T\d W_t^2=\left\langle\int_0^T\d W_t^2\right\rangle=\int_0^T\langle \d W_t^2\rangle=\int_0^T\d t,
\end{equation}
which leads to
\begin{equation}
\d W_t^2=\d t.
\end{equation}
This is a central property of stochastic increments known as It\^o's rule. We can use this result to change the time variable, {$t$,} in a stochastic differential equation {to another one, $s$} via
\begin{equation}
\d W_t=\sqrt{\frac{\d t}{\d s}}\d W_s.
\end{equation}
We use this result in Section \ref{sec:matrix} to change the time variable from cosmic time to $e$-folds.

It\^o's rule also {allows us to redefine the dependent variable of an stochastic differential equation.} Suppose, for instance, that we want to write Eq.~\eqref{ec:inc0} in terms of a certain function $z(y,t)$. Let us start by considering the Taylor expansion of $z(y,t)$ to {second order in} non-infinitesimal increments of the stochastic variable $y$ and time $t$,
\begin{equation}
\label{eq:wiener_z}
\Delta z=\frac{\partial z}{\partial t}\Delta t+\frac{\partial z}{\partial y}\Delta y+\frac{1}{2}\frac{\partial^2z}{\partial y^2}\Delta y^2+\mathcal{O}(\Delta t^2, \Delta y^3).
\end{equation}
Using Eq.~\eqref{ec:inc1} to substitute $\Delta y$ and $\Delta y^2$ into Eq.~(\ref{eq:wiener_z}), we get
\begin{equation}
\Delta z = \left(\frac{\partial z}{\partial t}+f\frac{\partial z}{\partial y}\right)\Delta t+g\frac{\partial z}{\partial y}\Delta W_t + \frac{g^2}{2}\frac{\partial^2 z}{\partial y^2}\Delta W_t^2+\mathcal{O}(\Delta t^2, \Delta W_t^3, \Delta W_t^2\Delta t).
\end{equation}
Dividing by $\Delta t$,
\begin{equation}
\frac{\Delta z}{\Delta t} = \left(\frac{\partial z}{\partial t}+f\frac{\partial z}{\partial y}\right)+g\frac{\partial z}{\partial y}\frac{\Delta W_t}{\Delta t} + \frac{g^2}{2}\frac{\partial^2 z}{\partial y^2}\frac{\Delta W_t^2}{\Delta t}+\mathcal{O}\left(\Delta t, \frac{\Delta W_t^3}{\Delta t},\Delta W_t^2\right).
\end{equation}
We can now take the continuum limit of this equation. The ratio ${\Delta z}{/\Delta t}$ becomes the derivative $\displaystyle{\frac{\d z}{\d t}}$. In the third term of the right-hand side, $\Delta W_t^2/\Delta t$ goes to $\displaystyle{\frac{\d W_t^2}{\d t}}$ which is equal to 1 by virtue of It\^o's rule. The terms containing $\Delta t$, $\Delta W_t^3/\Delta t$, $\Delta W_t^2$ and higher powers all vanish in the continuum limit. For $\Delta t$, this is immediate (indeed, $T/N\to 0$ as $N\to \infty$). For $\Delta W_t^3/\Delta t$, the fact that it vanishes follows from It\^o's rule: $ \Delta W_t^3/{\Delta t} = \Delta W_t \times \Delta W_t^2/ \Delta t \to 0 \times 1=0$. Invoking again  It\^o's rule, $\Delta W_t^2$ goes to $\Delta t$ in the continuum limit, and therefore it also vanishes. On the other hand, the ratio ${\Delta W_t}/{\Delta t}$ can be formally associated with the time derivative of the Wiener increment in the continuum limit, i.e.
\begin{equation}
\frac{\Delta W_t}{\Delta t} \to \frac{\d W_t}{\d t}\equiv \xi(t)
\end{equation}
This object is well defined. Moreover, it can be shown to have the following property~\cite{jacobs_2010}
\begin{equation}
\langle\xi(t)\xi(t')\rangle = 
\left\{
    \begin{array}{lr}
        0, & \text{if } t \neq t',\\
        \infty, & \text{if } t=t',
    \end{array}
\right.
\end{equation}
which allows to formally write
\begin{equation}
\langle\xi(t)\xi(t')\rangle = \delta(t-t')\,,
\end{equation}
i.e.~a standard white noise process. In the remaining of this appendix and everywhere else in this work we use the notation $\xi_t = \xi(t)$.
Taking the above considerations into account, the continuum limit version of \eqref{eq:wiener_z} reads
\begin{equation}
\label{eq:ito_lemma}
\frac{\d z}{\d t}=\left(\frac{\partial z}{\partial t}+f\frac{\partial z}{\partial y}+\frac{g^2}{2}\frac{\partial^2 z}{\partial y^2}\right)+g\frac{\partial z}{\partial y}\xi_t,
\end{equation}
or, in differential form,
\begin{equation}
\label{eq:ito_lemma_2}
\d z=\left(\frac{\partial z}{\partial t}+f\frac{\partial z}{\partial y}+\frac{g^2}{2}\frac{\partial^2 z}{\partial y^2}\right)\d t+g\frac{\partial z}{\partial y}\d W_t.
\end{equation}
This equation, valid for any $z(y,t)$, is known as It\^o's lemma, and is the stochastic equivalent of the chain rule. {As we will now see, It\^o's lemma has a very useful and direct application.}

The main quantity of interest when solving a stochastic differential equation is the probability distribution $P(y,t)$ for the stochastic variable to take the value $y$ at time $t$. This distribution can be found via the Fokker-Planck equation, which we now derive. Let us take the expectation value of both sides of Eq.~(\ref{eq:ito_lemma}) with respect to the probability distribution $P(y,t)$, assuming that $z$ is not explicitly time-dependent,
\begin{align}
\frac{\d\langle z\rangle_{\rm S}}{\d t}=\left\langle f\frac{\partial z}{\partial y}\right\rangle_{\rm S}+\left\langle\frac{g^2}{2}\frac{\partial^2 z}{\partial y^2}\right\rangle_{\rm S} \nonumber
&=\int\left(f\frac{\partial z}{\partial y}+\frac{g^2}{2}\frac{\partial^2 z}{\partial y^2}\right)P(y,t)\d y\\&=\int z\left[-\frac{\partial}{\partial y}(fP)+\frac{1}{2}\frac{\partial^2 }{\partial y^2}(g^2P)\right]\d y,
\end{align}
where we use $\langle\cdots\rangle_{\rm S}$ to denote the expectation value with respect to $P(y,t)$. Here we have used the fact that $\langle \d W_t\rangle_{\rm S}=0$ in the first step and integrated by parts in the last step assuming that the probability distribution vanishes at the boundaries. On the other hand, the time derivative of the mean of $z$ is also given by
\begin{equation}
\frac{\d\langle z\rangle_{\rm S}}{\d t}=\frac{\d}{\d t}\int z(y)P(y,t)\d y=\int z(y)\frac{\partial}{\partial t}P(y,t)\d y.
\end{equation}
Since this must be true for any $z(y)$, we can equate both expressions to find the Fokker-Planck equation for the probability density,
\begin{equation}
\label{eq:fokker_easy}
\frac{\partial}{\partial t}P(y,t)=-\frac{\partial}{\partial y}\Big[f(y,t)P(y,t)\Big]+\frac{1}{2}\frac{\partial^2}{\partial y^2}\Big[g^2(y,t)P(y,t)\Big].
\end{equation}
This is a deterministic equation and {therefore standard methods for partial differential equations may be applied to it.}

\subsection{Correlators in Fourier space}

Let us assume that the stochastic variable $y$ depends not only on the time $t$ but also on the spatial coordinates ${\bm x}$. As mentioned earlier, {the increments defining} Wiener processes are often written as $\d W_t=\xi_t\d t$. The quantity $\xi_t$ can be thought of as a distribution, with correlation function given by
\begin{equation}
\langle\xi_t({\bm x})\xi_{t'}({\bm x}')\rangle_{\rm S}=\delta(t-t')\delta^3({\bm x}-{\bm x}').
\end{equation}
The noise correlators in Fourier space are then
\begin{align}
\langle\xi_t({\bm k})\xi_{t'}^\star({\bm k}')\rangle_{\rm S}&=\int \d^3{\bm x}\int \d^3{\bm x}'\langle\xi_t({\bm x})\xi_{t'}({\bm x}')\rangle_{\rm S}\, e^{i{\bm k}{\bm x}}e^{-i{\bm k}'{\bm x}'}=(2\pi)^3\delta(t-t')\delta^3_-,
\end{align}
where we have defined the shorthand $\delta^3_\pm=\delta^3({\bm k}\pm{\bm k}')$. Following a similar procedure for the other entries, the entire correlation matrix can be computed
\begin{equation}
\bigg\langle\begin{pmatrix}
\label{eq:xi_star}
\xi_t({\bm k}) \\ \xi^\star_t({\bm k})
\end{pmatrix}
\begin{pmatrix}
\xi_{t'}({\bm k}') & \xi^\star_{t'}({\bm k}')
\end{pmatrix}\bigg\rangle_{\rm S}=(2\pi)^3
\begin{pmatrix}
\delta^3_+ & \delta^3_- \\
\delta^3_- & \delta^3_+
\end{pmatrix}\delta(t-t').
\end{equation}
For the real and imaginary parts of the noise ${\rm Re}(\xi_t)\equiv\xi^r_t$ and ${\rm Im}(\xi_t)\equiv\xi^i_t$, we find
\begin{equation}
\label{deltastructure}
\bigg\langle\begin{pmatrix}
\xi^r_t({\bm k}) \\ \xi^i_t({\bm k})
\end{pmatrix}
\begin{pmatrix}
\xi^r_{t'}({\bm k}') & \xi^i_{t'}({\bm k}')
\end{pmatrix}\bigg\rangle_{\rm S}=
\frac{1}{2}(2\pi)^3\begin{pmatrix}
\delta^3_-+\delta^3_+ & 0 \\
0 & \delta^3_--\delta^3_+
\end{pmatrix}\delta(t-t').
\end{equation}
In computing these entries it is necessary to use the parity of the $\delta$-function, $\delta^3({\bm k})=\delta^3(-{\bm k})$. We therefore find that the real and imaginary parts of the noise are uncorrelated, but $\xi_t$ and $\xi_t^\star$ are not.

\subsection{Derivation of the matrix equation}

Throughout the rest of this appendix we will focus on a specific multivariate version of Eq.~(\ref{sdeom}) which is close to the type of equations we deal with {in our scenario.}
Let us consider\footnote{Stochastic equations of this form are useful at linear order in perturbation theory and in Fourier space, where spatial derivatives are effectively decoupled. In particular, the system in Eqs.~(\ref{eq:phieom}) -- (\ref{eq:qeom}) is of this form.}
\begin{equation}
\label{eq:matrixeom}
\frac{\d\bm \Phi	}{\d t}+{\bm A}\bm \Phi	=\frac{1}{\sqrt{2}}{\bm \sigma}{\bm \xi}_t,
\end{equation}
where the stochastic time-dependent variable $\bm \Phi	(t)$ is an $n$-dimensional column complex vector, ${\bm A}$ is an $n\times n$ real matrix, ${\bm \sigma}$ is an $n\times m$ complex matrix, and ${\bm \xi}_t$ is an $m$-dimensional column complex noise vector whose real and imaginary parts have correlation functions given by the multivariate analogue of Eq.~(\ref{deltastructure}). We have absorbed the overall $1/2$ factor from Eq.~(\ref{deltastructure}) into the definition of ${\bm \xi}_t$, leading to the $1/\sqrt{2}$ factor in Eq.~(\ref{eq:matrixeom}).

We want to find the probability distribution $P(\bm \Phi	,\bm \Phi	^\star,t)$, where $\bm \Phi	^\star$ obeys the equation of motion
\begin{equation}
\frac{\d\bm \Phi	^\star}{\d t}+{\bm A}\bm \Phi	^\star=\frac{1}{\sqrt{2}}{\bm \sigma}{\bm \xi}_t^\star.
\end{equation}
The noise vectors ${\bm \xi}_t$ and ${\bm \xi}_t^\star$ are not independent as per Eq.~(\ref{eq:xi_star}). In order to use a generalized multivariate version of the Fokker-Planck equation (\ref{eq:fokker_easy}) for the probability distribution $P(\bm \Phi	,\bm \Phi	^\star,t)$, we need to rewrite Eq.~(\ref{eq:matrixeom}) in terms of uncorrelated noise sources. To this end, it is convenient to define the vectors $\bm \Psi	\equiv(\bm \Phi	^{\rm T},\bm \Phi	^{\dagger})^{\rm T}$ and ${\bm \chi}_c\equiv({\bm \xi}_t^{\rm T},{\bm \xi}_t^{\dagger})^{\rm T}$. The equation of motion for $\bm \Psi	$ is then
\begin{equation}
\frac{\d\bm \Psi	}{\d t}+{\bm \alpha}\bm \Psi	=\frac{1}{\sqrt{2}}{\bm \Sigma}{\bm \chi}_c,
\end{equation}
where
\begin{equation}
{\bm \alpha}=
\begin{pmatrix}
{\bm A} & 0 \\
0 & {\bm A}
\end{pmatrix},\qquad
{\bm \Sigma}=
\begin{pmatrix}
{\bm \sigma} & 0 \\
0 & {\bm \sigma}
\end{pmatrix}\,,
\end{equation}
{where $\bm{\alpha}$ and $\mathbf{\Sigma}$ do not depend on $\mathbf{\Psi}$.} This notation might make ${\bm \alpha}$ and ${\bm \Sigma}$ look like they have the same shapes, but ${\bm \alpha}$ is a $2n\times 2n$ matrix, whereas ${\bm \Sigma}$ is a $2n\times 2m$ matrix (the $0$ matrices inside have different shapes). The final step is to write this equation with correlated noises ${\bm \chi}_c$ (hence the subscript $c$) in terms of the uncorrelated noises ${\bm \chi}_u\equiv({\bm \xi}_t^{r{\rm T}},{\bm \xi}_t^{i{\rm T}})^{\rm T}$. We have
\begin{equation}
\label{psi_mat_eom}
\frac{\d\bm \Psi	}{\d t}+{\bm \alpha}\bm \Psi	=\frac{1}{\sqrt{2}}{\bm \Sigma}{\bm M}_\chi{\bm \chi}_u,
\end{equation}
with
\begin{equation}
{\bm M}_\chi=
\begin{pmatrix}
1 &  i \\
1 & -i
\end{pmatrix}
\quad\longrightarrow\quad
{\bm \Sigma}{\bm M}_\chi=
\begin{pmatrix}
{\bm \sigma} & i{\bm \sigma} \\
{\bm \sigma} & -i{\bm \sigma}
\end{pmatrix}.
\end{equation}
Since the noise sources in Eq.~(\ref{psi_mat_eom}) are uncorrelated, Eq.~(\ref{eq:fokker_easy}) can be generalized directly,
\begin{equation}
\frac{\partial P}{\partial t}=\sum_{k\ell}\bigg\{{\bm \alpha}_{k\ell}\frac{\partial}{\partial \bm \Psi	_k}(\bm \Psi	_{\ell} P)+\frac{1}{2}\bigg[\frac{1}{\sqrt{2}}{\bm \Sigma}{\bm M}_\chi\frac{1}{\sqrt{2}}{\bm M}_\chi^{\rm T}{\bm \Sigma}^{\rm T}\bigg]_{k\ell}\frac{\partial^2P}{\partial \bm \Psi	_k\partial \bm \Psi	_\ell}\bigg\},
\label{eq:Fokker-Planck}
\end{equation}
with
\begin{equation}
{\bm \Sigma}{\bm M}_\chi{\bm M}_\chi^{\rm T}{\bm \Sigma}^{\rm T}=
\begin{pmatrix}
{\bm \sigma} & i{\bm \sigma} \\
{\bm \sigma} & -i{\bm \sigma}
\end{pmatrix}\begin{pmatrix}
{\bm \sigma}^{\rm T} & {\bm \sigma}^{\rm T} \\ 
i{\bm \sigma}^{\rm T} & -i{\bm \sigma}^{\rm T}
\end{pmatrix}=
\begin{pmatrix}
0 & 2{\bm \sigma}{\bm \sigma}^{\rm T} \\
2{\bm \sigma}{\bm \sigma}^{\rm T} & 0
\end{pmatrix}.
\end{equation}

The sums in Eq.~(\ref{eq:Fokker-Planck}) can be expanded in terms of $\bm \Phi	$ and $\bm \Phi	^\star$ instead of $\bm \Psi	$. We find
\begin{equation}
\label{eq:fokker_full}
\frac{\partial P}{\partial t}=\sum_{k\ell}\bigg[{\bm A}_{k\ell}\frac{\partial}{\partial \bm \Phi	_k}(\bm \Phi	_{\ell} P)+{\bm A}_{k\ell}\frac{\partial}{\partial \bm \Phi	^\star_k}(\bm \Phi	^\star_{\ell} P)+({\bm \sigma}{\bm \sigma}^{\rm T})_{k\ell}\frac{\partial^2P}{\partial \bm \Phi	_k\partial \bm \Phi	^\star_\ell}\bigg]\,,
\end{equation}
{where the sum over each of $k$ and $l$ runs half the range than in \eq{eq:Fokker-Planck}.}

Finally, the equation of motion for the two-point statistical moments ${\bm Q}\equiv\langle\bm \Phi	\bm \Phi	^\dagger\rangle_{\rm S}$, defined via
\begin{equation}
\langle \bm \Phi	\bm \Phi	^\dagger\rangle_{\rm S}(t)\equiv \int \prod_i \diff \bm \Phi_i \int  \prod_j	\diff \bm \Phi^\star_j\;\bm \Phi	\bm \Phi	^\dagger\; P(\bm \Phi	,\bm \Phi^\star,t)\,,
\end{equation}
can be found by acting on this equation with a time derivative and using the Fokker-Planck equation (\ref{eq:fokker_full}). It is also necessary to integrate by parts and assume that the probability distribution vanishes at the boundaries. The result is
\begin{equation}
\frac{\d{\bm Q}}{\d t}=-{\bm A}{\bm Q}-{\bm Q}{\bm A}^{\rm T}+{\bm \sigma}{\bm \sigma}^{\rm T}.
\end{equation}
If there is only one source of noise then ${\bm \sigma}={\bm B}$ is a column vector and
\begin{equation}
\frac{\d{\bm Q}}{\d t}=-{\bm A}{\bm Q}-{\bm Q}{\bm A}^{\rm T}+{\bm B}{\bm B}^{\rm T}.
\end{equation}

\section{Stochastic and quantum expectation values}
\label{app:average}

As shown in Section \ref{sec:analytical}, the solution to the simplified equation of motion for $\delta\phi$ (\ref{deltaphieom}) is of the form
\begin{equation}
\label{phi_sol_app}
\delta\phi_k(N)=\delta\phi_k^{(h)}(N)+\int^N_{-\infty}\mathcal{S}_{k}(N,\hat{N})\xi_{\hat{N}}({\bm k})\d \hat{N},
\end{equation}
where $\xi_{N}$ {defines} a white noise process satisfying (\ref{eq:xi_star}) and $\mathcal{S}_{k}$ is a function of time, the form of which is not relevant in what follows. We can quantize the field ($\delta\phi_k\rightarrow\delta\hat{\phi}_k$) by writing the homogeneous solution in terms of creation and annihilation operators
\begin{equation}
\label{a_operators}
\delta\hat{\phi}^{(h)}_k=\delta\phi^{(h)}_k\hat{a}_k+\delta\phi^{(h)\star}_k\hat{a}^\dagger_k.
\end{equation}
Since we do not quantize the noise, we make the second term in Eq.~(\ref{phi_sol_app}) proportional to the identity operator. The quantum expectation value is defined via $\langle\cdots\rangle_{\rm Q}\equiv\langle 0|\cdots|0 \rangle$, {where $|0 \rangle$ defines the vacuum of the system,}  so we have
\begin{equation}
\label{q_expect}
\langle\delta\hat{\phi}_k\delta\hat{\phi}_q\rangle_{\rm Q}=\langle\delta\hat{\phi}^{(h)}_k\delta\hat{\phi}^{(h)}_q\rangle_{\rm Q}+\int_{-\infty}^N\int_{-\infty}^N\mathcal{S}_k(N,\hat{N})\mathcal{S}_q(N,\tilde{N})\xi_{\hat{N}}({\bm k})\xi_{\tilde{N}}(\bm q)\d\hat{N}\d\tilde{N}.
\end{equation}
The power spectrum is defined via
\begin{equation}
\label{spectrum_stoch_app}
\langle\delta\hat{\phi}_k\delta\hat{\phi}_q\rangle_{\rm Q}\equiv(2\pi)^3\frac{2\pi^2}{k^3}\mathcal{P}_{\delta\phi}(k)\delta^3({\bm k}+{\bm q})\,.
\end{equation}
We can show from Eq.~(\ref{a_operators}) that
\begin{equation}
\langle\delta\hat{\phi}^{(h)}_k\delta\hat{\phi}^{(h)}_q\rangle_{\rm Q}=(2\pi)^3|\delta\phi^{(h)}_k|^2\delta^3({\bm k}+{\bm q}),
\end{equation}
and since $\delta\phi^{(h)}_k$ is the solution to the homogeneous equation of motion for $\delta\phi$, which does not involve the stochastic noise, $\langle\delta\hat{\phi}^{(h)}_k\delta\hat{\phi}^{(h)}_q\rangle_{\rm Q}$ is a deterministic quantity. Note, however, that the second term in Eq.~(\ref{q_expect}) is still stochastic, as it should be, since every realization of the system leads to a different power spectrum. To obtain a deterministic quantity we can take the stochastic expectation value $\langle\cdots\rangle_{\rm S}$, computed by averaging over many realizations. We denote the double expectation value by brackets without subindices, $\langle\cdots\rangle\equiv\langle\langle\cdots\rangle_{\rm Q}\rangle_{\rm S}$. Thus,
\begin{align}
\langle\delta\hat{\phi}_k\delta\hat{\phi}_q\rangle&=(2\pi)^3|\delta\phi^{(h)}_k|^2\delta^3({\bm k}+{\bm q})+\int_{-\infty}^N\int_{-\infty}^N\mathcal{S}_k(N,\hat{N})\mathcal{S}_q(N,\tilde{N})\langle\xi_{\hat{N}}({\bm k})\xi_{\tilde{N}}({\bm q})\rangle_{\rm S}\d\hat{N}\d\tilde{N}\nonumber\\
&=(2\pi)^3|\delta\phi^{(h)}_k|^2\delta^3({\bm k}+{\bm q})+\int_{-\infty}^N\int_{-\infty}^N\mathcal{S}_k(N,\hat{N})\mathcal{S}_q(N,\tilde{N})(2\pi)^3\delta(\hat{N}-\tilde{N})\delta^3({\bm k}+{\bm q})\d\hat{N}\d\tilde{N}\nonumber\\
&=(2\pi)^3\bigg(|\delta\phi^{(h)}_k|^2+\int_{-\infty}^N\mathcal{S}_k(N,\hat{N})^2\d\hat{N}\bigg)\delta^3({\bm k}+{\bm q}).
\end{align}
By combining this with Eq.~(\ref{spectrum_stoch_app}), we find
\begin{equation}
\langle\mathcal{P}_{\delta\phi}(k)\rangle_{\rm S}=\frac{k^3}{2\pi^2}\bigg(|\delta\phi^{(h)}_k|^2+\int_{-\infty}^N\mathcal{S}_k(N,\hat{N})^2\d\hat{N}\bigg).
\end{equation}
This formula holds as long as we assume that the noise term is not quantized and the stochastic and quantum expectation values are independent from each other. 

The four-point function for $\delta\phi$ can be found in a similar manner. Since these results bear direct relation to the calculation of Section \ref{sec:gw_main}, we present them in conformal time. If we assume that $\delta\phi^{(h)}_k$ is Gaussian with respect to the quantum expectation value, and $\xi_\eta$ is Gaussian with respect to the stochastic one, then the following identities follow from Wick's theorem
\begin{align}
\langle\delta\hat{\phi}^{(h)}_{q}(\eta)\delta\hat{\phi}^{(h)}_{k-q}(\eta)\delta\hat{\phi}^{(h)}_{l}(\eta')\delta\hat{\phi}^{(h)}_{p-l}(\eta')\rangle&=(2\pi)^6\delta^3_{k+p}\Big(\delta^3_{q+l}+\delta^3_{q+p-l}\Big)\nonumber\\
&\qquad\delta\phi_q^{(h)}(\eta)^\star\delta\phi_l^{(h)}(\eta')\delta\phi^{(h)}_{k-q}(\eta)^\star\delta\phi^{(h)}_{p-l}(\eta'),\\
\langle\xi_\eta({\bm q})\xi_{\eta'}({\bm k}-{\bm q})\xi_{\eta''}({\bm l})\xi_{\eta'''}({\bm p}-{\bm l})\rangle&=(2\pi)^6\delta^3_{k+p}\delta_{\eta+\eta'-\eta''-\eta'''}\Big(\delta^3_{q+p-l}\delta_{\eta-\eta'''}+\delta^3_{q+l}\delta_{\eta-\eta''}\Big),\\
\langle\xi_\eta({\bm q})\delta\hat{\phi}^{(h)}_{k-q}(\eta')\xi_{\eta''}({\bm l})\delta\hat{\phi}^{(h)}_{p-l}(\eta''')\rangle&=(2\pi)^6\delta^3_{k+p}\delta^3_{q+l}\delta_{\eta-\eta''}\delta\phi^{(h)}_{k-q}(\eta')^\star\delta\phi^{(h)}_{p-l}(\eta'''),
\end{align}
where we have introduced the shorthand notation $\delta^3({\bm k})=\delta_k^3$ for readability, and we have once again assumed that the quantum and stochastic expectation values are independent from each other.
The four-point function {of $\delta \phi$ we are interested in} is given by the sum of four terms,
\begin{equation}
\langle\delta\hat{\phi}_{q}(\eta')\delta\hat{\phi}_{k-q}(\eta')\delta\hat{\phi}_{l}(\eta'')\delta\hat{\phi}_{p-l}(\eta'')\rangle=(2\pi)^6\bigg(\sum_{i=1}^4F_{\delta\phi}^{(i)}\bigg)\delta^3({\bm k}+{\bm p}),
\end{equation}
where\footnote{
We have ignored some contact interaction terms that are not proportional to $\delta^3({\bm k}+{\bm p})$. The upper integration limit arises because, if $a<c<b$, then
\begin{equation}
\int_a^b\int_a^cf(\eta)\delta(\eta-\eta')\d\eta \d\eta'=\int_a^c\int_a^cf(\eta)\delta(\eta-\eta')\d\eta \d\eta'+\underbrace{\int_c^b\int_a^cf(\eta)\delta(\eta-\eta')\d\eta \d\eta'}_{0}=\int_a^cf(\eta)\d\eta,
\end{equation}
since $a<\eta<c$ but $c<\eta'<b$, so that the regions in the second double integral after the first equality do not overlap, making it impossible to satisfy the constraint $\eta=\eta'$ imposed by the $\delta$ function and forcing the integral to vanish. An analogous result holds for $a<b<c$.
}
\begin{align} \displaybreak[0]
F_{\delta\phi}^{(1)}&=\Big(\delta^3_{q+l}+\delta^3_{q+p-l}\Big)\delta\phi^{(h)}_q(\eta')^\star\delta\phi^{(h)}_q(\eta'')\delta\phi^{(h)}_{k-q}(\eta')^\star\delta\phi^{(h)}_{k-q}(\eta''),\\ \displaybreak[0]
F_{\delta\phi}^{(2)}&=\Big(\delta^3_{q+l}+\delta^3_{q+p-l}\Big) \delta\phi^{(h)}_{q}(\eta')^\star\delta\phi^{(h)}_{q}(\eta'')\int_{-\infty}^{{\rm min}(\eta',\eta'')}\mathcal{S}_{k-q}(\eta'',\hat{\eta})\mathcal{S}_{k-q}(\eta',\hat{\eta})\d\hat{\eta},\\ \displaybreak[0]
F_{\delta\phi}^{(3)}&=\Big(\delta^3_{q+l}+\delta^3_{q+p-l}\Big)\delta\phi^{(h)}_{k-q}(\eta')^\star\delta\phi^{(h)}_{ k-q}(\eta'')\int_{-\infty}^{{\rm min}(\eta',\eta'')}\mathcal{S}_{q}(\eta'',\hat{\eta})\mathcal{S}_{q}(\eta',\hat{\eta})\d\hat{\eta},\\ \displaybreak[0]
F_{\delta\phi}^{(4)}&=\Big(\delta^3_{q+l}+\delta^3_{q+p-l}\Big)\int_{-\infty}^{{\rm min}(\eta',\eta'')}\mathcal{S}_{q}(\eta',\hat{\eta})\mathcal{S}_{q}(\eta'',\hat{\eta})\d\hat{\eta}\int_{-\infty}^{{\rm min}(\eta',\eta'')}\mathcal{S}_{k-q}(\eta',\tilde{\eta})\mathcal{S}_{k-q}(\eta'',\tilde{\eta})\d\tilde{\eta}.
\end{align}

\section{Numerically stable system of equations}
\label{app:eqs}

As discussed in Section \ref{sec:matrix}, we find that ignoring the constraint in Eq.~(\ref{qconstraint}) and keeping Eq.~(\ref{eq:qeom}) for $\delta q_r=4\rho_r\delta v_r/3$ is often numerically more stable. Here we present the matrices for the corresponding $5\times 5$ system in the language of Section \ref{sec:matrix}. We have checked that this system of equations yields the same results as the $4\times4$ system.

Let $\bm \Phi	=(\delta\rho_r,\delta q_r,\psi,\delta\phi,\d\delta\phi/\d N){^{\rm T}}$. One should be careful with the order of the variables when comparing with the results of Section \ref{sec:matrix}. The matrix ${\bm A}$ is
\begin{equation}
{\bm A}=
\begin{pmatrix}
G_\rho +4\rho_r f_\rho& -H k^2/(a^2H^2) & G_\psi +4\rho_r f_\psi & G_\phi +4\rho_r f_\phi& G_{\d\phi} +4\rho_r f_{\d\phi}\\

1/(3H) & 3 & 4\rho_r/(3H) & \Gamma (\d\phi/\d N) & 0 \\

f_\rho & 0 & f_\psi & f_\phi & f_{\d\phi} \\

0 & 0 & 0 & 0 & -1 \\

h_\rho +4(\d\phi/\d N)f_\rho& 0 & h_\psi +4(\d\phi/\d N)f_\psi& h_\phi +4(\d\phi/\d N)f_\phi& h_{\d\phi} +4(\d\phi/\d N)f_{\d\phi}
\end{pmatrix},
\end{equation}
where
\begin{align}
G_\rho&=g_\rho+\frac{k^2}{3a^2H^2},\\
G_\psi&=g_\psi+\frac{k^2}{3a^2} \bigg[2M_p^2\frac{k^2}{a^2H^2}-\bigg(\frac{\d\phi}{\d N}\bigg)^2\bigg],\\
G_\phi&=g_\phi+\frac{k^2}{3a^2H^2}\bigg[3H^2 \frac{\d\phi}{\d N}+V_\phi\bigg],\\
G_{\d\phi}&=g_{\d\phi}+\frac{k^2}{3a^2}\frac{\d\phi}{\d N},
\end{align}
and the $f$, $g$ and $h$ functions are defined as in Section \ref{sec:matrix}. The vectors ${\bm B}$ and ${\bm C}$ are
\begin{align}
{\bm B}&=\begin{pmatrix}
-(\d\phi/\d N)\sqrt{2\Gamma T H/a^3} \\
0 \\
0 \\
0 \\
\sqrt{2\Gamma T/(aH)^3} \\
\end{pmatrix},
&
{\bm C}&=\frac{1}{3H^2(\d\phi/\d N)^2+4\rho_r}
\begin{pmatrix}
0 \\
3H \\
-3H^2(\d\phi/\d N)^2-4\rho_r \\
-3H^2(\d\phi/\d N)\\
0
\end{pmatrix}.
\end{align}

Finally, we assume $\delta q_r=0$ initially, so the initial conditions matrix ${\bm Q}_i \equiv {\bm Q} (N_\text{ini})$ is
\begin{equation}
{\bm Q}_i=
\frac{1}{2ka^2(N_\text{ini})}\begin{pmatrix}
 0 & 0 & 0 & 0 & 0\\
 0 & 0 & 0 & 0 & 0\\
 0 & 0 & 0 & 0 & 0\\
 0 & 0 & 0 & 1 & -1+i(k/k_i)\\
 0 & 0 & 0 & -1-i(k/k_i) & 1+(k/k_i)^2
\end{pmatrix}.
\end{equation}

\section{Fluctuation-dissipation theorem}
\label{app:FD}

A rigorous derivation of the stochastic term in Eq.~\eq{fdf} should be worked out in the context of the Schwinger-Keldysh formalism of non-equilibrium QFT. However, it is possible to gain some intuition on it from a classical toy model \cite{CALDEIRA1983587,Calzetta:2008iqa} in which a massive particle $P$ moving along a coordinate $q$ with momentum $p$ is coupled to a system of $N\gg 1$ harmonic oscillators with coordinates $x_i$, momenta $p_i$ and natural frequencies $\omega_i$ through coupling constants $g_i$. The Hamiltonian of the system is
\begin{equation}\label{ec:fd1}
H=\underbrace{\frac{p^2}{2M}+V(q)}_{\text{particle Hamiltonian}}
+ \underbrace{\sum_{i=1}^N\left(\frac{p_i^2}{2m_i}+\frac{1}{2}m_i\omega_i^2 x_i^2\right)}_{\text{oscillators Hamiltonian}} 
-\underbrace{ q\sum_{i=1}^Ng_ix_i + q^2\sum_{i=1}^N\frac{g_i^2}{2m_i\omega_i^2}}_{\text{interaction Hamiltonian}}.
\end{equation}
The last term of the Hamiltonian has been introduced \textit{ad hoc} to simplify forthcoming calculations. It can be physically interpreted as a correction to the potential $V(q)$ which ensures that the Hamiltonian can be written in a translationally invariant form when $V=0$ (although these details are not relevant for us). The corresponding equations of motion are\footnote{From now on, we drop the sum limits since they are always the same as in \eqref{ec:fd1}.}
\begin{align}
M\ddot{q}+V'(q)+q\sum_{i}\frac{g_i^2}{m_i\omega_i^2}&=\sum_{i}g_ix_i, \label{ec:fd2}\\
\ddot{x_i}+\omega_i^2x_i  &= \frac{g_i}{m_i}q.\label{ec:fd3}
\end{align}
Should we know the initial conditions for both $q$, $p$ and $x_i$, $p_i$, we could, in principle, integrate the full system as a mechanical problem by solving \eqref{ec:fd2} and \eqref{ec:fd3}. However, since $N\gg 1$, a more practical approach is a statistical one, in which we know $q(0)$, $p(0)$ and only some general properties of $x_i(0)$, $p_i(0)$. This suggests an interpretation of the harmonic oscillators as an environment with which the particle of interest $P$ interacts. With this approach in mind, let us look for an effective description of the motion of $P$. We start by formally solving \eqref{ec:fd3}. The solution of the homogeneous equation is spanned by the functions $\sin(\omega_i x)$ and  $\cos(\omega_ix)$. The solution to the full equation can be obtained computing a Green's function. Doing so, we get
\begin{equation}
x_i = x_i(0)\cos(\omega_i t)+\frac{p_i(0)}{m_i\omega_i}\sin(\omega_i t) + \frac{g_i}{m_i\omega_i}\int_0^t \ \text{d}s  \ \sin[\omega_i(t-s)]q(s)\,.
\end{equation}
Substituting this solution into Eq.~\eq{ec:fd2} and integrating by parts we obtain
\begin{equation}\label{ec:fd6}
M\ddot{q}+V'(q)+\int_0^t \ \text{d}s  \ \gamma(t-s)\dot{q}(s)=\xi(t),
\end{equation}
where
\begin{align}
\gamma(t) & =\sum_{i}\frac{g_i^2}{m_i\omega_i^2}\cos(\omega_i t)\,,\\
\xi(t) & =\sum_i g_i\left[\left(x_i(0)-\frac{g_i}{m_i\omega_i^2}q(0)\right)\cos(\omega_i t)+\frac{p_i(0)}{m_i\omega_i}\sin(\omega_i t)\right].\label{ec:fd7.5}
\end{align}
The function $\gamma(t)$ comes from the primitive of the sine inside the integral when integrating by parts. The term with $q(0)$ inside $\xi(t)$ comes from the boundary term of the integration by parts evaluated in the lower integration limit. Finally, the boundary term evaluated in the upper limit exactly cancels the third term in \eqref{ec:fd2}. Notice that $\gamma$ only depends on the intrinsic properties of the environment (i.e.\ the natural frequencies of the oscillators, $\omega_i$), and the interaction between the particle $P$ and the environment (i.e.\ the coupling constants $g_i$). We identify this function with the dissipative coefficient, since it appears multiplying the velocity of $P$ in the equation of motion (even if it is under the integral sign). On the other hand, $\xi$ depends on the initial conditions of the environment, whose only available description is, by fiat, statistical, so that $x_i(0)$ and $p_i(0)$ can be treated as random variables.  Hence, we can identify $\xi$ with a stochastic force. This simple example illustrates how deterministic dissipation ($\gamma$) and random fluctuations ($\xi$) arise together when coupling a system to a complicated environment.  

Furthermore, this model allows to extract information about the power spectrum of the noise $\xi$. Let us assume that the random variables $x_i(0)$ and $p_i(0)$ are homogeneously distributed along the classical orbits in phase space, and that the initial conditions of different oscillators are independent. Under these conditions, the classical virial theorem ensures that
\begin{equation}\label{ec:fd8}
m_i\omega_i^2\langle	x_i(0)x_j(0)\rangle=\frac{1}{m_i}\langle	p_i(0)p_j(0)\rangle=\delta_{ij}\langle E_i\rangle, \quad \langle	x_i(0) p_j(0)\rangle = 0,
\end{equation}
where $\langle E_i\rangle$ is the expectation value of the energy of the $i$-th harmonic oscillator at $t=0$. Furthermore, assuming the environment of oscillators is at thermal equilibrium at $t=0$, the equipartition theorem implies that
\begin{equation}\label{ec:fd9}
\langle E_i\rangle = T\,.
\end{equation}	
Let us now compute the correlation of $\xi(t)$ with $\xi(s)$. From \eqref{ec:fd7.5}, we have
\begin{equation}
\langle\xi(t)\xi(s)\rangle = \sum_{i,j}g_ig_j \langle x_i(0)x_j(0)\rangle\cos(\omega_i t)\cos(\omega_j s) + \frac{g_i}{m_i\omega_i}\frac{g_j}{m_j\omega_j}\langle p_i(0)p_j(0)\rangle\sin(\omega_i t)\sin(\omega_j s).
\end{equation}
The rest of the terms are zero either because they involve the average of one single random variable or because they involve the correlation of position and momentum. Substituting the correlators of initial conditions by their values, summing over one index and using \eqref{ec:fd9}, we get
\begin{equation}\label{ec:fd10}
\langle\xi(t)\xi(s)\rangle = T\sum_i\frac{g_i^2}{m_i\omega_i^2} \cos[\omega_i(t-s)]\,,
\end{equation}
which implies
\begin{equation}\label{ec:fd15}
\langle\xi(t)\xi(s)\rangle = T \gamma(t-s).
\end{equation}
This illustrates the main idea of the fluctuation-dissipation theorem: \textit{the power of random fluctuations $\langle\xi(t)\xi(t')\rangle$ associated to a dissipative process increases with the strength of the deterministic dissipative coefficient (and the temperature of the system)}.

In the $N\to\infty$ limit, it is common to write $\gamma$ as
\begin{equation}\label{ec:fd20}
\gamma(t)=\frac{2}{\pi}\int_0^\infty\frac{\text{d}\omega}{\omega}J(\omega)\cos(\omega t), \qquad J(\omega)=\pi\sum_i\frac{g_i^2}{2m_i\omega_i}\delta(\omega-\omega_i),
\end{equation}
where $J(\omega)$ is the so-called \emph{spectral density of modes} of the environment. Let us consider the lowest order in a Taylor expansion of the spectral density, $J(\omega)=\Gamma_0\omega$, with $\Gamma_0$ a constant. In this case $\gamma(t)= 2\Gamma_0 \delta(t)$.
For this form of the dissipative coefficient, \eqref{ec:fd6} reduces to
\begin{equation}
M\ddot{q}+V'(q)+\Gamma_0\dot{q}(t)=\xi(t),
\end{equation} 
in which $\Gamma_0$ appears explicitly as a friction term. Similarly, the correlation in \eqref{ec:fd15} yields
\begin{equation}
\langle\xi(t)\xi(s)\rangle = \Gamma_0 T \delta(t-s).
\end{equation}

\addcontentsline{toc}{section}{References}
\bibliographystyle{utphys}
\bibliography{references}

\end{document}